\documentclass[prd,showpacs,superscriptaddress]{revtex4}

\usepackage{graphicx}
\usepackage[normalem]{ulem}	
\usepackage[usenames]{color}    
\usepackage{epsfig}
\usepackage{amsmath}

\def\lsim{\raise0.3ex\hbox{$\;<$\kern-0.75em\raise-1.1ex
\hbox{$\sim\;$}}}
\def\gsim{\raise0.3ex\hbox{$\;>$\kern-0.75em\raise-1.1ex
\hbox{$\sim\;$}}}

\def\be{\begin{equation}}
\def\ee{\end{equation}}
\def\ba{\begin{eqnarray}}
\def\ea{\end{eqnarray}}

\begin{document}
\title{Neutrino Mass Matrix Textures: A Data-driven Approach}
\author{E.~Bertuzzo}
\email{enrico.bertuzzo@cea.fr}
\affiliation{Institut de Physique Th\'eorique, CEA-Saclay, 91191 Gif-sur-Yvette, France}
\author{P.~A.~N.~Machado}
\email{accioly@fma.if.usp.br}
\affiliation{Institut de Physique Th\'eorique, CEA-Saclay, 91191 Gif-sur-Yvette, France}
\affiliation{
Instituto de F\'{\i}sica, Universidade de S\~ao Paulo, 
 C.\ P.\ 66.318, 05315-970 S\~ao Paulo, Brazil}
\affiliation{TH Division, Physics Department, CERN, CH-1211 Geneva 23, 
Switzerland}
\author{R.~Zukanovich Funchal} 
\email{zukanov@if.usp.br} 
\affiliation{Institut de Physique Th\'eorique, CEA-Saclay, 91191 Gif-sur-Yvette, France}
 \affiliation{
Instituto de F\'{\i}sica, Universidade de S\~ao Paulo, 
 C.\ P.\ 66.318, 05315-970 S\~ao Paulo, Brazil}
\date{January 28, 2013}

\begin{abstract} 
We analyze the neutrino mass matrix entries and their correlations in
a probabilistic fashion, constructing probability
distribution functions using the latest results from neutrino
oscillation fits. Two cases are considered: the
standard three neutrino scenario as well as the inclusion of a new
sterile neutrino that potentially explains the reactor and gallium
anomalies. We discuss the current limits and future perspectives on
the mass matrix elements that can be useful for model building.
\end{abstract} 

\maketitle

\section{Introduction}
\label{sec:intro}
The year 2012 represents a milestone in neutrino physics.  Thanks to
the measurement of the last mixing angle of the standard neutrino
oscillation scenario, $\theta_{13}$, by the reactor experiments
Double-CHOOZ~\cite{Abe:2011fz}, Daya-Bay\cite{An:2012eh} and
RENO~\cite{Ahn:2012nd}, after the first positive evidence from
accelerators~\cite{Abe:2011sj,Adamson:2011qu}, the mixing in the
leptonic sector is starting to shape up.  The impact of a rather
unexpectedly large mixing angle $\theta_{13}$ is two fold: it promotes
the discovery of CP violation in the neutrino sector to a yet daunting
but conceivable task, and at the same time it proves that the
description of neutrino oscillation data must involve all three
Standard Model neutrino flavors.

Hints of the sensitivity to CP phases are already showing their first
signs when we combine accelerator $\nu_\mu \to \nu_e$ with reactor
$\bar \nu_e \to \bar \nu_e$ data~\cite{Machado:2011ar} or when we
perform global fits~\cite{Fogli:2012ua,GonzalezGarcia:2012sz}.
Furthermore, the combined fit of neutrino oscillation data shows for
the first time a very precise and almost complete determination of the
parameters that enter the standard neutrino oscillation scheme. In
fact, in spite of the unknowns (neutrino mass hierarchy, absolute
neutrino mass scale, CP phases and the correct octant for
$\theta_{23}$) all measured parameters are now so well determined that
it is enough to quote them by giving the best fit value with the 1
$\sigma$ uncertainty.

However, not all neutrino data can be explained by this standard scenario
of three flavor neutrinos. In fact, along the years a number of 
so-called {\em anomalies} have crept into the picture. 
First, the excess of $\bar{\nu}_e$ events in the $\bar{\nu}_\mu \to \bar{\nu}_e$
mode observed by the short baseline LSND~\cite{LSND} experiment, 
now also supported by MiniBOONE data~\cite{MiniBOONE}, gave rise to the 
long-standing {\em LSND anomaly}.
Second, the deficit of $\nu_e$ compared to expectations observed by 
the source calibration experiments performed in   
the gallium radiochemical solar neutrino detectors 
GALLEX~\cite{gallex} and SAGE~\cite{sage}. This is the so-called
{\em gallium anomaly}.
Third, and more recently, a re-evaluation of the reactor $\bar \nu_e$
flux~\cite{Mueller:2011nm,Huber:2011wv} resulted in an increase of the 
total flux by 3.5\%. While this increase has essentially no impact on
the results of long baseline experiments, it induces a deficit of about 
5.7\% in the observed event rates for short baseline ($ < 100$ m) reactor 
neutrino experiments. This problem has been referred to as 
the {\em reactor antineutrino anomaly}~\cite{reactor-anomaly}.

There are attempts in the literature that try to explain some or all
of these anomalies by extending the standard picture to include one or
more sterile neutrinos~\cite{Kopp:2011qd,Giunti:2011gz,Machado:2011kt}.  These
extensions, as a rule, cannot make appearance and disappearance
experiments compatible. However, if one disregards the anomaly connected with
the appearance experiments LSND/MiniBOONE (for instance, assuming it is not 
due to oscillations), it is possible to construct
a coherent picture of all solar, atmospheric, reactor and accelerator
neutrino oscillation data adding one extra sterile neutrino to the 
standard framework. This constitutes what has been known as the 3+1 
scenario. 

Given the current status of the mixing parameters measurements, and in
view of the progress expected in the near future, we think it is
timely to analyze the possible structures and correlations among the
neutrino mass matrix elements that are compatible with data. In this
sense, we update Refs.~\cite{Frigerio:2002rd, Frigerio:2002fb} using
the most recent available data. However, our analysis will be
probabilistic, since we will construct probability distribution
functions for each element of the neutrino mass matrix.  We also
discuss how this might change with better determination of the
presently known oscillation parameters, as well as, the Dirac CP-phase
$\delta$. We hope this can be helpful to understand better the
patterns statistically preferred by data, and serve as a guide for
model builders.

We organize our paper as follows. In Sec.~\ref{sec:general-method} we
describe how we will proceed for the construction of probability
density functions (PDF) for each element of the neutrino mass matrix
and what are the assumptions in each case.  In Sec.~\ref{sec:standard}
we analyze the possible textures of the mass matrix in the standard scenario,
discussing the correlations among matrix elements in the hierarchical
and almost degenerate cases. We also discuss the future prospects
for better determining these matrix elements with neutrino oscillation
and non-oscillation data, and the possible impact on the theory.  In
Sec.~\ref{sec:3+1} we extend our analysis to include the possibility
of a sterile neutrino with mass and mixings allowed by the
reactor and gallium anomalies. We discuss what is the mass matrix
pattern in this scenario and how different it will be from the
standard case. Finally, in Sec.\ref{sec:conclusions}, we make our last
comments and draw our conclusions.

\section{Reconstructing the Neutrino Mass Matrix from Experimental Data}
\label{sec:general-method}

To access the impact of the progress on the determination of the
neutrino oscillation parameters in the last year on the knowledge of
the low energy effective neutrino mass matrix, in a probabilistic way,
we will construct a PDF for each element of the mass matrix in the
gauge basis,

\begin{equation}
m_{\alpha \beta} = \sum_i m_i \, e^{-i \lambda_{i}} \; U^{*}_{\alpha i} U^{*}_{\beta i}
\label{eq:matrix-elem}
\end{equation}
with $\alpha,\beta = e,\mu,\tau$,  $U_{\alpha i}$ being the elements of 
the mixing matrix, $m_i$ the neutrino masses and $\lambda_i$ the 
Majorana-type CP phases, using the most recent available information from 
the combination of neutrino oscillation data. Without loss of generality 
we will take $\lambda_2=0$ in our parametrization. 

We will use the following best fit points for the standard mixing parameters~\cite{GonzalezGarcia:2012sz}
\begin{eqnarray}
\Delta m^2_{21} &=& (7.50 \pm 0.185) \times 10^{-5} \, {\rm eV^{2}} \nonumber\\
\sin^2 \theta_{12}& =& 0.30 \pm 0.013 \nonumber\\
\Delta m^2_{31}  &=& (+2.47 \pm 0.07) \times 10^{-3} \, {\rm eV^{2}} \; \; {(\rm normal \; \; ordering)}\nonumber\\
\Delta m^2_{32}  &=& (-2.43 \pm 0.06) \times 10^{-3} \, {\rm eV^{2}} \; \; {(\rm inverted \; \; ordering)}\nonumber\\
\sin^2 \theta_{13}& =& 0.023 \pm 0.0023 \nonumber\\
\sin^2 \theta_{23} &=& \left \{ \begin{array}{c}
 0.41 \pm 0.037 \; \; {(\rm 1^{st} \; octant)}  \\
 0.59 \pm 0.022 \; \; {(\rm 2^{nd} \; octant)}  
\end{array} \right . \nonumber
\label{eq:data}
\end{eqnarray}
and assume these parameters to follow a normal distribution with mean at the
best fit point and standard deviation equal to the 1 $\sigma$
uncertainty.  We will take the unknown Dirac-type and Majorana-type 
CP phases to be flat distributed between 0 and $2 \pi$.

For the 3+1 scenario we will fix the squared mass difference between
the sterile and the lightest state to two different experimentally
allowed values, $\Delta m_{41}^2 = 1.71$ eV$^2$ and $\Delta m_{41}^2 =
0.95$ eV$^2$, allowing the corresponding mixing to vary with a flat
distribution inside the ranges $|U_{e4}|^2 = 8\times 10^{-3} - 4\times
10^{-2}$ and $|U_{e4}|^2 = 8\times 10^{-3} - 2.5\times 10^{-2}$,
respectively.

To construct the PDF of each $m_{\alpha \beta}$ we use a Monte Carlo method. 
For all the mixing parameters, we generate random numbers according to their   
assumed distribution,  and compute all the elements of $m_{\alpha\beta}$ in 
each case. In this manner their distribution will be naturally correlated.

Since we do not know the neutrino mass hierarchy,  the correct octant 
for $\theta_{23}$ and the absolute neutrino mass scale, we will have to 
analyze each case separately.

\section{The Standard Scenario}
\label{sec:standard}

In the standard scenario we use the standard 
parametrization for the mixing matrix,
 \begin{equation}
U = \left(\begin{array}{ccc}
c_{12} \, c_{13} & s_{12} \, c_{13} & s_{13} e^{-i \delta} \\
-s_{12}\,c_{23}-c_{12}s_{13}s_{23}\,e^{i \delta} & 
c_{12}\,c_{23}-s_{12}s_{13}s_{23}\,e^{i \delta} & c_{13}\,s_{23}\\
s_{12}\,s_{23}-c_{12}s_{13}c_{23}\,e^{i \delta} & 
-c_{12}\,s_{23}-s_{12}s_{13}c_{23}\,e^{i \delta} & c_{13}\,c_{23}
\end{array} \right),
\label{eq:matrix}
\end{equation}
with $s_{ij}=\sin \theta_{ij}$, $c_{ij}=\cos \theta_{ij}$ and $\delta$
the Dirac-type CP phase. The two additional Majorana-type CP phases
are denoted by $\lambda_1$ and $\lambda_3$, and the three neutrino
mass eigenstates are ordered either as $m_1<m_2<m_3$ (normal ordering)
or as $m_3<m_1<m_2$ (inverted ordering).

We will study six different cases: hierarchical with $m_1 \to
0$ and $\theta_{23}$ both in the first and second octant; 
hierarchical with $m_3 \to 0$ and $\theta_{23}$ in the first and
second octant; quasi-degenerate with $m_1 \sim m_2 \sim m_3 \sim 0.1$
eV and $\theta_{23}$ in the first and second octant.
There are two possible ordering also in the quasi-degenerate case; 
however, we have checked that the results are very similar.

\begin{figure}[htb]
\begin{center}
 \includegraphics[width=0.45\textwidth]{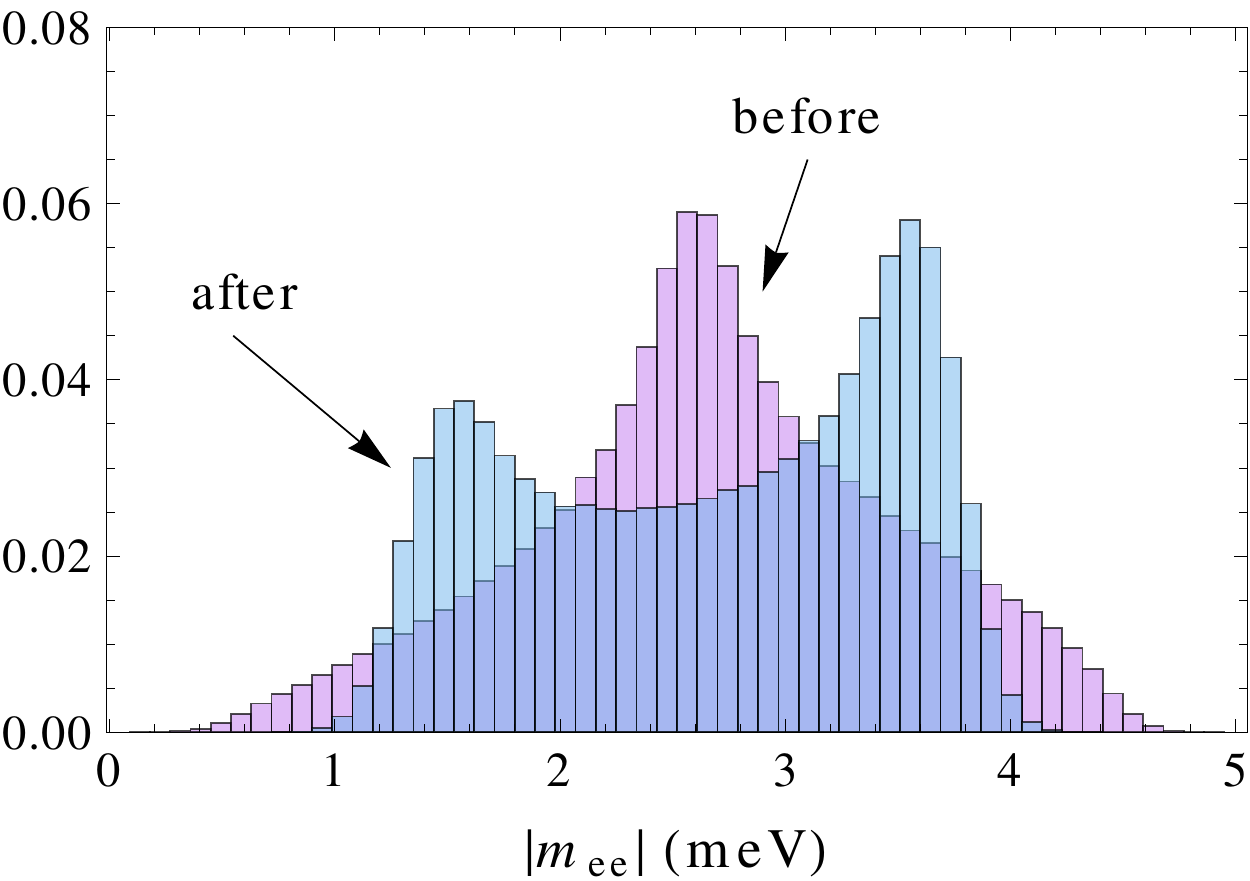}
\vspace{-2mm}
\end{center}
\vspace{-0.1cm}
\caption{PDF for $\vert m_{ee}\vert$
  when $m_1 \to 0$. The ``before'' (``after'') distribution corresponds 
 to the situation before (after) the determination of $\sin^2\theta_{13}$
  by the reactor experiments.}
\label{fig:mee-s13}
\end{figure}

In Fig.~\ref{fig:mee-s13} we illustrate, for the case $m_1 \to 0$, the
impact of the determination of $\sin^2\theta_{13}$ on the PDF of
$|m_{ee}|$.  The distribution labeled ``before'' (magenta) is
  obtained assuming $\sin^2\theta_{13}$ to be flat distributed between
  0 and 0.04 (CHOOZ limit \cite{Apollonio:2002gd}), while the one
  labeled ``after'' (blue) shows the current situation.  The two peaks in
the ``after'' distribution are due to the interference
  between the real $U_{e2}^2 m_2$ term and the complex $U_{e3}^2 m_3$
  term (see Appendix~\ref{appendixA} for detailed expressions), which
depends on the cosine of the randomly distributed CP-phases. This term
depends on $\theta_{13}$ which is not compatible with zero anymore and
thus gives a sizable contribution. The distance between the peaks
depends on $m_3$: a larger $m_3$ would place the peaks further apart.

\begin{figure}[htb]
\begin{center}
 \includegraphics[width=0.45\textwidth]{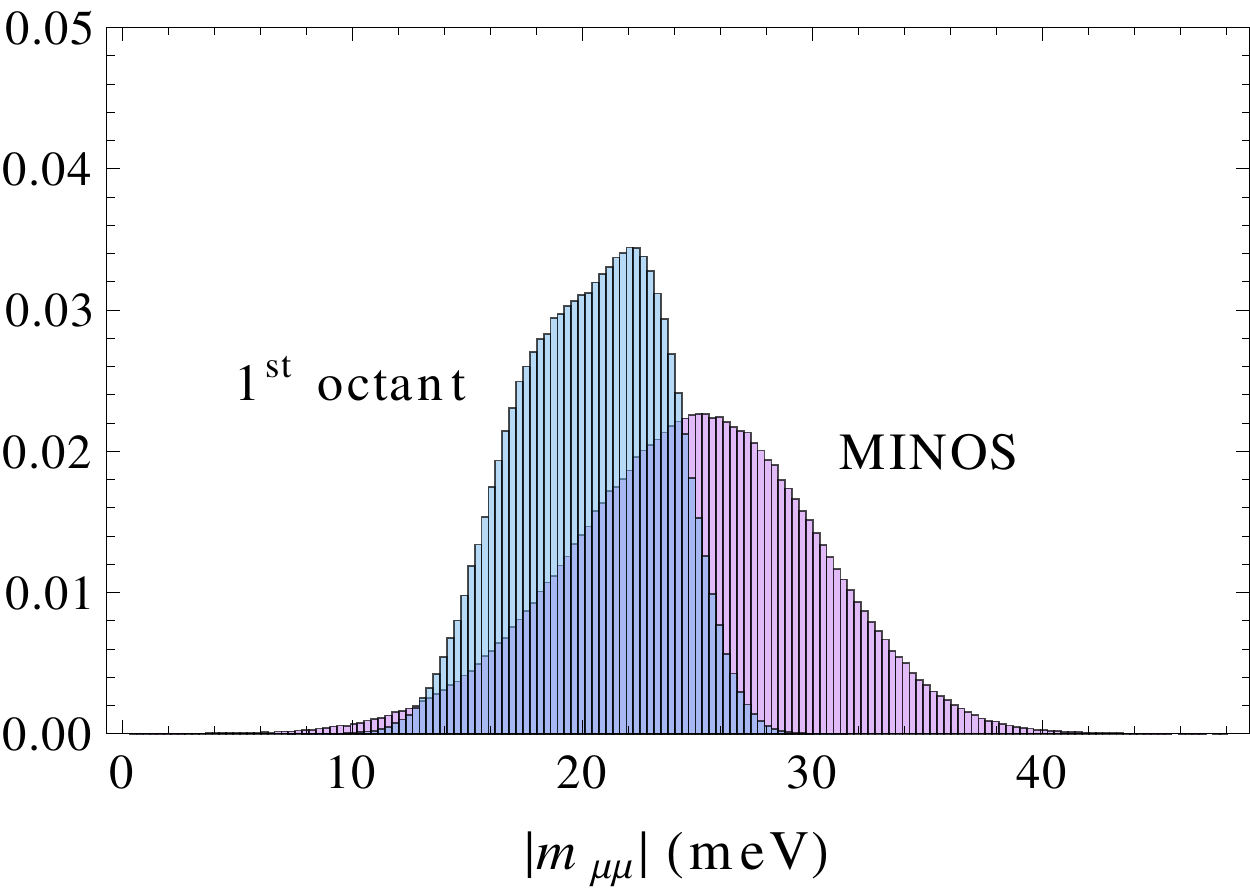}
 \includegraphics[width=0.45\textwidth]{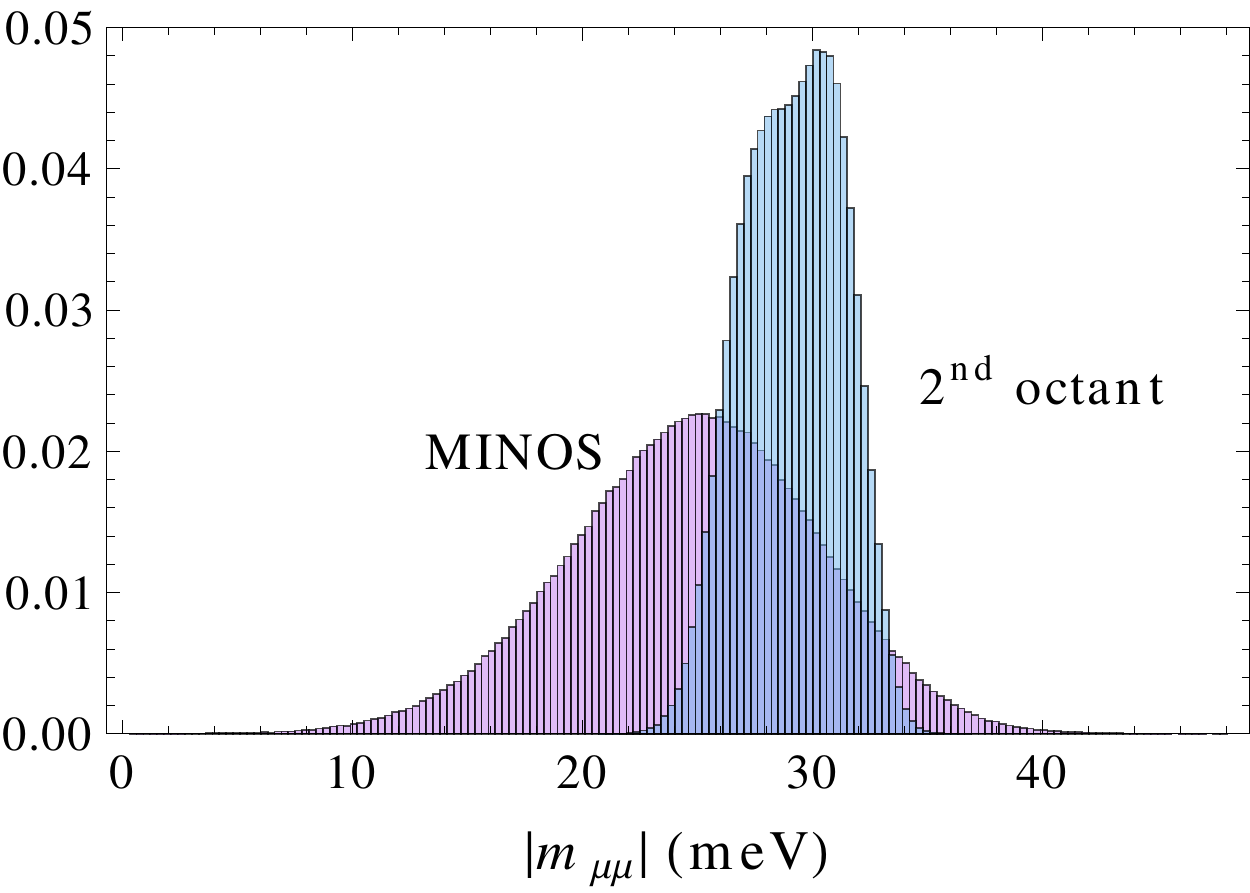}
\vspace{-2mm}
\end{center}
\vspace{-0.1cm}
\caption{PDF for $\vert m_{\mu\mu}\vert$ when $m_1 \to 0$. The left (right) 
panel corresponds to the distribution for $\theta_{23}$ in the first (second)
octant. We also show the distribution for the case of maximal mixing with 
MINOS uncertainty~\cite{Adamson:2011ig}.}
\label{fig:mmm-s23}
\end{figure}

In Fig.~\ref{fig:mmm-s23} we illustrate, again for the case $m_1 \to
0$, the effect of the determination of the $\theta_{23}$ octant on the
PDF of $\vert m_{\mu\mu}\vert$.  On the left (right) panel we show the
current distribution of $\vert m_{\mu\mu} \vert$ for the best fit
value of $\sin^2 \theta_{23}$ in the first (second) octant.  The
asymmetric two peaks structure is due to the fact that $\theta_{12}$
is not maximal. A larger $m_3$ would shift the right endpoint of
  the distribution to higher values of $\vert m_{\mu\mu}\vert$, while
a larger $m_2$ would separate the two peaks.  We observe that the
distribution on the right panel is thinner than the one on the left
panel, but this is simply because the uncertainty on $\sin^2
  \theta_{23}$ is smaller in the second octant.  For comparison, in
both panels we also show the PDF taking the MINOS result
$\sin^2\theta_{23}=0.5\pm 0.1$~\cite{Adamson:2011ig}.

\subsection{Correlations Among Matrix Elements}
\label{subsec:corr}

\subsubsection{Hierarchical Case}

In the case $m_1 \to 0$, $m_3 \approx 0.05$ eV $\gg m_2 \approx
0.009$ eV and only two CP-phases, $\delta$ and $\lambda_3$, are
relevant.  Due to the $\mu \to \tau$ symmetry, 
accomplished by $s_{23} \to c_{23}$ and $c_{23} \to -
s_{23}$, the PDFs for the solution in the first $\theta_{23}$ octant
are basically the same as for the solution in the second octant (apart
from the uncertainty in the determination of $\sin^2 \theta_{23}$,
which is smaller for the second octant) as long as we replace: $\vert
m_{e\tau}\vert \leftrightarrow \vert m_{e\mu} \vert$, $\vert
m_{\tau\tau}\vert \leftrightarrow \vert m_{\mu\mu} \vert$.

\begin{figure}[bth!]
\begin{center}
 \includegraphics[width=0.3\textwidth]{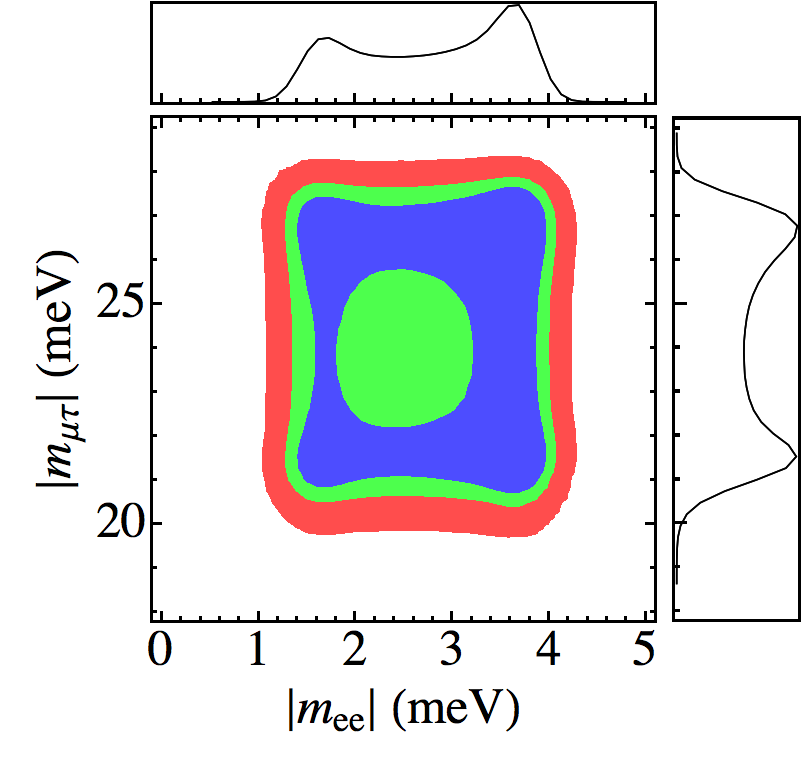}
 \includegraphics[width=0.3\textwidth]{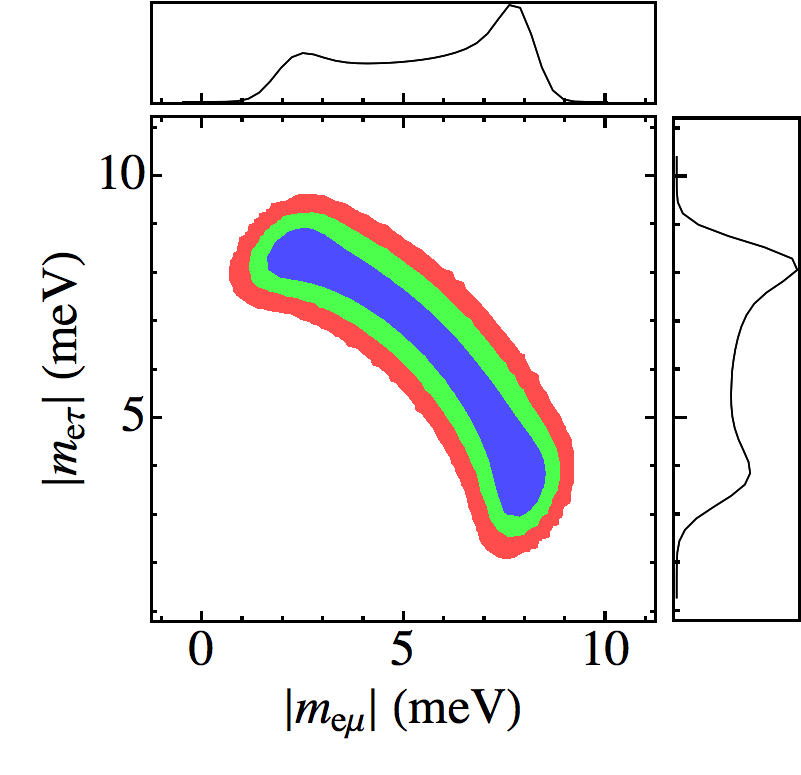}
 \includegraphics[width=0.3\textwidth]{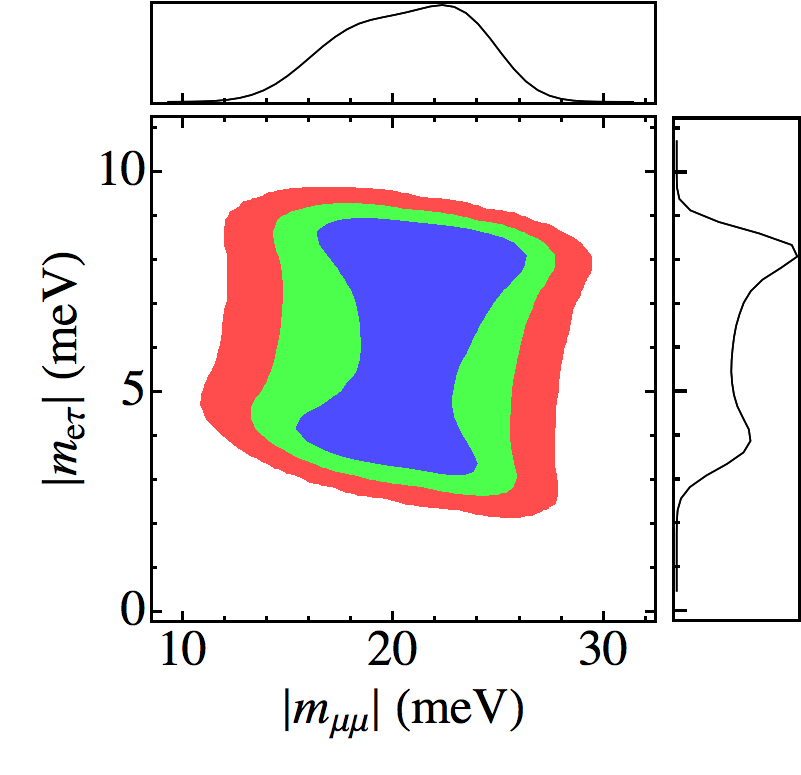}
 \includegraphics[width=0.3\textwidth]{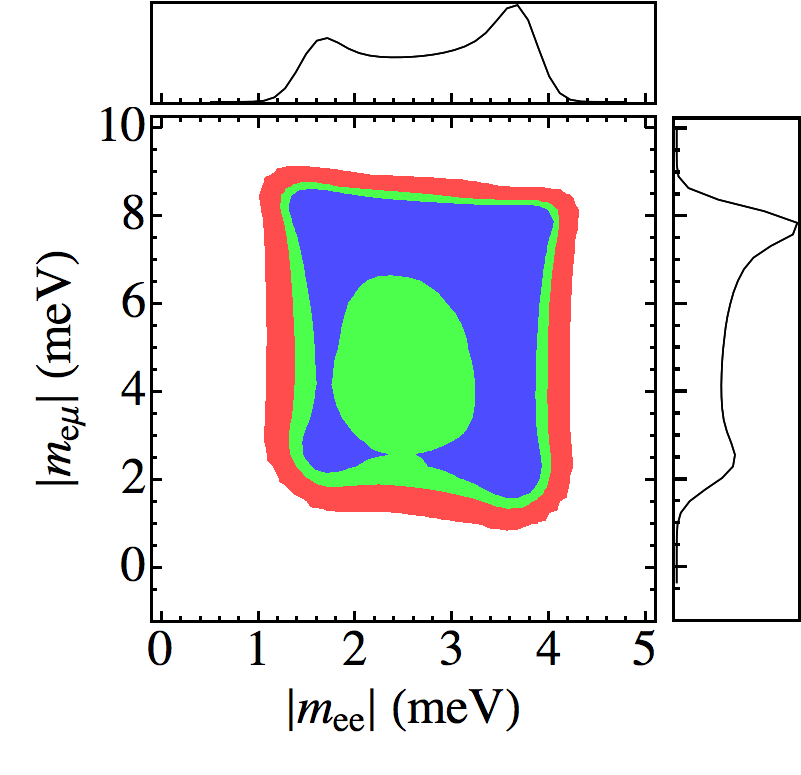}
 \includegraphics[width=0.3\textwidth]{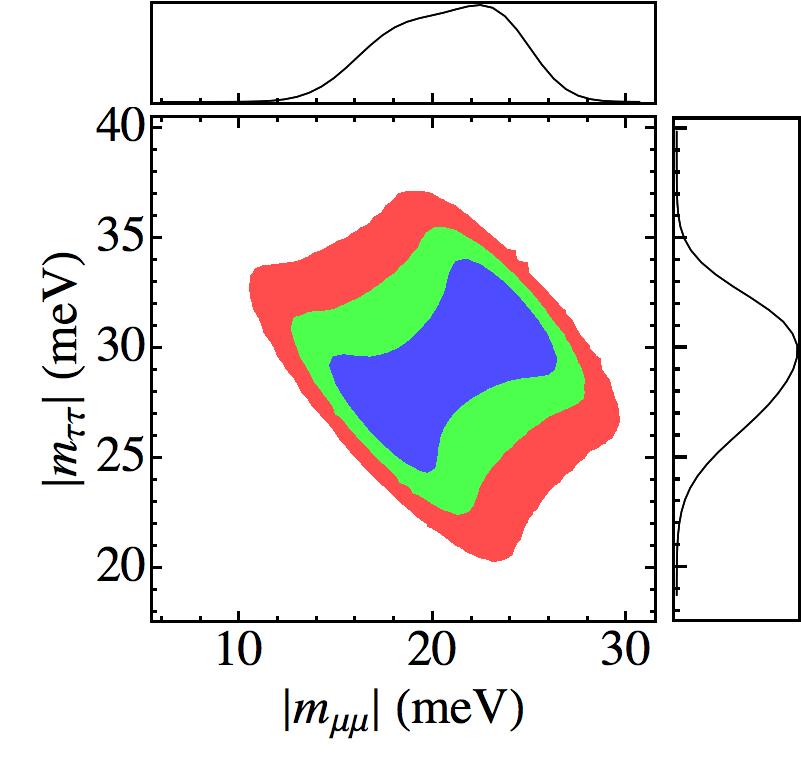}
 \includegraphics[width=0.3\textwidth]{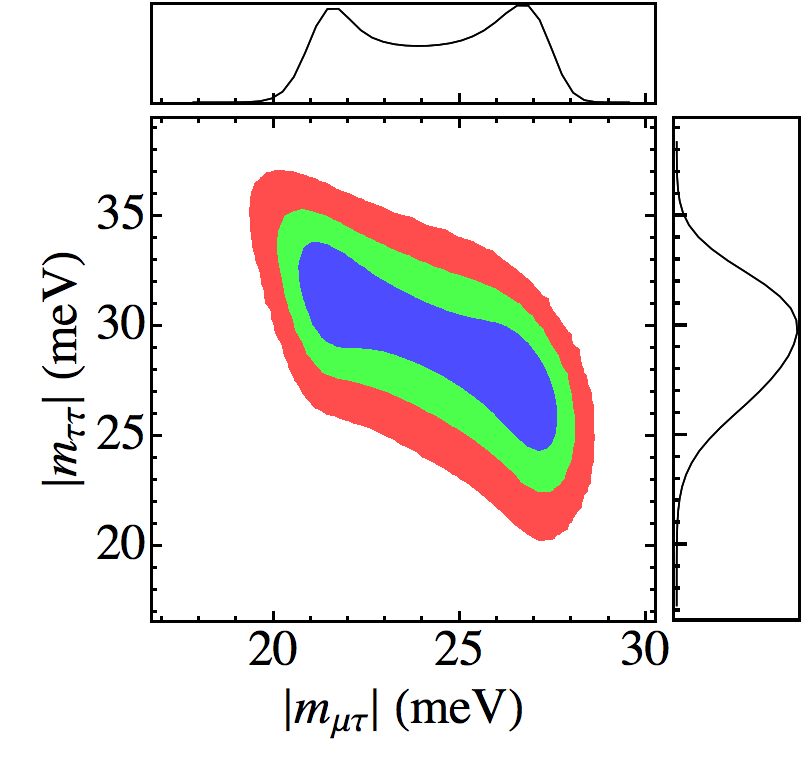}
\vspace{-2mm}
\end{center}
\vspace{-0.1cm}
\caption{PDFs for the distribution of the absolute value of several
  pairs of matrix elements. Top panels: $\vert m_{ee}\vert \times
  \vert m_{\mu \tau}\vert$ (left), $\vert m_{e\mu}\vert \times \vert
  m_{e \tau}\vert$ (center) and $\vert m_{\mu\mu}\vert \times \vert
  m_{e \tau}\vert$ (right).  Bottom panels: $\vert m_{ee}\vert \times
  \vert m_{e \mu}\vert$ (left), $\vert m_{\mu\mu}\vert \times \vert
  m_{\tau \tau}\vert$ (center) and $\vert m_{\mu\tau}\vert \times
  \vert m_{\tau \tau}\vert$ (right).  At the top and right of each two
  dimensional PDF we show the PDF of the absolute value of the
  corresponding matrix element.  In these plots we use blue, green and
  red for the allowed region at 68.27\%, 95.45\% and 99.73\% CL,
  respectively.  Here $m_1 \to 0$ and $\theta_{23}$ is assumed to be
  in the first octant.}
\label{fig:nh1st}
\end{figure}

In Fig.~\ref{fig:nh1st} we show the correlations among the absolute
values of some of the matrix elements $m_{\alpha \beta}$ for $m_1 \to
0$ and $\theta_{23}$ in the first octant. In appendix \ref{appendixB}
we present a complete set of those plots, as well as the corresponding
plots for $\theta_{23}$ taken to be in the second octant.  This was
done by constructing a two-dimensional PDF for each pair of elements.
We also present in each case on the top and to the
right of each two-dimensional distribution the projected PDFs.
In these plots we use blue, green and red for the allowed region at
68.27\%, 95.45\% and 99.73\% CL, respectively.  The range of the
values allowed at 95.45\% CL are given in Table ~\ref{tab:allowed}.

From Figs.~\ref{fig:nh1st}, \ref{fig:nh-1st} and \ref{fig:nh-2nd}
we observe that $\vert m_{ee}\vert$ is not very correlated to any other 
element. However, due to $\theta_{23}$ the pairs
$\vert m_{e\mu}\vert \times \vert m_{e\tau} \vert$,
$\vert m_{e\tau}\vert \times \vert m_{\mu\mu} \vert$,
$\vert m_{\mu\mu}\vert \times \vert m_{\mu\tau} \vert$,
$\vert m_{\mu\tau}\vert \times \vert m_{\tau\tau} \vert$,
$\vert m_{\mu\mu}\vert \times \vert m_{\tau\tau} \vert$ and 
$\vert m_{e\mu}\vert \times \vert m_{\tau\tau} \vert$ are very correlated.
So, for instance, if a model predicts $\vert m_{e\mu}\vert \approx  6$ meV, 
$\vert m_{e \tau} \vert$ must be in the range 5--8 meV, while without taking 
into account this correlation the allowed range would be 2.7--9.3 meV at 
95.45\% CL.

From the expressions for the matrix elements given in appendix~\ref{appendixA}
 it is easy to show that in this case:
 $$\vert m_{e\mu} \vert^2 + \vert m_{e \tau}\vert^2 \sim m_3^2 y^2 + m_2^2 x^2,$$ 
$$\vert m_{\mu \mu} \vert \sim m_3 z^2,$$
$$\vert m_{\tau \tau}\vert^2 \sim m_3^2 (1-2z^2) + \vert m_{\mu \mu}\vert^2 , $$
and 
$$ \vert m_{\mu \tau}\vert^2 + \vert m_{\tau \tau}\vert^2 \sim m_3^2 (1-z^2),$$
where $x=\sin\theta_{12}$, $y = \sin\theta_{13}$ and $z = \sin\theta_{23}$.

Due to the prevalence of the $m_3$ mass, the determination
of $\sin^2\theta_{13}$ with the present uncertainty of
10\% by the reactor experiments not only affected the range of $\vert
m_{e\alpha}\vert$, $\alpha = e, \mu, \tau$, but also their PDFs.
Also, the determination of $\sin^2\theta_{23}$ with an
uncertainty of 9\% (4\%) in the first (second) octant changes the
range and the form of the PDFs of all matrix elements except for
$\vert m_{ee}\vert$.

\begin{table}[t!]
\begin{tabular}{|l|c|c|c|c| }
\hline
\multicolumn{5}{|c|}{in meV}\\
\hline
Element & $m_1 \to 0$ (1$^{\rm st}$ oct.) & $m_1 \to 0$ (2$^{\rm nd}$ oct.) & $m_3 \to 0$ (1$^{\rm st}$ oct.) & $m_3 \to 0$ (2$^{\rm nd}$ oct.) \\
\hline
$\vert m_{ee} \vert$ & 1.3 -- 4.2 & 1.4 -- 4.2 & 18 -- 53 & 18 -- 54 \\
$\vert m_{e\mu} \vert$ & 1.5 -- 8.8 & 2.5 -- 9.4 & 2 -- 40 & 2 -- 35 \\
$\vert m_{e\tau} \vert$ & 2.7 -- 9.3 & 1.5 -- 9 & 2 -- 35 & 2 -- 40 \\
$\vert m_{\mu\mu} \vert$ & 14 -- 27 & 24 -- 34 & 6 -- 35 & 2 -- 25 \\
$\vert m_{\mu\tau} \vert$ & 21 -- 28 & 20.5 -- 28 & 9-- 27 & 9 -- 27 \\
$\vert m_{\tau\tau} \vert$ & 23 -- 35 & 14 -- 26 &2 -- 26 & 6 -- 34 \\
\hline
\end{tabular}
\caption{\label{tab:allowed} Range of allowed values of $\vert
  m_{\alpha \beta}\vert$ at 95.45 \% CL for the very hierarchical
  cases.  }
\end{table}

\begin{figure}[htb]
\begin{center}
 \includegraphics[width=0.3\textwidth]{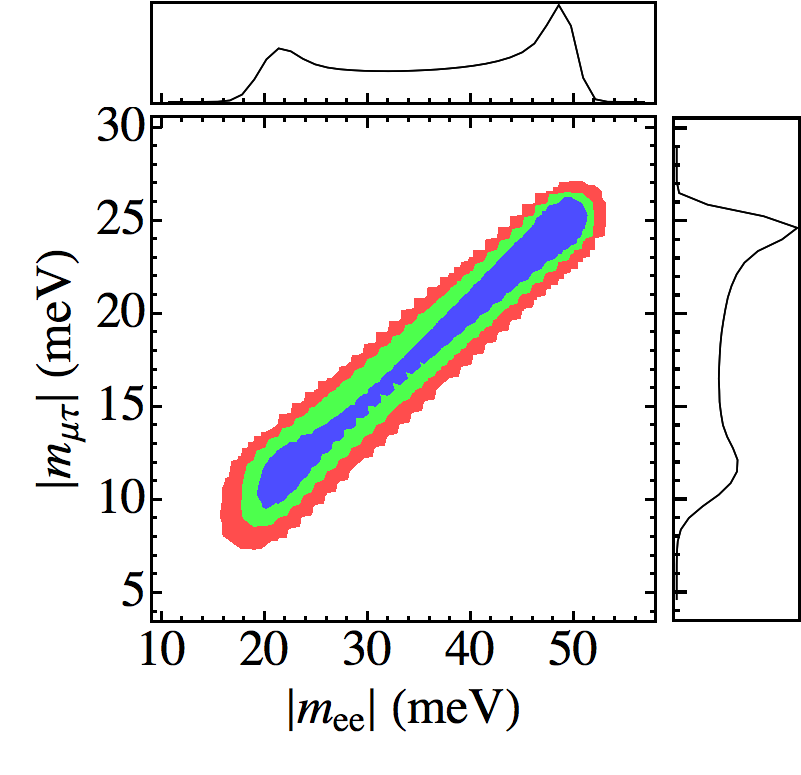}
 \includegraphics[width=0.3\textwidth]{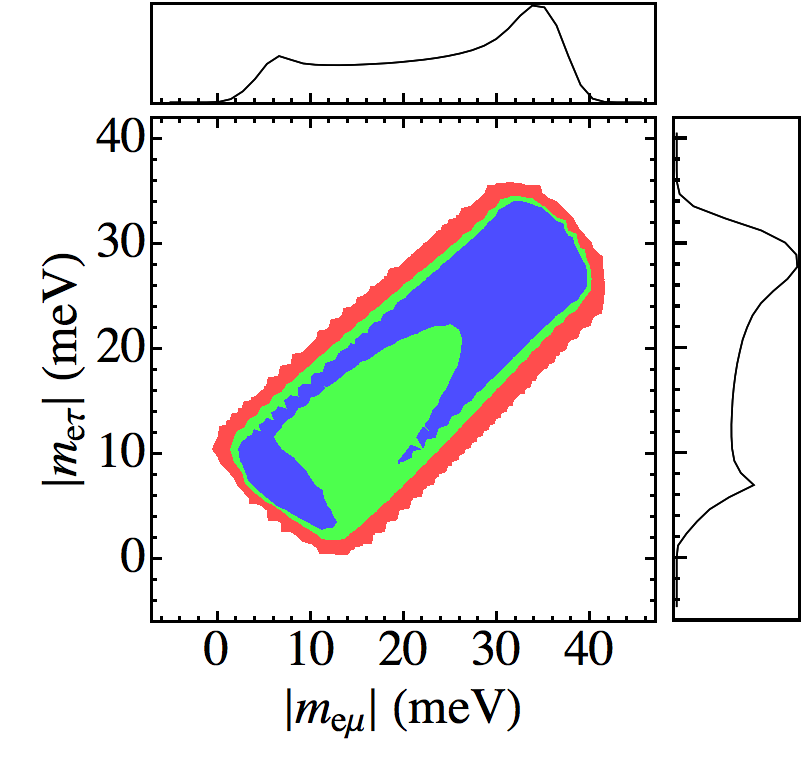}
 \includegraphics[width=0.3\textwidth]{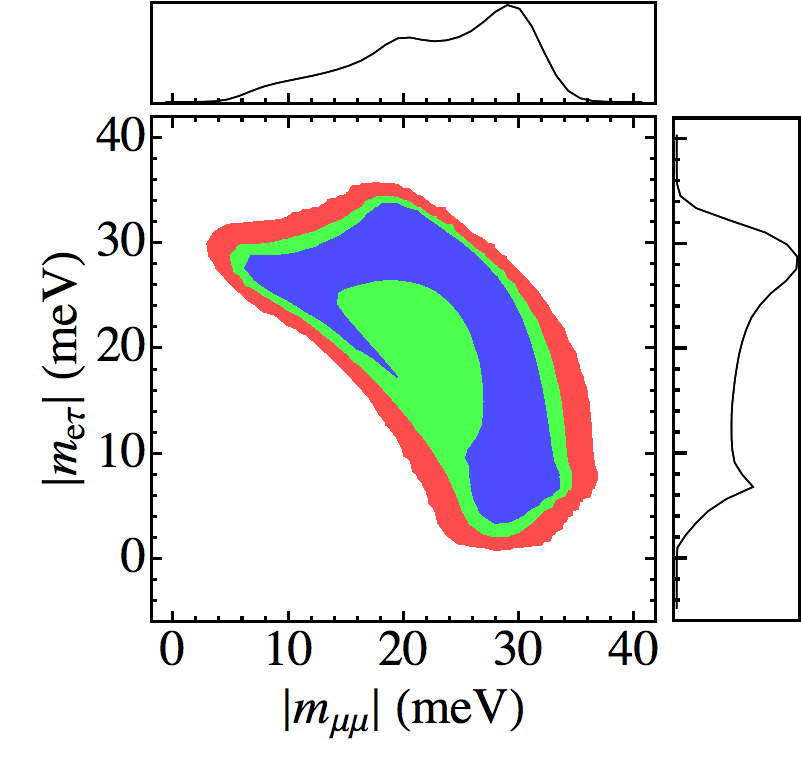}
 \includegraphics[width=0.3\textwidth]{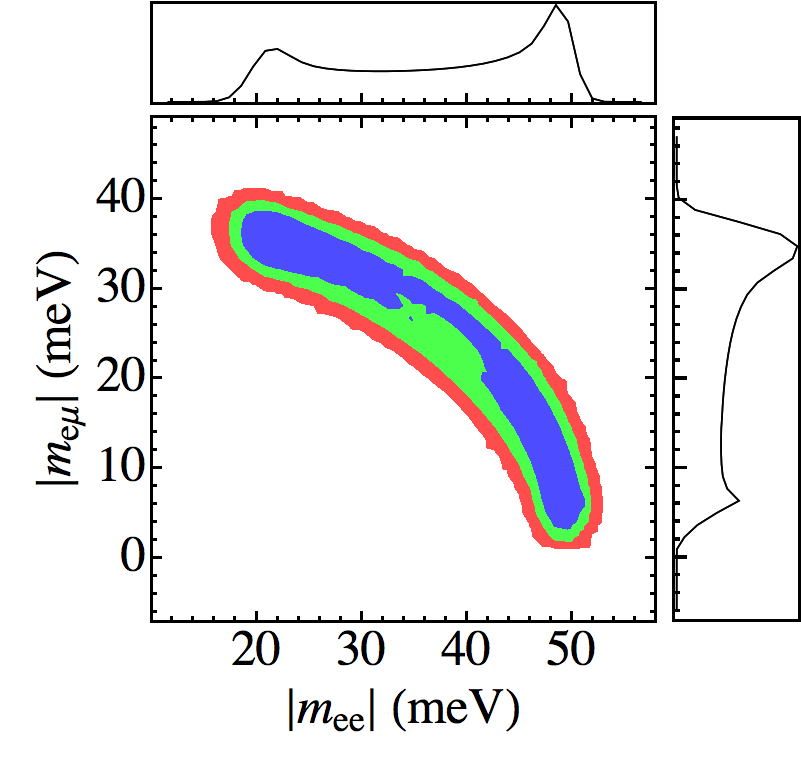}
 \includegraphics[width=0.3\textwidth]{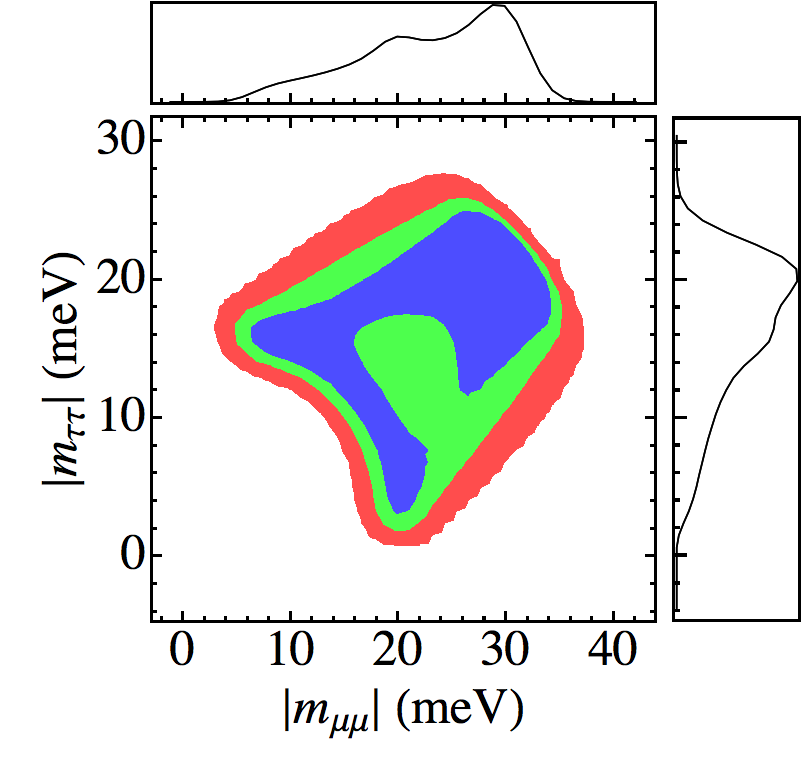}
 \includegraphics[width=0.3\textwidth]{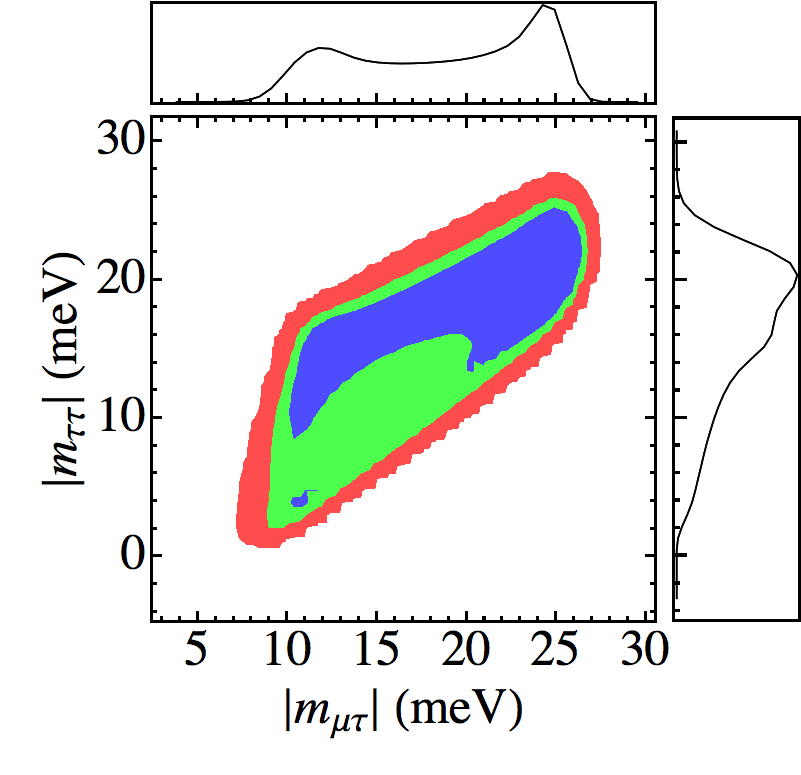}
\vspace{-2mm}
\end{center}
\vspace{-0.1cm}
\caption{Same as Fig.~\ref{fig:nh1st} but for $m_3 \to 0$ and 
$\theta_{23}$ in the first octant.}
\label{fig:ih1st}
\end{figure}

In the case $m_3 \to 0$, $m_1 \approx m_2 \approx 0.05$ eV and only
two CP-phases, $\delta$ and $\lambda_1$, are important.  Again the PDFs
obtained for the absolute value of the matrix elements in the first
and second octant of $\theta_{23}$ are basically the same as long as
we replace $\mu \leftrightarrow \tau$.  In Fig.~\ref{fig:ih1st} we
show two-dimensional PDFs for some pairs of elements of the matrix
$m_{\alpha \beta}$ in the case $m_3 \to 0$ and $\theta_{23}$ in the
first octant.  A more complete set of plots can be found in
appendix ~\ref{appendixB}, while in Table~\ref{tab:allowed} we present
the 95.45\% CL allowed ranges.

Generally, the predominant terms comprise $m_1$ or $m_2$, which have
similar sizes, and their contributions involve $\theta_{12}$ and
$\theta_{23}$, which are not maximal, without being suppressed by
$\theta_{13}$. There are at least three consequences of such a fact.
First, the determination of $\sin^2\theta_{13}$ by the reactor
experiments basically did not affect the range of the matrix elements
but changed the shape of some of their PDFs. Second, the determination of
$\sin^2\theta_{23}$ with an uncertainty of 9\% (4\%) in the first
(second) octant changes the range $\vert m_{e\mu}\vert$, $\vert
m_{e\tau}\vert$, $\vert m_{\mu\mu}\vert$ and $\vert
m_{\tau\tau}\vert$, while the shape of the PDFs remain basically the
same except in the case $\vert m_{\mu\mu}\vert$ and $\vert
m_{\tau\tau}\vert$. Last, the mass matrix entries are very
  correlated.

We observe the strong correlations between all pairs of elements
  in Figs.~\ref{fig:ih1st}, \ref{fig:ih-1st} and \ref{fig:ih-2nd}.
So in this case it is even more important to take into account these
correlations in model building.  For instance, we can easily see, from
the expressions in appendix~\ref{appendixA}, that in this case

$$ \vert m_{\mu \tau} \vert \sim \sqrt{z^2 (1-z^2)} \, \vert m_{ee}\vert ,$$
$$ \vert m_{e e} \vert^2 \sim m_2^2 -  (1-z^2)^{-1}\vert m_{e \mu}\vert^2 ,$$
and 
$$ \vert m_{e\mu}\vert \sim \sqrt{\frac{(1-z^2)}{z^2}} \vert m_{e \tau}\vert .$$
This behavior is confirmed by Fig.~\ref{fig:ih1st}.

\subsubsection{Quasi-Degenerate  Case}

In the quasi-degenerate case, $m_1 \sim m_2 \sim m_3$ and the effect
of the ordering is very small.  In this case, all masses and CP  
phases play a role.

\begin{figure}[htb]
\begin{center}
 \includegraphics[width=0.3\textwidth]{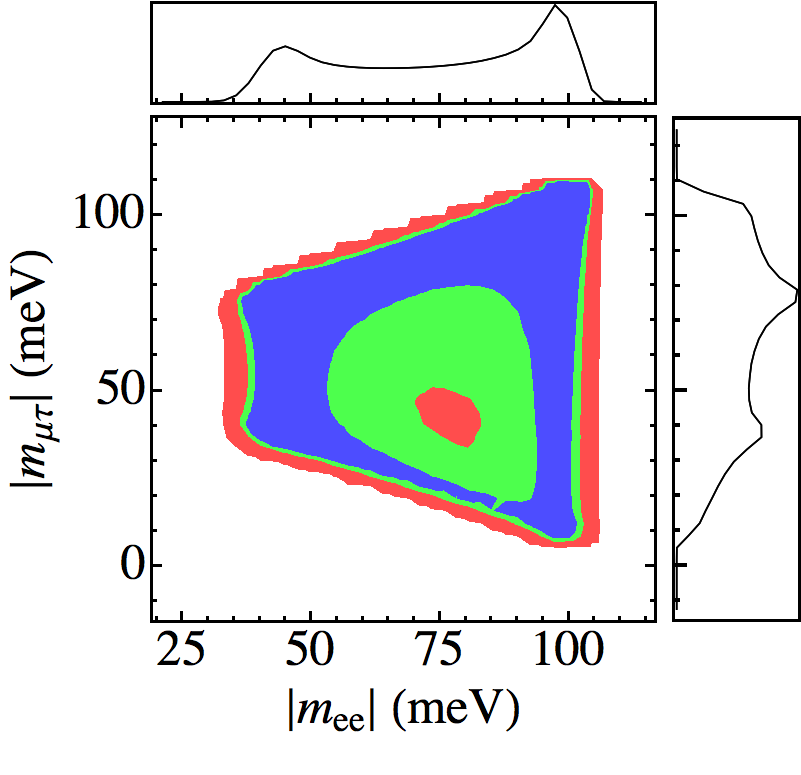}
 \includegraphics[width=0.3\textwidth]{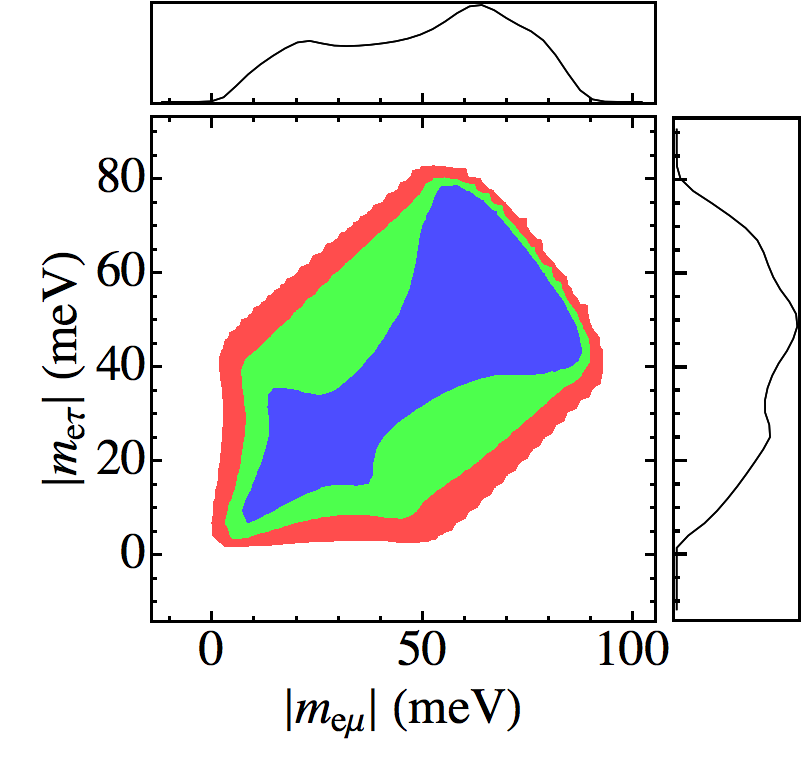}
 \includegraphics[width=0.3\textwidth]{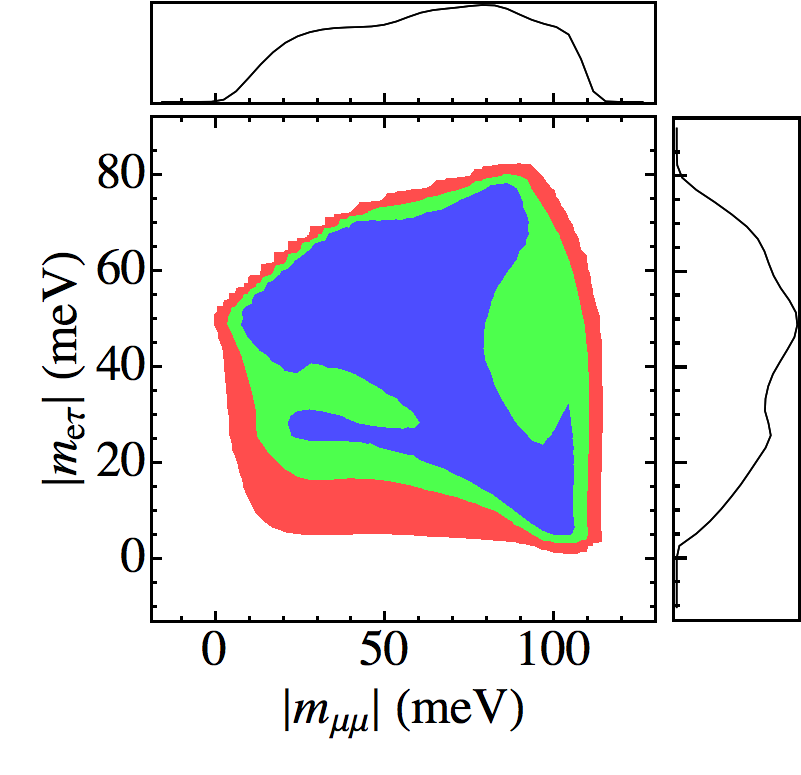}
 \includegraphics[width=0.3\textwidth]{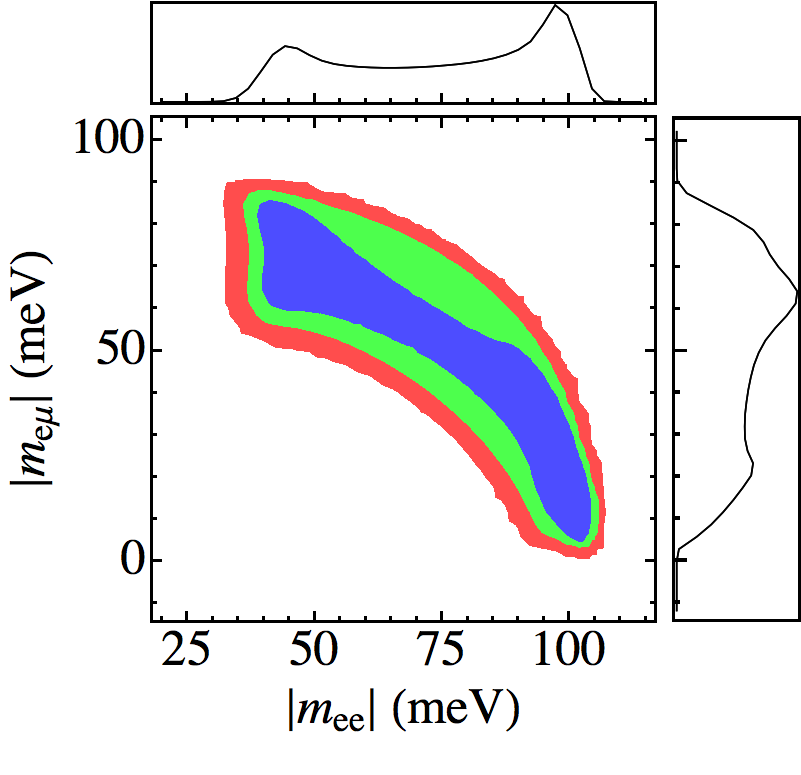}
 \includegraphics[width=0.3\textwidth]{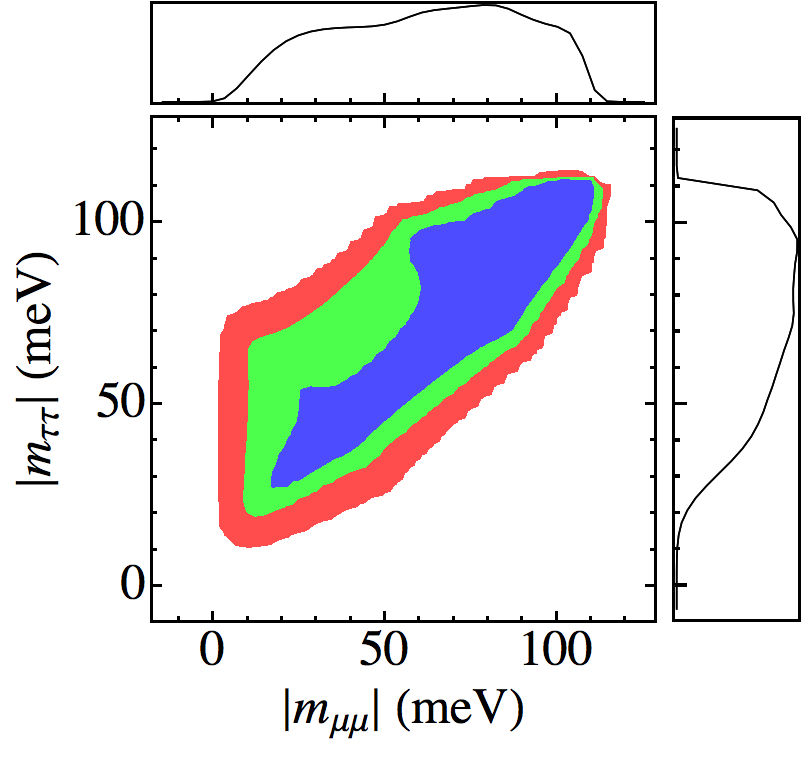}
 \includegraphics[width=0.3\textwidth]{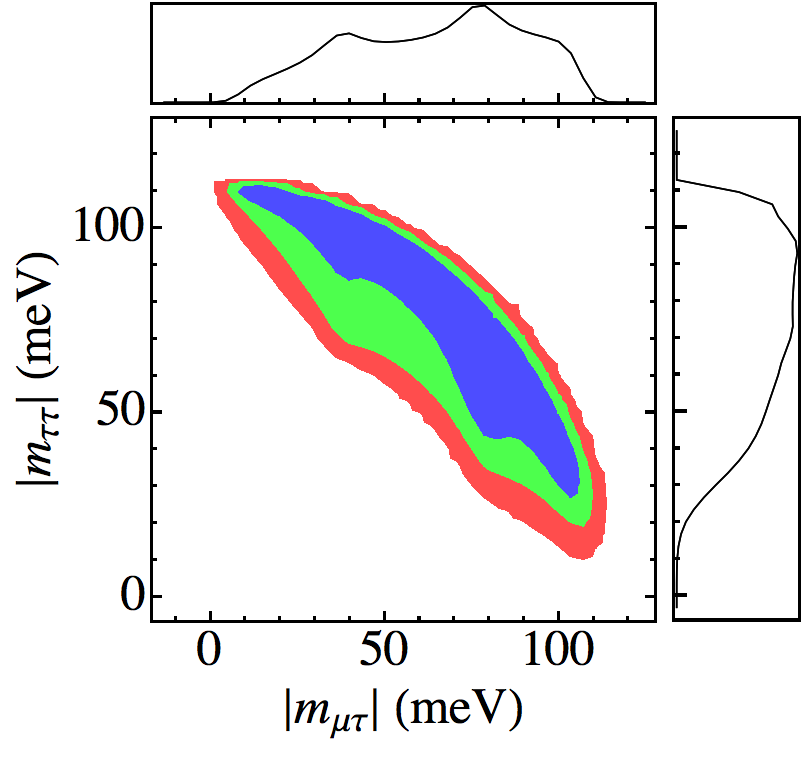}
\vspace{-2mm}
\end{center}
\vspace{-0.1cm}
\caption{Same as Fig.~\ref{fig:nh1st} but for $m_1 = 0.1$ eV and 
$\theta_{23}$ in the first octant.}
\label{fig:deg1st}
\end{figure}

As an example, we take $m_1 = 0.1$ eV. In Fig.~\ref{fig:deg1st} we
show the correlations among the absolute values of some of the matrix
elements $m_{\alpha \beta}$ for the normal mass ordering and
$\theta_{23}$ in the first octant. In appendix \ref{appendixB} one can
find a complete set of plots for this case, Fig.~\ref{fig:nh-1st-m01},
and for $\theta_{23}$ in the second octant, Fig.~\ref{fig:nh-2nd-m01}. 
We  use the same color coding as in previous figures.  The range of the
values allowed at 95.45\% CL are given in Table~\ref{tab:allowed2}.

The correlations here are either similar to the very hierarchical case 
in normal ordering or in the inverted ordering. For example, the PDFs for
$\vert m_{ee}\vert \times \vert m_{e\mu} \vert$,
$\vert m_{\mu\mu}\vert \times \vert m_{\tau\tau} \vert$,
$\vert m_{\mu\mu}\vert \times \vert m_{e\tau} \vert$ and
$\vert m_{e\mu}\vert \times \vert m_{e\tau} \vert$, are correlated like 
in the inverted ordering, while 
$\vert m_{\mu\tau}\vert \times \vert m_{\tau\tau} \vert$ and
$\vert m_{ee}\vert \times \vert m_{\mu\tau} \vert$ are more like the 
normal ordering.

\begin{table}[t!]
\begin{tabular}{|l|c|c| }
\hline
\multicolumn{3}{|c|}{in meV}\\
\hline
Element & 1$^{\rm st}$ oct. & 2$^{\rm nd}$ oct.  \\
\hline
$\vert m_{ee} \vert$ & 35 -- 108 & 36 -- 107  \\
$\vert m_{e\mu} \vert$ & 5 -- 90 &   5 -- 80 \\
$\vert m_{e\tau} \vert$ & 5 -- 80 &  5 -- 90\\
$\vert m_{\mu\mu} \vert$ & 10 -- 115 &  22 --110 \\
$\vert m_{\mu\tau} \vert$ & 10 -- 115 &  9 -- 110\\
$\vert m_{\tau\tau} \vert$ & 20 -- 115 &  12 -- 115\\
\hline
\end{tabular}
\caption{\label{tab:allowed2} Range of allowed values of $\vert m_{\alpha \beta}\vert$ at 95.45 \% CL for the quasi-degenerate case with $m_1=0.1$ eV.}
\end{table}

\subsubsection{Future Perspectives}

To evaluate the effect of a better determination of the mixing
parameters on our knowledge of the mass matrix we have studied the
effect of reducing by half the uncertainty of a parameter at a time,
keeping the other parameters at their current uncertainties.

\begin{figure}[htb]
\begin{center}
\includegraphics[width=0.49\textwidth]{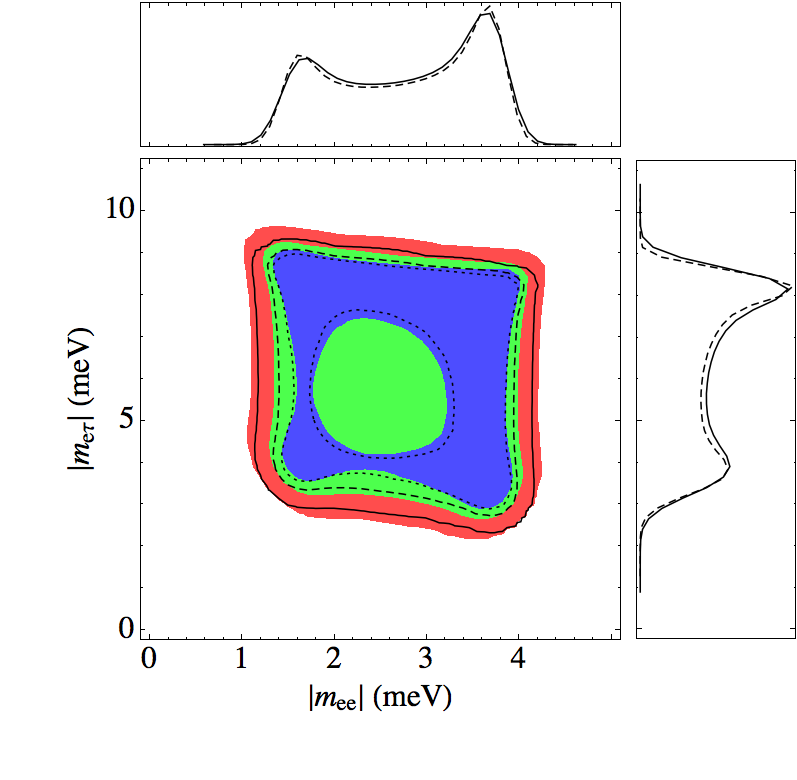}
\includegraphics[width=0.49\textwidth]{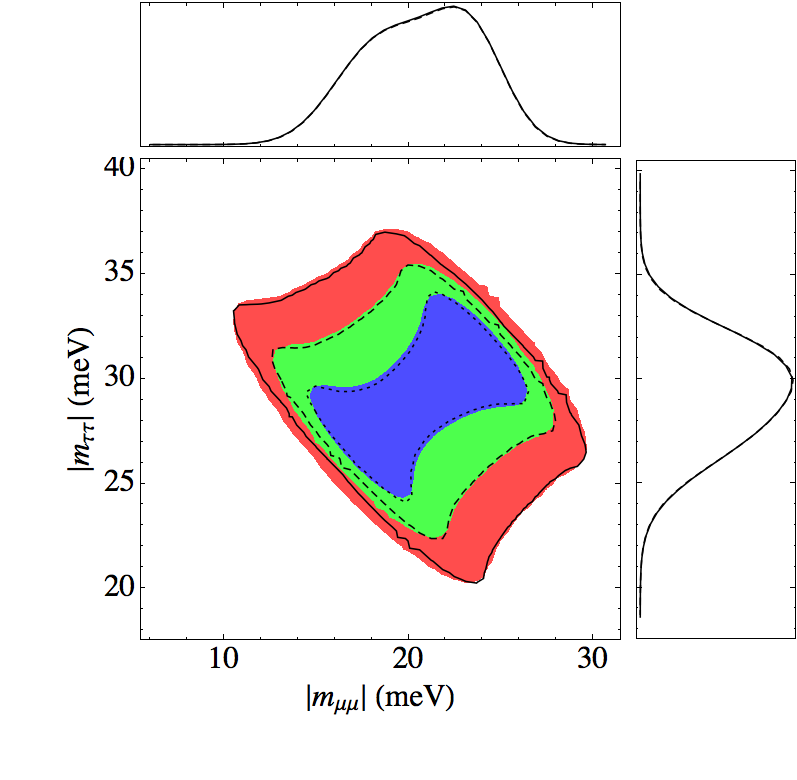}
\includegraphics[width=0.49\textwidth]{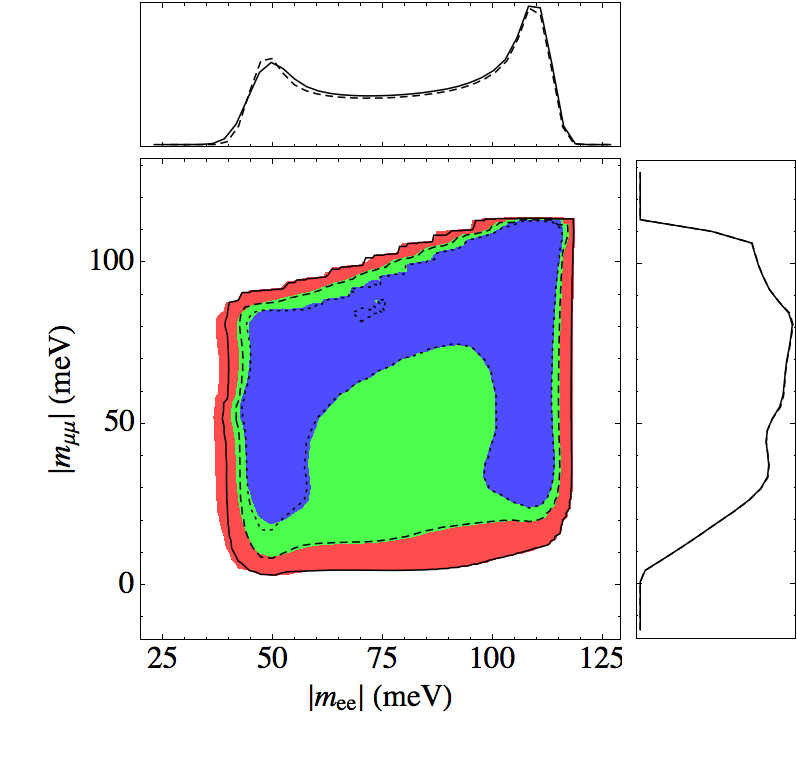}
\includegraphics[width=0.49\textwidth]{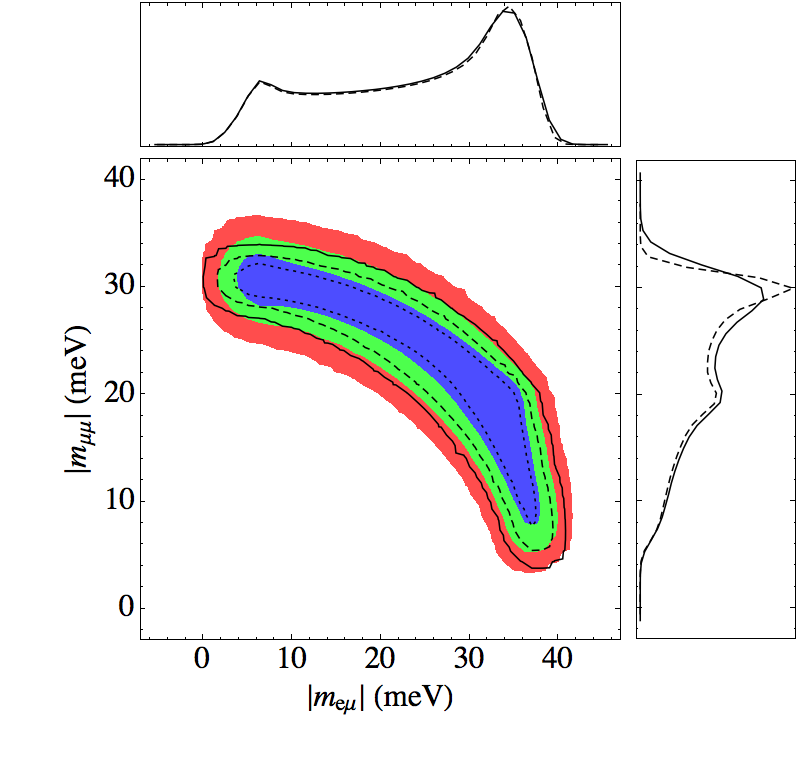}
\vspace{-2mm}
\end{center}
\vspace{-0.1cm}
\caption{PDFs for the distribution of:  
$\vert m_{ee} \vert \times 
\vert m_{e\tau}\vert$ for the the normal hierarchy  
with $\sin^2 \theta_{13}$ uncertainty reduced to 
a half of its present value (top left panel); 
$\vert m_{\mu\mu} \vert \times 
\vert m_{\tau\tau}\vert$ for the the normal hierarchy with 
$\Delta m_{31}^2$ uncertainty reduced to 
a half of its present value (top right panel); 
$\vert m_{ee} \vert \times 
\vert m_{\mu \mu}\vert$ for the the quasi-degenerate case   
with $\sin^2 \theta_{12}$ uncertainty reduced to 
a half of its present value (bottom left panel) and
$\vert m_{e\mu} \vert \times 
\vert m_{\mu\mu}\vert$ for the the inverted hierarchy with 
$\sin^2 \theta_{23}$ uncertainty reduced to 
a half of its present value (bottom right panel).
In all cases $\theta_{23}$ was taken to be in the first octant.
The colored areas are for the present uncertainties of the 
oscillation parameters, whereas the back lines are for the assumed future 
reduced uncertainty of one of the parameters.}
\label{fig:future1}
\end{figure}

In Fig.~\ref{fig:future1} we illustrate the effect of a better 
determination of $\sin^2 \theta_{13}$ (top left panel), 
$\Delta m_{31}^2$ (top right panel), $\sin^2 \theta_{21}$ (bottom 
left panel) and $\sin^2 \theta_{23}$ (bottom right panel).
We observe that the effect of a better determination of any 
of these parameters is very small. We have verified that this is true 
for both mass hierarchies and $\theta_{23}$ octants. The  biggest 
effect comes from a better determination of $\sin^2 \theta_{23}$,  
as one could have guessed, but still this only reduces significantly 
the 3 $\sigma$ region. We do not show the effect of a better determination 
of $\Delta m^2_{21}$ because it is even smaller than for the other 
parameters.

\begin{figure}[htb]
\begin{center}
\includegraphics[width=0.3\textwidth]{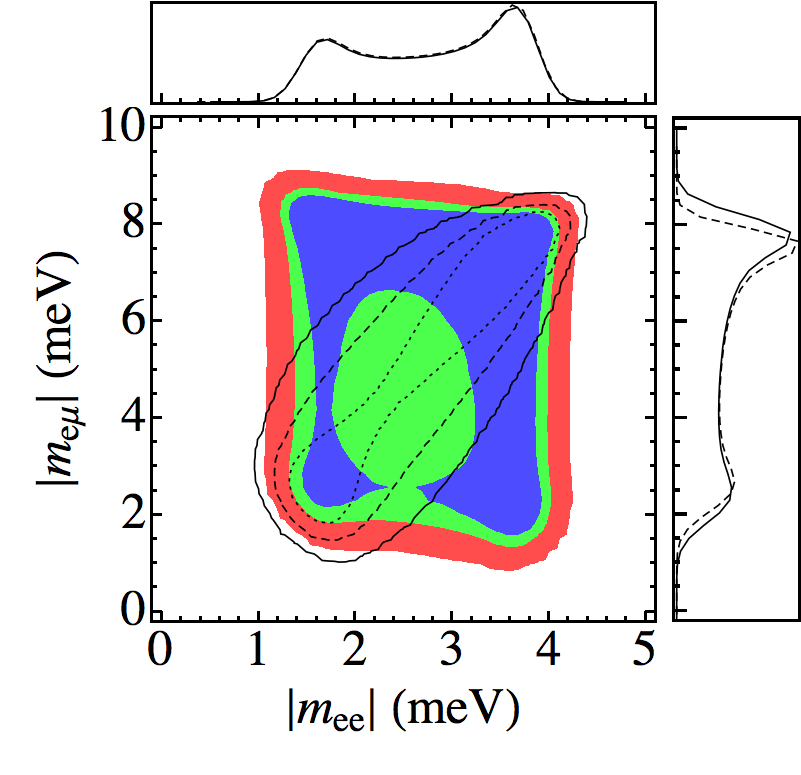}
\includegraphics[width=0.3\textwidth]{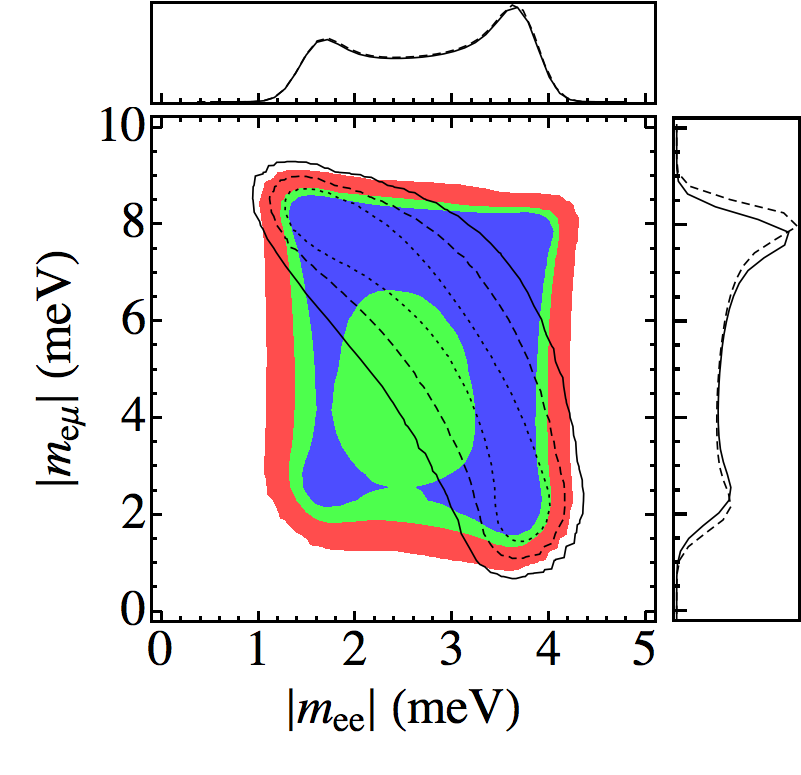}
\includegraphics[width=0.3\textwidth]{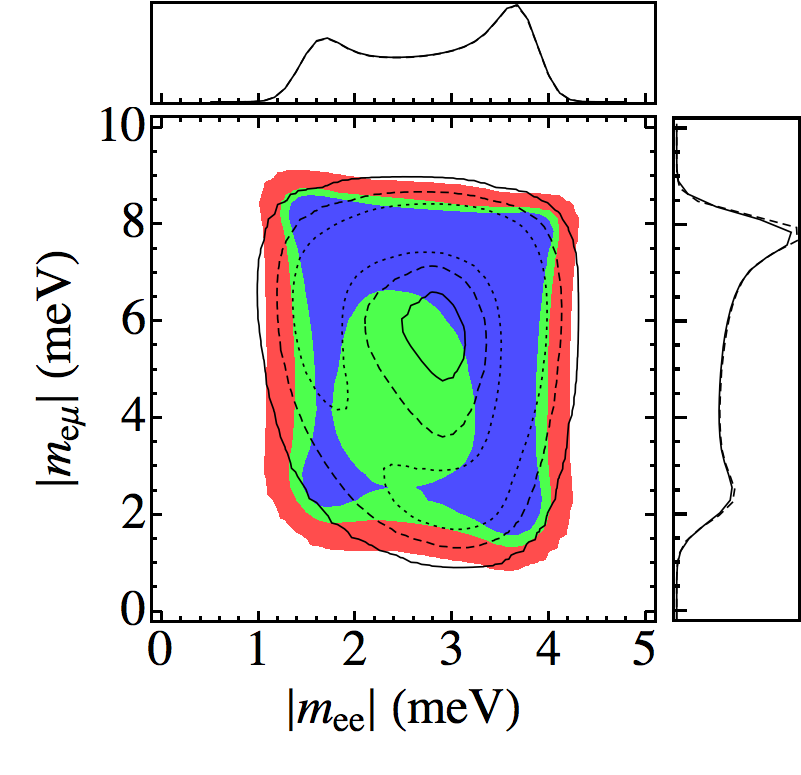}
\includegraphics[width=0.3\textwidth]{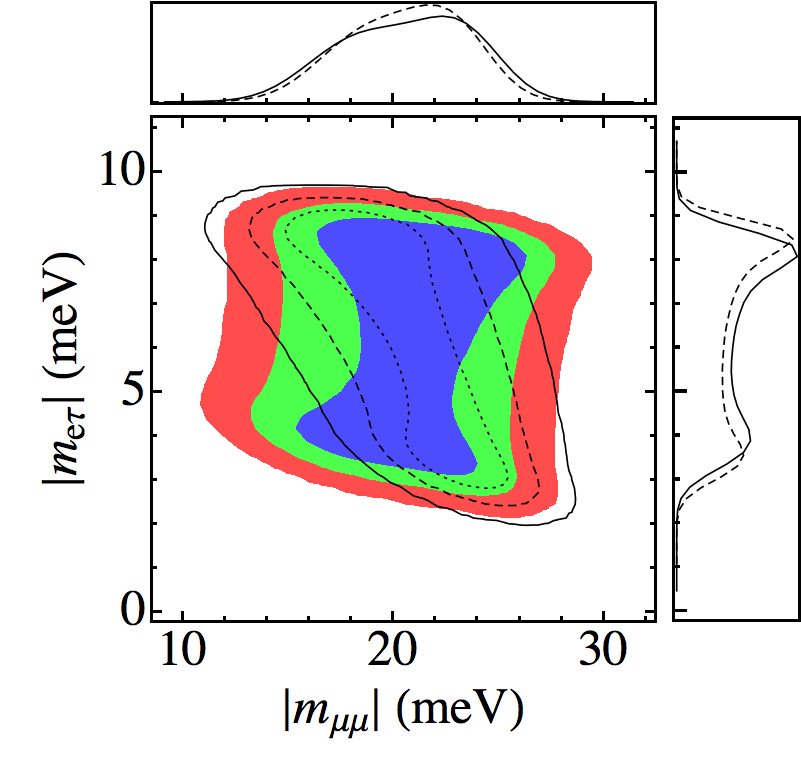}
\includegraphics[width=0.3\textwidth]{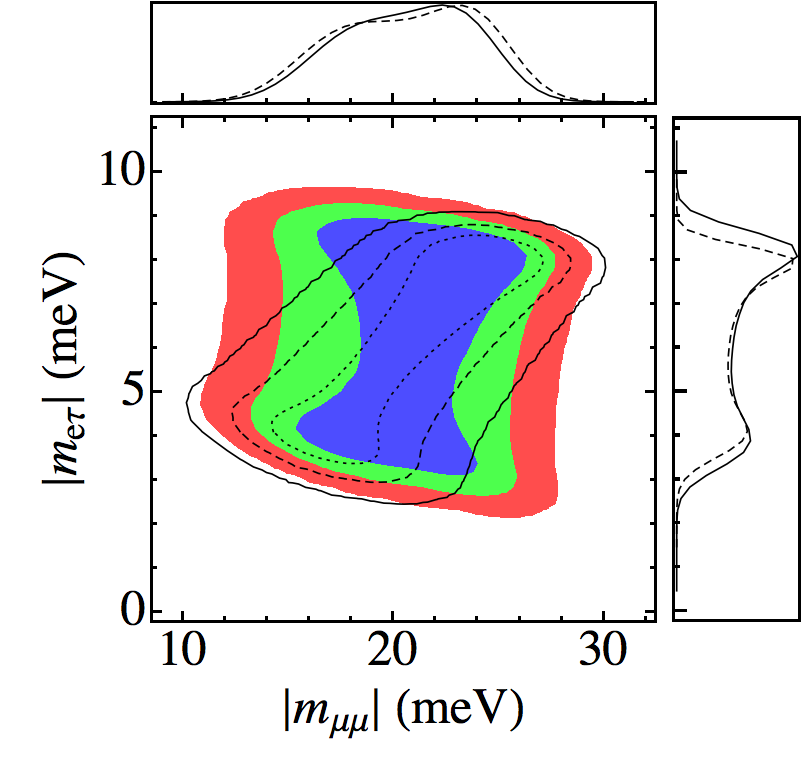}
\includegraphics[width=0.3\textwidth]{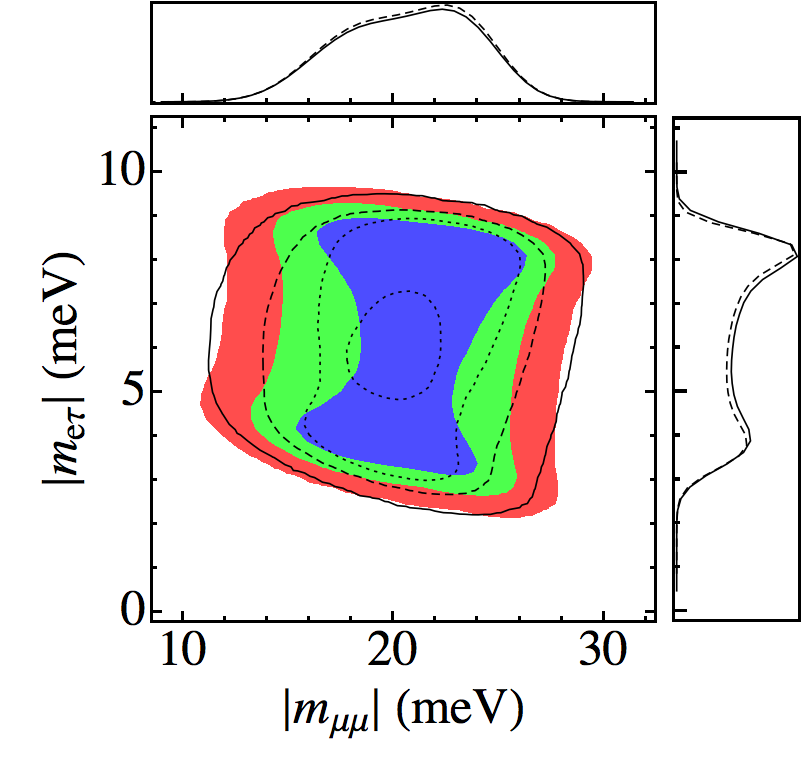}
\vspace{-2mm}
\end{center}
\vspace{-0.1cm}
\caption{
PDFs for the distribution of:  
$\vert m_{ee} \vert \times 
\vert m_{e\mu}\vert$ (top panels)  and 
$\vert m_{\mu\mu} \vert \times \vert m_{e\tau}\vert$ (bottom panels) 
for the normal ordering.
From left to right $\delta=0^\circ, 180^\circ$ and $270^\circ$,   
assumed to be determined within 20$^\circ$.}
\label{fig:future2a}
\end{figure}

\begin{figure}[htb]
\begin{center}
\includegraphics[width=0.3\textwidth]{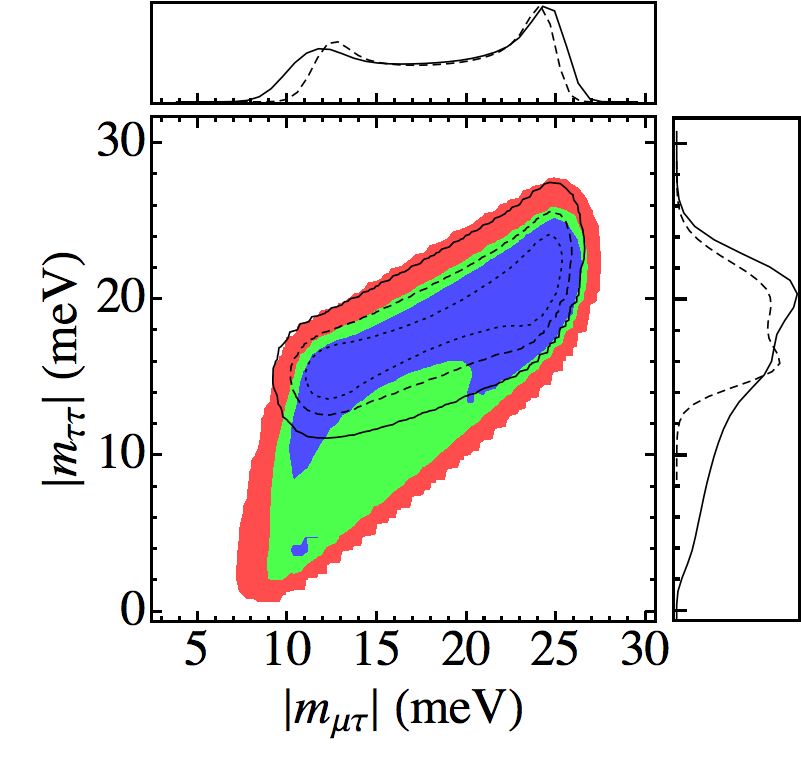}
\includegraphics[width=0.3\textwidth]{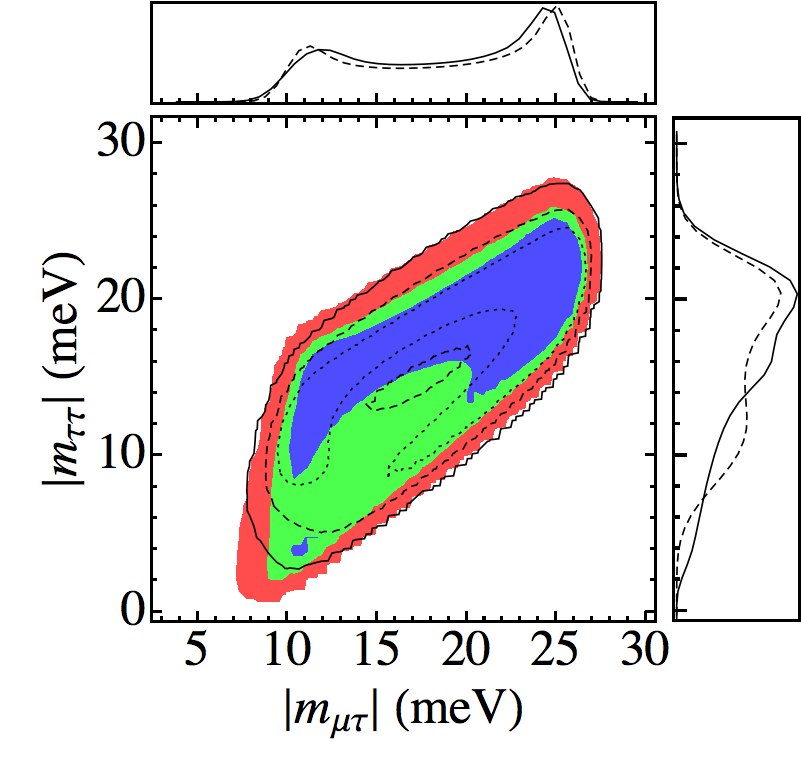}
\includegraphics[width=0.3\textwidth]{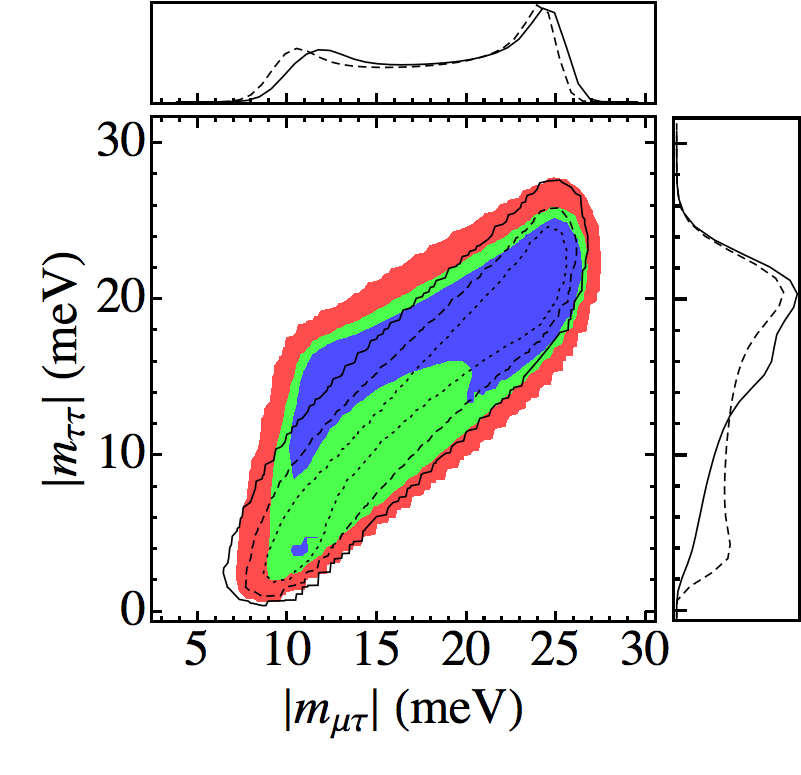}
\includegraphics[width=0.3\textwidth]{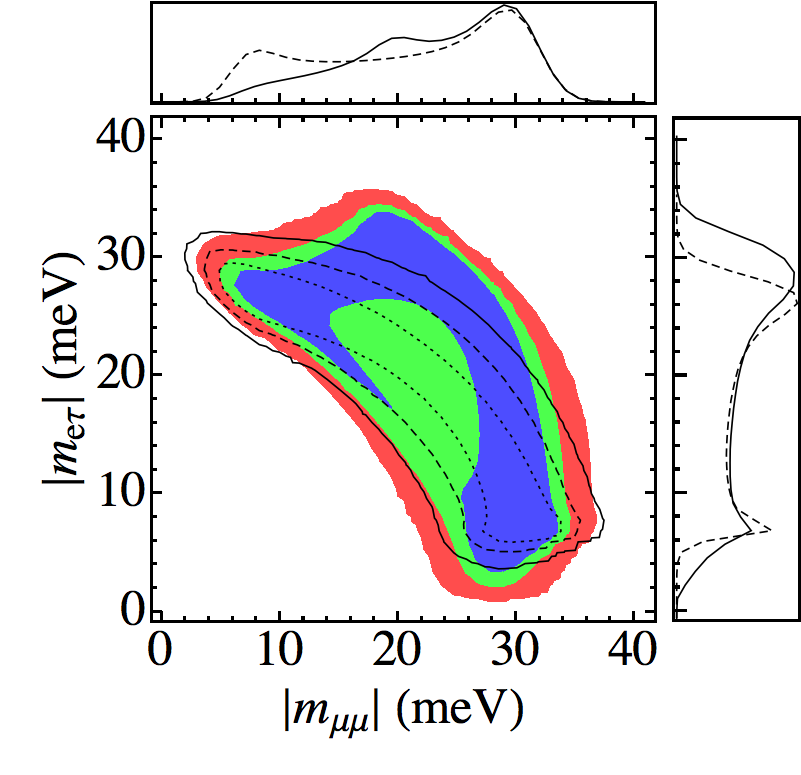}
\includegraphics[width=0.3\textwidth]{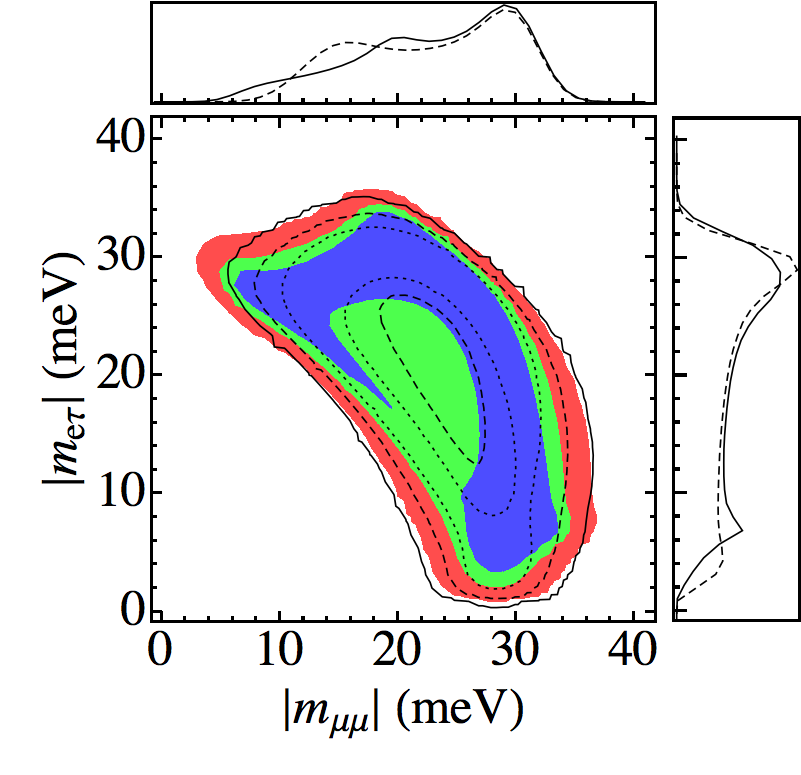}
\includegraphics[width=0.3\textwidth]{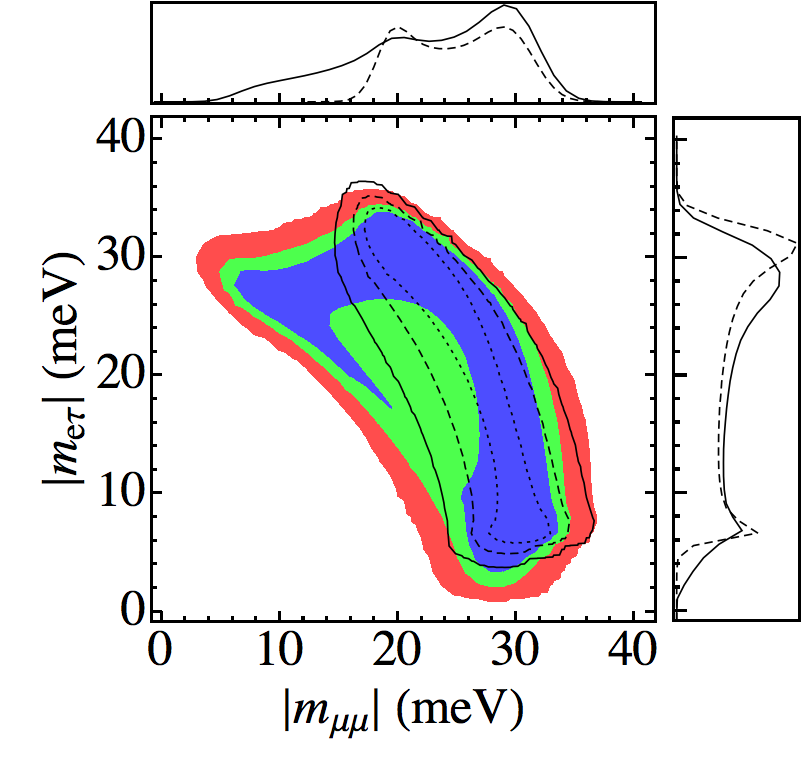}
\vspace{-2mm}
\end{center}
\vspace{-0.1cm}
\caption{
PDFs for the distribution of:  
$\vert m_{\mu \tau} \vert \times 
\vert m_{\tau\tau}\vert$ (top panels)  and 
$\vert m_{\mu\mu} \vert \times \vert m_{e\tau}\vert$ (bottom panels) 
for the inverted ordering. 
From left to right $\delta=0^\circ, 90^\circ$ and $180^\circ$,  
assumed to be determined within 20$^\circ$.}
\label{fig:future2b}
\end{figure}

On the other hand, a measurement of $\delta$ with an uncertainty of
20$^\circ$, that can be envisaged according to
Ref.~\cite{Coloma:2012ji} for long baseline neutrino oscillation
experiments, could be significant.  To illustrate this we show in
Figs.~\ref{fig:future2a}-\ref{fig:future2b} the effect of a determination of
$\delta$ with an uncertainty of 20$^\circ$ for some of the
correlations between pairs of mass matrix elements.  
On the top (bottom) panels of Fig.~\ref{fig:future2a} we can see this for 
$\vert m_{ee} \vert \times \vert
m_{e\mu}\vert$ ($\vert m_{\mu\mu} \vert \times \vert
m_{e\tau}\vert$) in the normal ordering for 
$\delta = 0^\circ$ (left), $180^\circ$ (center) and $270^\circ$ (right), 
while on the top (bottom) panels of Fig.~\ref{fig:future2b} we show 
$\vert m_{\mu \tau} \vert
\times \vert m_{\tau\tau}\vert$ ($\vert m_{\mu\mu} \vert
\times \vert m_{e\tau}\vert$) in the inverted ordering for
$\delta = 0^\circ$ (left), $90^\circ$ (center) and $180^\circ$ (right), 
These cases for normal (inverted) ordering and $\delta= 90^\circ$ 
($\delta= 270^\circ$) are not shown because they are very similar to 
$\delta = 270^\circ$ ($\delta = 90^\circ$).

For the normal ordering, with $m_1 \to 0$, the determination of 
$\delta$ will play an important role in the correlation of $\vert
m_{ee}\vert$ with all other mass matrix elements, but will be more
significant for $\vert m_{e\mu}\vert$ or $\vert m_{e\tau}\vert$.  This
is due to the fact that for these mass matrix elements the leading
phase term is the one that accompanies $\cos
\left[2(\delta+\lambda_3)\right]$, just as for $\vert m_{ee}\vert$. This
will also affect the correlations involving $\vert m_{\mu\mu}\vert$
and $\vert m_{\tau\tau}\vert$, since for them the leading phase terms
are, in order of importance, the ones that go with $\cos(2\lambda_3)$
and $\cos(\delta + 2 \lambda_3)$. However, the correlations with 
$\vert m_{\mu\tau}\vert$ will only slightly change because the 
leading phase term for this element does not depend on $\delta$.

For the inverted ordering, with $m_3 \to 0$, the determination
of $\delta$ will play a bigger role in the PDFs of 
$\vert m_{\tau \tau}\vert$ and $\vert m_{\mu \mu}\vert$. 
This is because, as we can see from their 
expressions in appendix~\ref{appendixA}, the leading coefficients of 
$\cos (2 \lambda_1)$, $\cos(\delta \pm 2 \lambda_1)$ 
and $\cos \delta$ are all of the same order.
The PDFs of $\vert m_{e\mu} \vert$ and $\vert m_{e\tau} \vert$ 
will also be affected because the terms that depend on $\delta$ are not 
negligible in comparison to the leading term that depends on 
$\cos(2\lambda_1)$, however their relative importance will depend on 
the $\theta_{23}$ octant. The PDFs for $\vert m_{ee}\vert$ and
$\vert m_{\mu\tau} \vert$ are basically independent of $\delta$, the first 
because $m_3 \to 0$, the second because these terms are suppressed 
by $\sin^2\theta_{13}$ or factors of this order.

We also have checked that the effect of the determination of 
$\delta$ for the quasi-degenerate case with $m_1 = 0.1$ eV is 
smaller than for the hierarchical cases because there are more 
phases involved.

\begin{figure}[htb]
\begin{center}
 \includegraphics[width=0.9\textwidth]{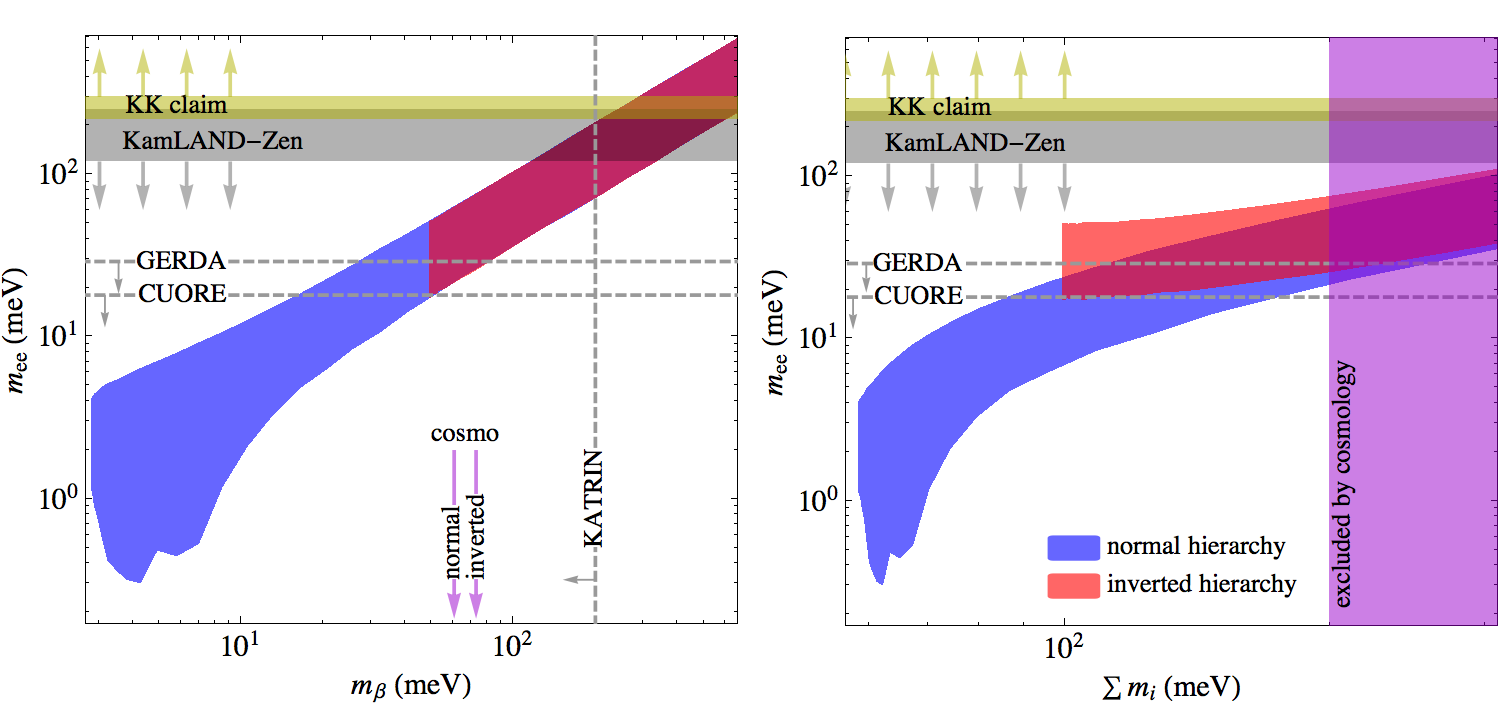}
\vspace{-2mm}
\end{center}
\vspace{-0.1cm}
\caption{We show the current allowed regions for $\vert m_{ee}\vert$
  at 99\% CL as a function of the effective electron neutrino mass,
  $m_\beta$, on the left panel and as a function of the sum of the
  neutrino masses, $\sum m_i$, on the right panel. The region allowed
  by the normal (inverted) mass ordering is in blue (red), the recent
  limit on $\vert m_{ee}\vert$ given by
  KamLAND-Zen~\cite{Gando:2012zm} in gray and the region excluded by
  cosmology~\cite{cosmo} in magenta. We also show the reach expected
  for the beta decay experiment Katrin~\cite{katrin}, as well as the
  ultimate reach aimed by the neutrinoless double beta decay
  experiments GERDA and CUORE according to
  Ref.~\cite{Sarazin:2012ct}. }
\label{fig:futurenus}
\end{figure}

In the future we also expect to have information from three different
sources: neutrinoless double beta decay experiments, beta decay
experiments and cosmology.  In Fig.~\ref{fig:futurenus} we show the
current allowed region for $\vert m_{ee}\vert$ as a function of the
effective electron neutrino mass, $m_\beta$ and of the sum of the
neutrino masses
\footnote{There is no one-to-one correspondence between $m_{ee}$
    and $m_\beta$ or $\sum_i m_i$. Hence, in order to plot
    Fig.~\ref{fig:futurenus}, for each value of $m_0$ we extracted the
    allowed interval of $m_{ee}$ and plotted it against $m_\beta$ and
    $\sum_i m_i$ \emph{calculated at the best fit values} of the oscillation
    parameters (as the Majorana phases do not play a role in these last
    two quantities).},
 $\sum m_i$ at 99\% CL.  The region allowed by the normal
(inverted) mass ordering is in blue (red) and the recent limit on
$\vert m_{ee}\vert$ given by KamLAND-Zen~\cite{Gando:2012zm}, $\vert
m_{ee}\vert<(120-250)$ meV, is shown in gray.  Cosmology excludes the
magenta region $\sum m_i>(0.2-0.6)$ eV~\cite{cosmo}. 
Notice that the allowed regions were built from the pdfs constructed
from data (except for the CP phases, which we assumed to be flat
distributed). This is why the $m_{ee}\to 0$ region is absent, as it is
very unlikely to have the necessary degree of cancellations.

The forecast sensitivity on $\vert m_{ee}\vert$ of the most ambitious
neutrinoless double beta decay experiments, after 5 years of exposure,
is 29--73 meV (GERDA phase-3)  and 18--39 meV (CUORE)~\cite{Sarazin:2012ct}. 

The KArlsruhe TRitium Neutrino mass experiment (KATRIN)
will have a sensitivity on the electron neutrino effective mass,
$m_\beta= \sqrt{\sum_i \vert U_{ei}\vert^2 \, m_i^2}$, of 0.2 eV at 95\%
CL~\cite{katrin}.

Cosmological limits today still allow for quasi-degenerate neutrino masses;
however, this possibility will soon be confirmed or ruled out.
According to Ref.~\cite{cosmo}, many cosmological probes, with
different systematics, will reach a sensitivity on $\sum m_i$ of 0.1~eV or
lower. Lyman $\alpha$ forest can reach 0.1~eV, lensing of Cosmic
Microwave Background 0.2--0.05~eV, lensing of galaxies 0.07~eV, observations
of the redshifted 21 cm neutral hydrogen line 0.1--0.006~eV, galaxy
clusters and galaxy distribution surveys 0.1~eV.

If cosmology will point to a quasi-degenerate neutrino spectrum:

\begin{enumerate}
 \item the determination of $\delta$ by future experiments will 
not modify much the current correlation among mass matrix elements;
\item $\vert m_{ee}\vert$ should be in the reach of most proposed
  neutrinoless double beta decay experiments, if neutrinos are of
  Majorana nature, ergo the non-observability of the
    $0\nu\beta\beta$ would point to Dirac neutrinos;
\item $m_\beta$ may be in the reach of KATRIN;
\item the ordering of neutrino masses can be settled by future 
neutrino oscillation experiments but will not have great impact 
on the neutrino mass matrix;
\item if $\vert m_{ee}\vert$ is measured we will also be able to 
constrain much more $\vert m_{e\mu}\vert$ and $\vert m_{e\tau}\vert$. However, 
it will be very difficult to say something about the Majorana-CP phases.
\end{enumerate}

On the other hand, cosmology can also place a limit on $\sum m_i$, such that  
we will know if we have normal ordering with $m_1 \to 0$. In this case:

\begin{enumerate}
 \item $\vert m_{ee}\vert$ will be out of the reach of the proposed 
$0\nu\beta\beta$ experiments;
\item $m_\beta$ will be out of the reach of KATRIN;
\item the experimental determination of $\delta$ will 
increase the correlation among mass matrix elements and 
help to determine the structure of the neutrino mass matrix;
\item the ordering of neutrino masses should be confirmed by future 
neutrino oscillation experiments.
\end{enumerate}

It may be the case that cosmology will rule out a quasi-degenerate
spectrum, but not the inverted ordering.  If this happens:

\begin{enumerate}
 \item $\vert m_{ee}\vert$ may be in the reach of the proposed 
$0\nu\beta\beta$ experiments;
\item $m_\beta$ will be out of the reach of KATRIN;
\item the ordering of neutrino masses should be determined by future 
neutrino oscillation experiments;
\item the experimental determination of $\delta$ will 
increase the correlation among mass matrix elements and 
help to determine the structure of the neutrino mass matrix, specially 
if the mass ordering is known;
\item we may be able to say something about one of the  Majorana
CP phases, if $\vert m_{ee}\vert$ is measured.
\end{enumerate}

These future advances may thus provide new clues for the understanding of 
the flavor problem in the lepton sector. Some models of neutrino mixing based on 
discrete flavor groups have predictions that can be tested in the 
future. For instance, the Lin model~\cite{linmodel}, where the $A_4$
symmetry is broken by additional $Z_n$ parities, predicts 
$$\sin^2 \theta_{23} = \frac{1}{2} + \frac{1}{\sqrt{2}} \sin
\theta_{13} \cos \delta. $$

Another example is the SUSY model based on the flavor symmetry group 
$S_4 \times Z_4 \times U(1)$ discussed in \cite{Altarelli:2012bn} where 
the relation
$$ \sin^2 \theta_{12} =\frac{1}{2} + \sin \theta_{13} \cos \delta + {\cal O}
(\sin^2\theta_{13})$$
appears. Both  relations can, in principle, be experimentally tested and, 
if true, they will impose new correlations among the neutrino mass 
matrix elements.

Other models, such as the one presented in Ref.~\cite{Chen:2012st}, 
are even more predictive. This model, which is
based on a type-I seesaw framework with an underlying $A_4$ flavor symmetry, 
can, given a set of vacuum expectation value alignments for the flavon 
fields which break the $A_4$ symmetry, predict neutrino masses, the mass 
hierarchy, $\theta_{23}$ and $\delta$.

There are also model-independent approaches to the flavor problem in
the neutrino sector. In Ref.~\cite{Hernandez:2012ra,Hernandez:2012sk},
relations among the mixing parameters were obtained, in the context of
discrete flavor symmetries, under general assumptions that the flavor
symmetry group is of the von Dyck type. Again these relations can, in
principle, be experimentally tested, and if ratified by experiment
induce more correlations among the mixing matrix entries.

\section{The 3+1 Scenario}
\label{sec:3+1}
Whether or not one deems this to be a 
plausible scenario, we still believe it is important to examine what are 
its consequences to the possible textures of the neutrino mass matrix. 

In this case the mixing matrix can be parametrized as 
 \begin{equation}
U = \left(\begin{array}{cccc}
c_{12} \, c_{13} \, c_{14} & s_{12} \, c_{13} \, c_{14} & s_{13} \, c_{14} e^{-i \delta} & s_{14}\\
-s_{12}\,c_{23}-c_{12}s_{13}s_{23}\,e^{i \delta} & 
c_{12}\,c_{23}-s_{12}s_{13}s_{23}\,e^{i \delta} & c_{13}\,s_{23} & 0\\
s_{12}\,s_{23}-c_{12}s_{13}c_{23}\,e^{i \delta} & 
-c_{12}\,s_{23}-s_{12}s_{13}c_{23}\,e^{i \delta} & c_{13}\,c_{23} & 0 \\
-c_{12} \, c_{13} \, s_{14} & -s_{12} \, c_{13} \, s_{14} & -s_{13} \, s_{14} e^{-i \delta} & c_{14}
\end{array} \right),
\label{eq:matrix4}
\end{equation}
where we use the same notation as in Eq.~(\ref{eq:matrix}). This expression can be readily 
derived from \cite{Cirelli:2004cz} once we identify 
$\theta_s \equiv \theta_{14}$ and we assume $\vec{n} = (1,0,0)$, \emph{i.e.} the sterile 
state mixes only with the electron neutrino. With this assumptions, there are no 
extra Dirac CP phases. Nevertheless, there is one extra Majorana phase, 
$\lambda_4$. The expression of all the squared matrix elements, in the limit 
$c_{13} \sim c_{14} \to 1$, can be found in Appendix~\ref{appendixA}.

For simplicity, we examine here two cases: (a) $\Delta m_{41}^2 =1.71$ eV$^2$ and
$\sin^2 \theta_{14} = (0.8-4.2) \times 10^{-2}$,
(b) $\Delta m_{41}^2 =0.95$ eV$^2$ and
$\sin^2  \theta_{14} = (0.8-2.5) \times 10^{-2}$,
which are two possible solutions to the reactor and gallium 
anomalies~\cite{reactor-anomaly}. These solutions seem at first glance to be at 
odds with cosmology, but we will ignore this fact at this point.
Since $\sin^2 \theta_{14}$ is small, we do not expect big 
changes in the PDFs of  $\vert m_{\alpha\beta}\vert$, 
$\alpha,\beta=e, \mu, \tau$, except for the case $\vert m_{ee}\vert$.
We have explicitly checked that this is the case. 

\begin{figure}[htb]
\begin{center}
 \includegraphics[width=0.39\textwidth]{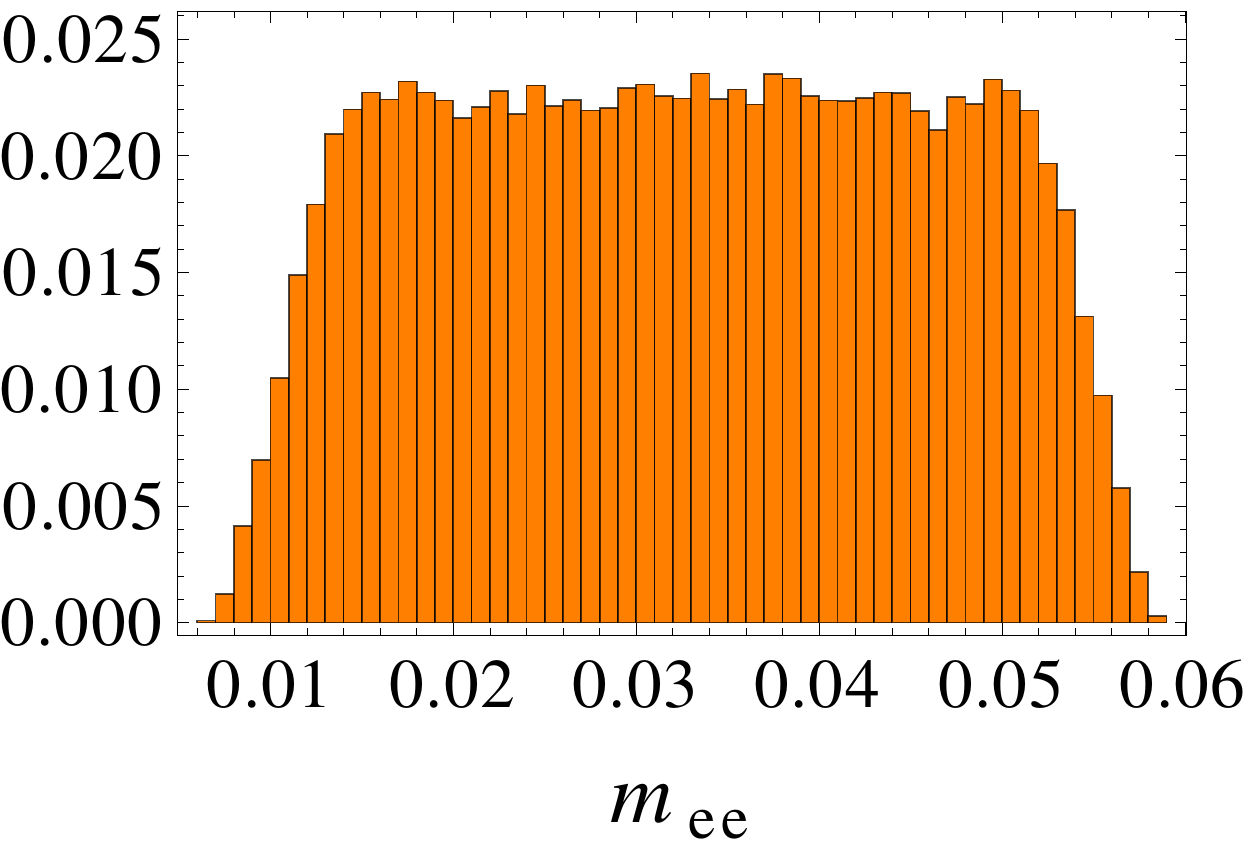}
 \includegraphics[width=0.39\textwidth]{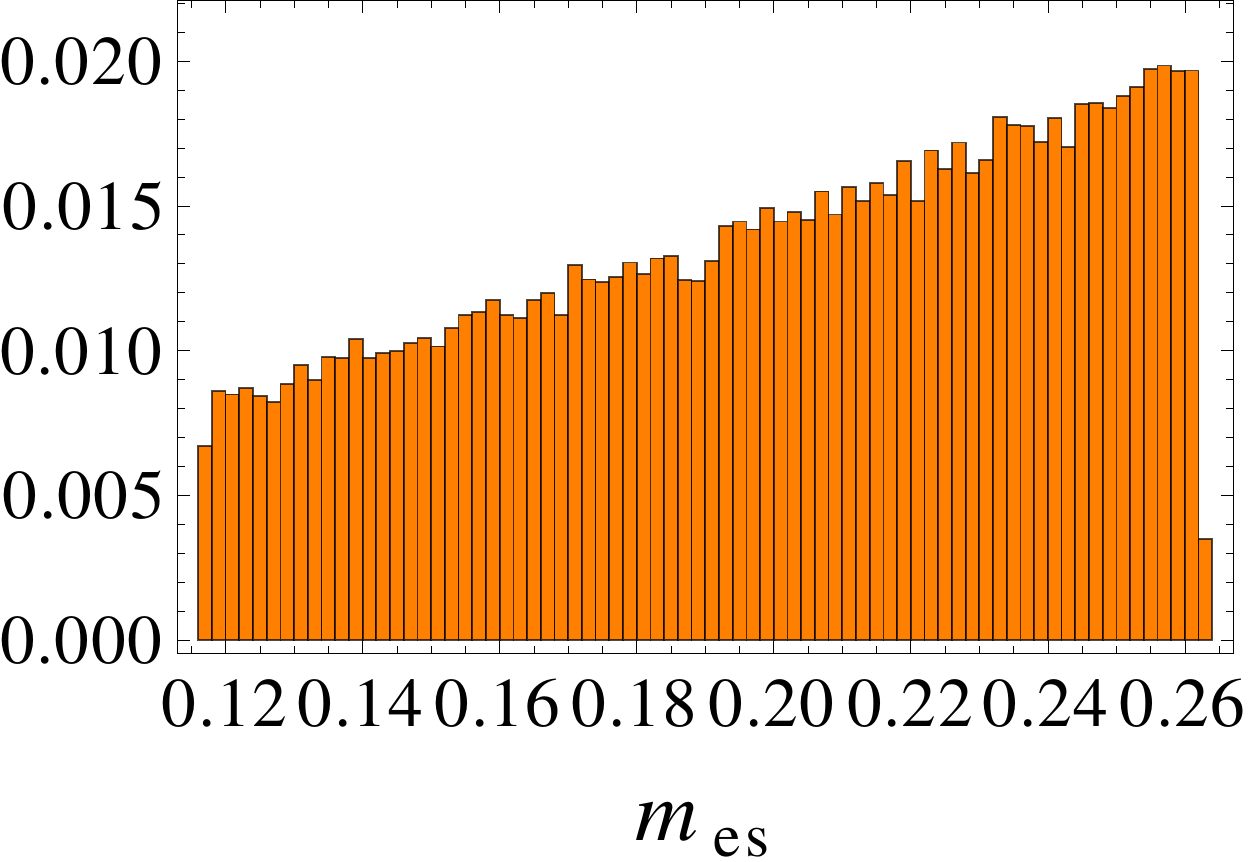}
\includegraphics[width=0.39\textwidth]{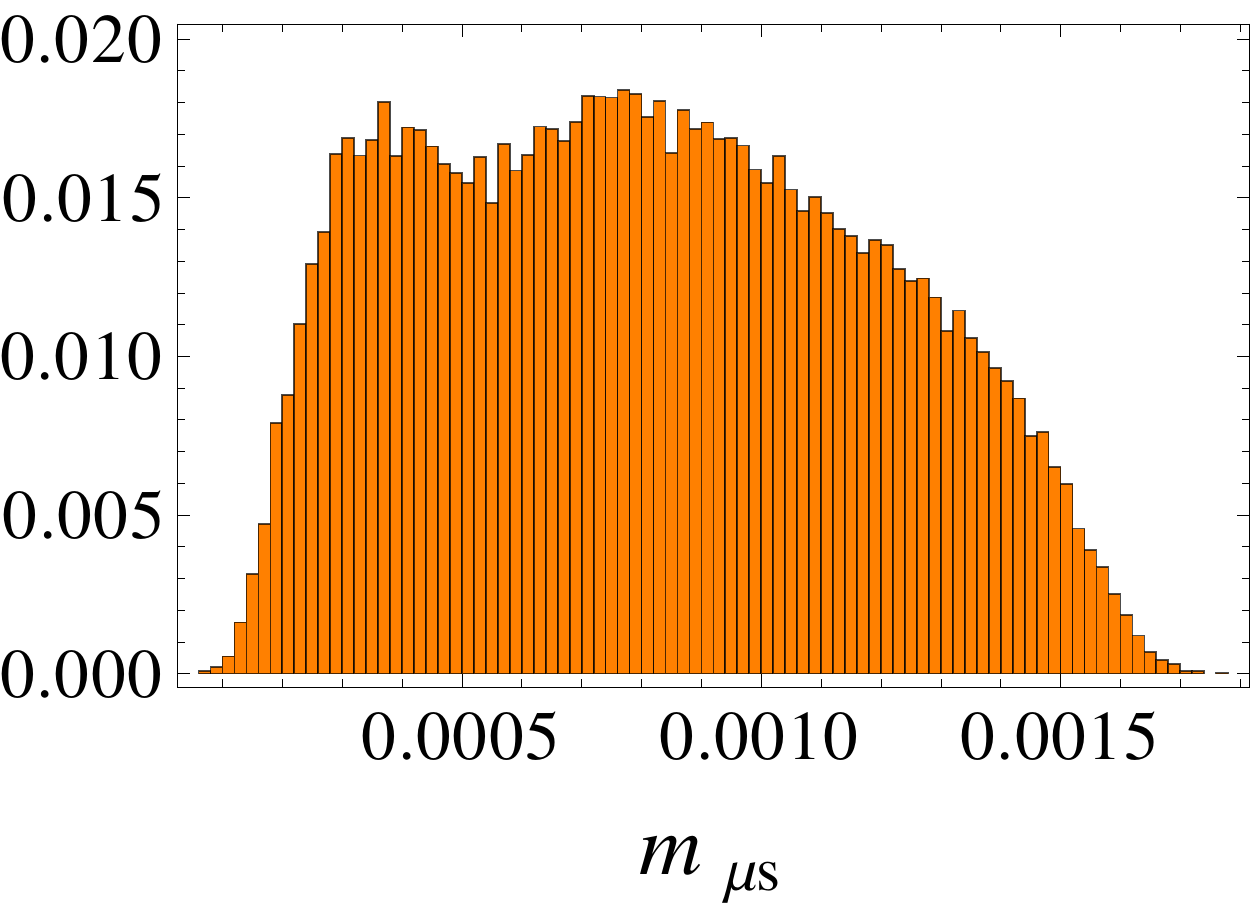}
\includegraphics[width=0.39\textwidth]{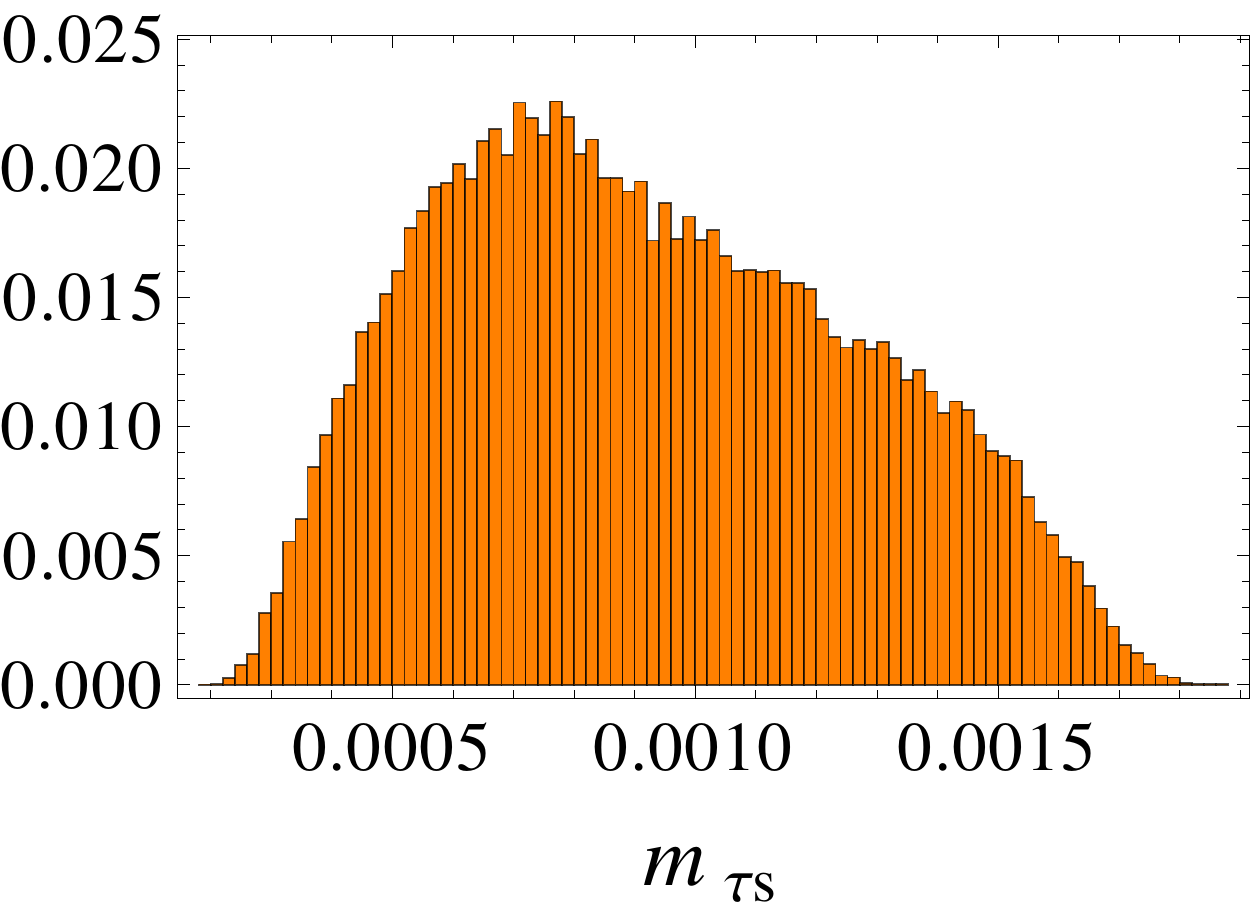}
\includegraphics[width=0.39\textwidth]{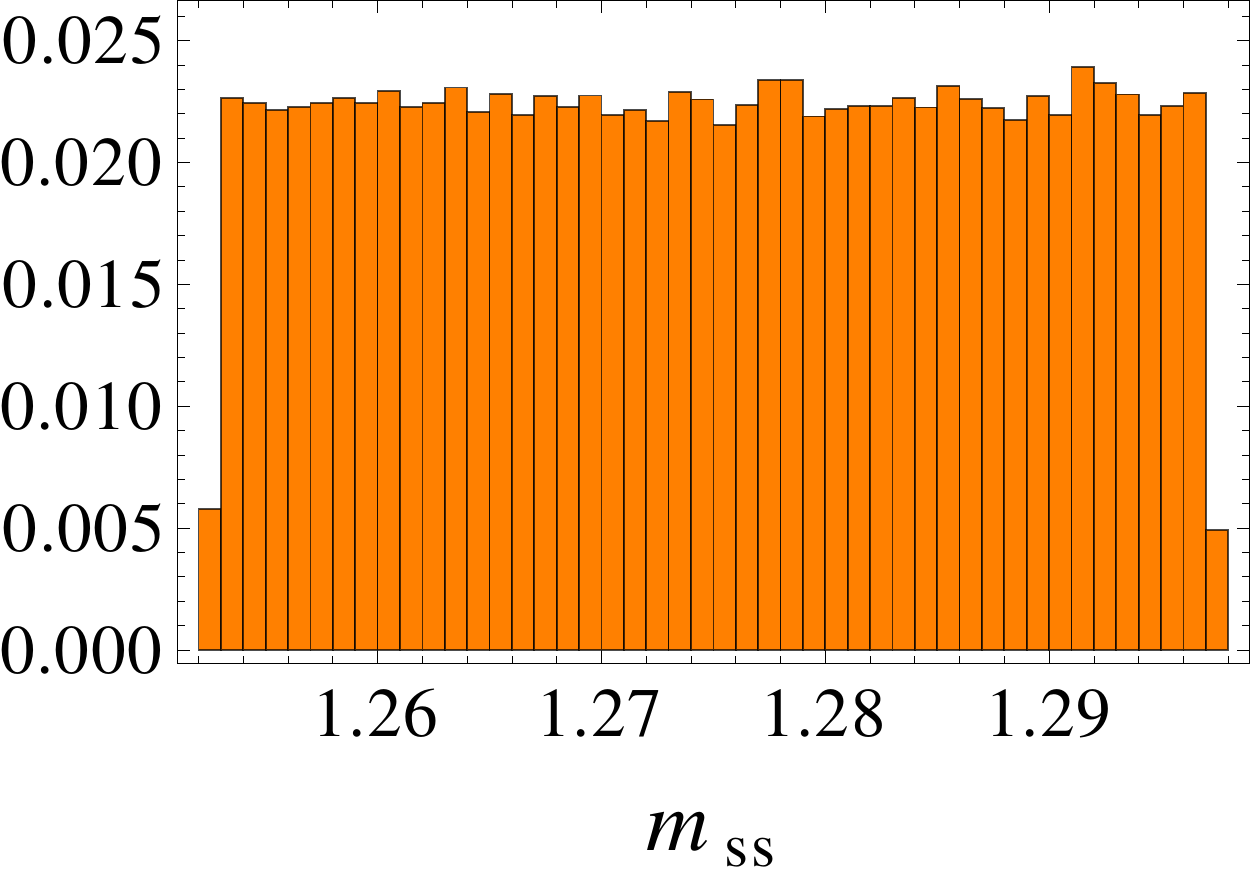}
\vspace{-2mm}
\end{center}
\vspace{-0.1cm}
\caption{PDFs for the distributions of the absolute value of 
$\vert m_{ee} \vert$, $\vert m_{es} \vert$,
$\vert m_{\mu s} \vert$, $\vert m_{\tau s}\vert$ 
and $\vert m_{s  s} \vert$, for the normal hierarchy
 and $\Delta m^2_{41}= 1.71$ eV$^2$.}
 \label{fig:4nus}
\end{figure}

In Fig.~\ref{fig:4nus} we show $\vert m_{ee}\vert$ and the new entries
$\vert m_{\alpha s}\vert, \quad \alpha = e,\mu,\tau$ for the normal
hierarchy and $\Delta m^2_{41}= 1.71$ eV$^2$.  For the case $\Delta
m^2_{41}= 0.95$ eV$^2$, the plots would be similar in shape, however 
with different scale. For $\vert m_{ee}\vert$ the largest value goes 
down from 0.06 eV to 0.045 eV. For $\vert m_{es}\vert$  the range 
changes from $\sim$(0.12 -- 0.26) eV to  $\sim$(0.09 -- 0.20) eV.
For $\vert m_{\mu s}\vert,\vert m_{\tau s}\vert$ there is basically no difference and 
for $\vert m_{ss}\vert$ there is again a shift in the range from 
$\sim$(1.25 -- 1.30) eV to  $\sim$(0.935 -- 0.97) eV.

\begin{figure}[htb]
\begin{center}
 \includegraphics[width=0.44\textwidth]{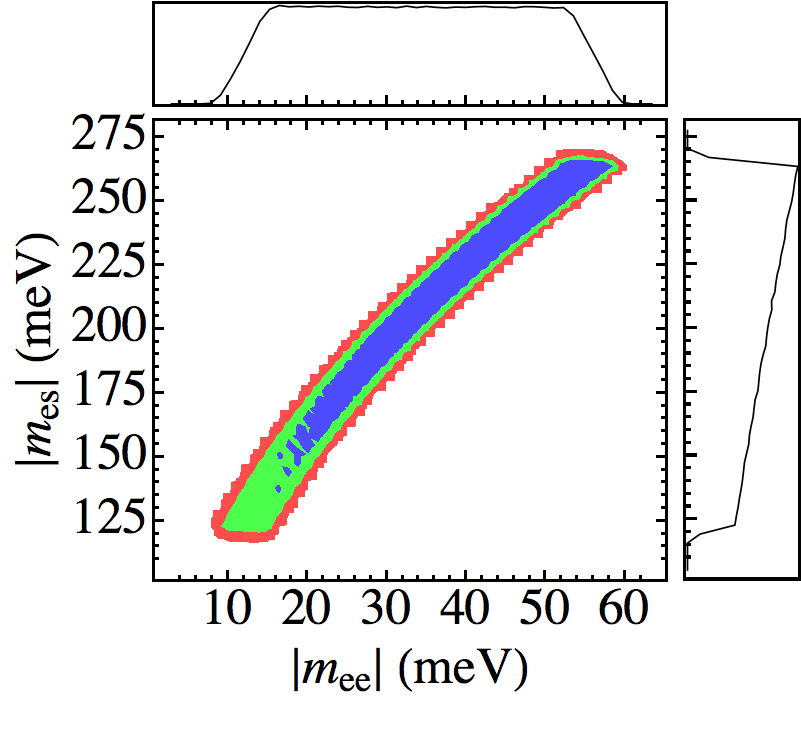}
 \includegraphics[width=0.44\textwidth]{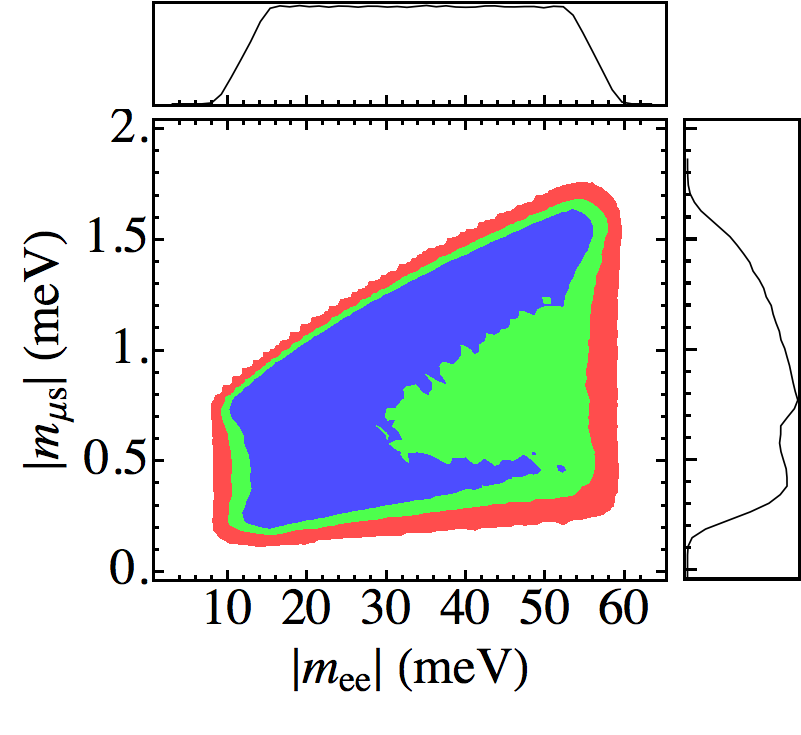}
\includegraphics[width=0.44\textwidth]{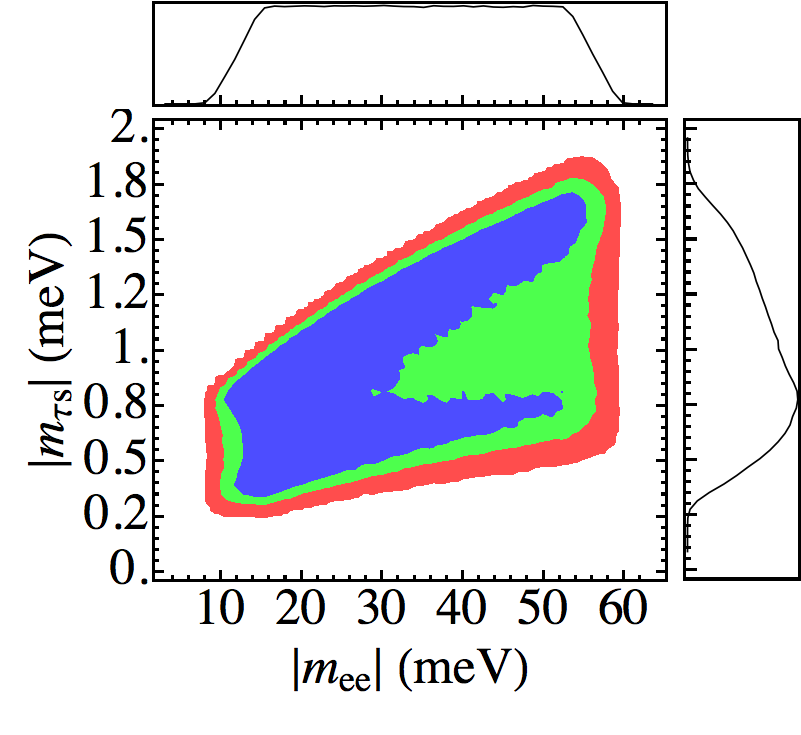}
\includegraphics[width=0.44\textwidth]{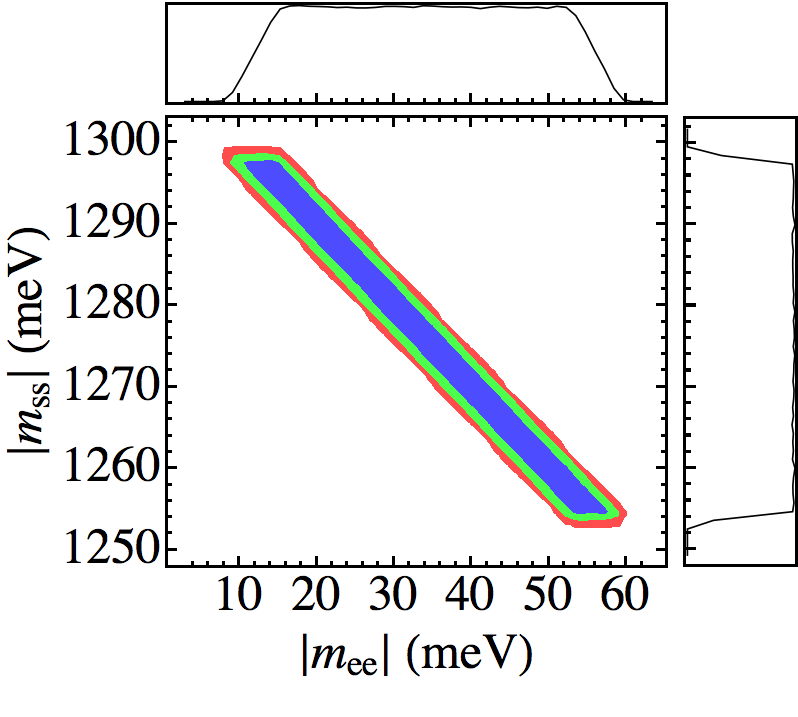}
\vspace{-2mm}
\end{center}
\vspace{-0.1cm}
\caption{PDFs for the distributions of the absolute values  
$\vert m_{ee} \vert \times \vert m_{es} \vert$,
$\vert m_{ee} \vert \times \vert m_{\mu s} \vert$,
$\vert m_{ee} \vert \times \vert m_{\tau s} \vert$ and 
$\vert m_{ee} \vert \times \vert m_{ss} \vert$, for the normal hierarchy and 
$\Delta m^2_{41}= 1.71$ eV$^2$.}
 \label{fig:4nus-cor}
\end{figure}

In Fig.~\ref{fig:4nus-cor} we show the correlations among the PDF's of $\vert m_{ee} \vert$ and 
$\vert m_{es} \vert$, $\vert m_{\mu s} \vert$, $\vert
m_{\tau s} \vert$, $\vert m_{ss} \vert$,
for the normal hierarchy and $\Delta m^2_{41}= 1.71$ eV$^2$.  Since
the difference between the case $\Delta m^2_{41}= 1.71$ eV$^2$ and
$\Delta m^2_{41}= 0.95$ eV$^2$ is basically the scale, as commented
above, we do not show the correlations in this case either.

We observe that $\vert m_{ee}\vert$ is very correlated with $\vert
m_{es}\vert$. This is easy to understand from the formulae in Appendix
\ref{appendixA}, as $\vert m_{ee} \vert$ goes as $s_{14}^2$ while
$\vert m_{es} \vert$ goes like $s_{14}$ producing a squared root
behavior.  The thickness is driven by the CP phases.  The linear
behavior between $\vert m_{ee}\vert$ and $\vert m_{ss}\vert$ can be
explained by noting that $\vert m_{ss}\vert \approx m_4 - \vert
m_{ee}\vert$. Again due to the approximate $\mu$-$\tau$ symmetry we
get similar ranges and behaviors for $\vert m_{ee} \vert \times \vert
m_{\mu s} \vert$ and $\vert m_{ee} \vert \times \vert m_{\tau s}
\vert$.

Also in this case it is interesting to consider future perspectives,
in particular considering the interplay between oscillation
experiments and cosmology. From the cosmology point a view, new data
by Planck should be soon released, with a measurement of the number of
light species. From the oscillation point of view, a part of the
currently allowed parameter space (that allows for a solution to the
reactor and Gallium anomalies) will be probed in the next couple of
years by the NUCIFER~\cite{nucifer} and STEREO~\cite{stereo}
experiments.  A much broader portion should be explored with a
timescale of more than ten years by the updated version of KamLAND,
CeLAND~\cite{celand}, in principle allowing to confirm or rule out the
presence of a sterile neutrino independently from the cosmological
measurements.

\begin{figure}[htb]
\begin{center}
 \includegraphics[width=0.9\textwidth]{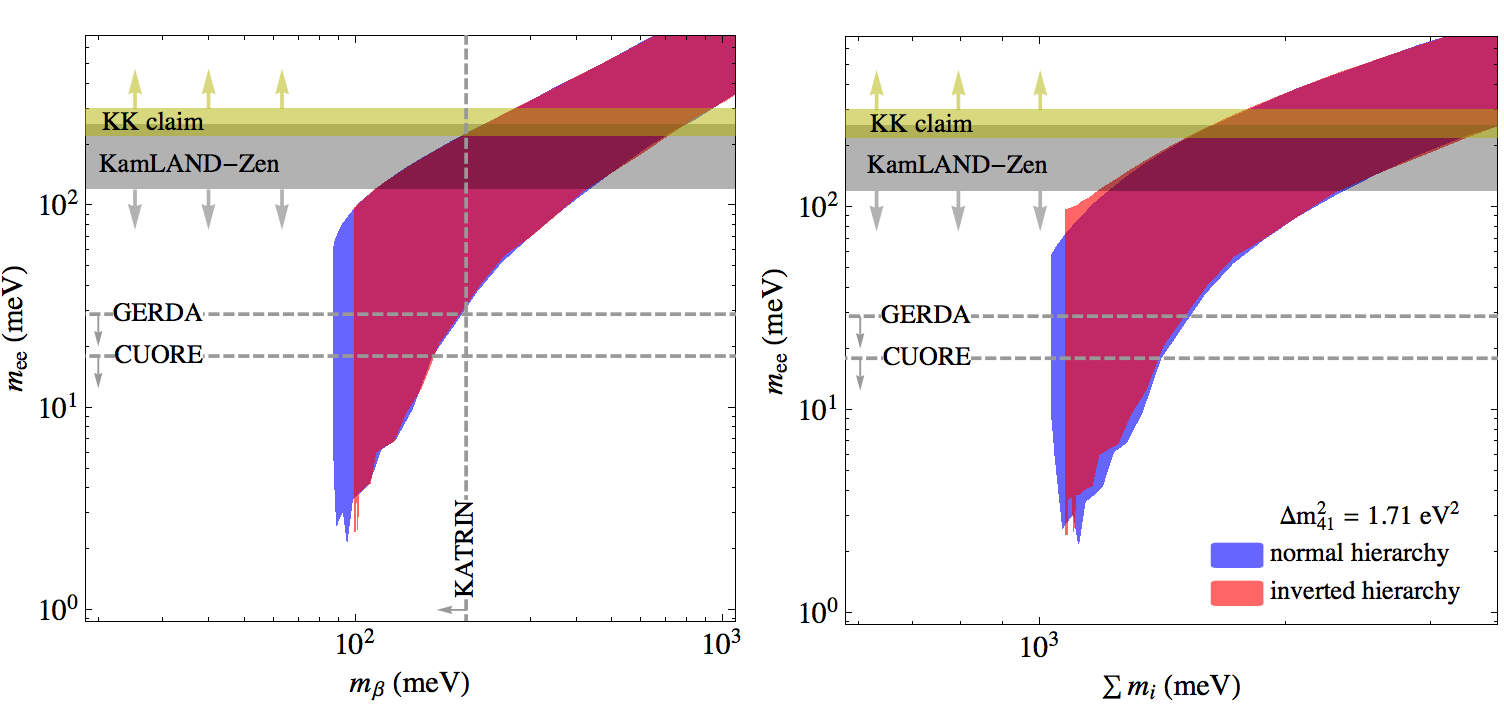}
\vspace{-2mm}
\end{center}
\vspace{-0.1cm}
\caption{We show the current allowed regions for $\vert m_{ee}\vert$
  at 99\% CL as a function of the effective electron neutrino mass,
  $m_\beta$, on the left panels and as a function of the sum of the
  neutrino masses, $\sum m_i$, on the right panels for the 3+1
  scenario. Here $\Delta m^2_{41} = 1.71$ eV$^2$.  The region
  allowed by the normal (inverted) mass ordering is in blue (red), and
  the recent limit on $\vert m_{ee}\vert$ given by
  KamLAND-Zen~\cite{Gando:2012zm} in gray. We also show the reach
  expected for the beta decay experiment Katrin~\cite{katrin}, as well
  as the ultimate reach aimed by the neutrinoless double beta decay
  experiments GERDA and CUORE according to Ref.~\cite{Sarazin:2012ct}. }
\label{fig:futurenus4}
\end{figure}

\section{Final Discussion and Conclusion}
\label{sec:conclusions}

In this paper we have analyzed in a probabilistic way the possible
structures and correlations among the neutrino mass matrix elements in
the standard neutrino oscillation framework, in view of the latest
global analysis of the neutrino oscillation data.  This is done by
constructing PDFs for each matrix element by assuming gaussian
distributions for the known oscillation parameters and flat
distributions for the unknown ones.

We analyzed the possible textures of the mass matrix allowed by data
in the hierarchical and almost degenerate cases and discussed the
future perspectives for better determining these matrix elements by
future neutrino oscillation and non-oscillation data.  The conclusion is 
that a better determination of the currently measured oscillation
parameters will, in general, have a small effect on the matrix
elements.  The biggest effect will come from a better determination of
$\sin^2 \theta_{23}$ solving the octant degeneracy, as one could have
guessed. A determination of $\delta$ would be significant, particularly for the normal
hierarchy.  In the inverted ordering, it would play a bigger role in
the determination of $\vert m_{\mu \tau} \vert$ and $\vert m_{\tau
  \tau} \vert$. For the quasi-degenerate case, the impact of the
determination of $\delta$ is rather small due to the presence of more
relevant CP phases. Future inputs from beta and neutrinoless double
beta decay experiments, as well as cosmology, seem to be the most
promising in providing new clues to understand the flavor structure.
A specially encouraging scenario would be to have $m_{\beta}$ in the
reach of KATRIN experiment, the neutrino mass hierarchy settled by
near future oscillation experiments, and also a possible cosmological
measurement of the sum of neutrino masses.

Models of neutrino mixing based on discrete flavor symmetries, as 
discussed at the end of in Sec.~\ref{sec:standard}, anticipate relations 
among mixing angles and the $\delta$ phase that can be tested in the 
future. If these relations turn out to be true, they will impose  
correlations among the mass matrix elements beyond the ones considered here.

We extend our analysis to include the possibility of a sterile
neutrino with mass and mixings allowed by the reactor and
gallium anomalies (disregarding current cosmological bounds). We
discuss what are the modifications to the mass matrix pattern in this
scenario, finding only relevant modifications for $m_{ee}$ as
expected. Despite the presence of more parameters than in the
standard scenario, the larger sterile neutrino mass would make easier
to measure both $\vert m_{ee} \vert$ and $m_\beta$. This may also have
a big impact in cosmological models, as well as in the future
strategies for oscillation experiments.

\begin{acknowledgments} 
  \vspace{-0.3cm} 
This work was supported by Funda\c{c}\~ao de Amparo
  \`a Pesquisa do Estado de S\~ao Paulo (FAPESP), Conselho Nacional 
de Desenvolvimento Cient\'ifico e Tecnol\'ogico (CNPq), by the European 
Commission under the contract PITN-GA-2009-237920 and by the Agence National 
de la Recherche under contract ANR 2010 BLANC 0413 01. 
R.Z.F. acknowledges partial support from the  European Union FP7  ITN INVISIBLES (Marie Curie Actions, PITN- GA-2011- 289442). 
\end{acknowledgments}

\appendix
\section{Matrix Elements Squared}
\label{appendixA}

Here we give some approximate expressions for the 
matrix elements $\vert m_{\alpha \beta} \vert^2$. We use the following notation:
$x^2 = \sin^2\theta_{12}$, $y^2 = \sin^2\theta_{13}$, 
$z^2 = \sin^2\theta_{23}$ and $w^2 = \sin^2\theta_{14}$. 
Here we set $\sqrt{1-y^2} \to 1$ and $\sqrt{1-w^2} \to 1$.
The standard case is recovered by taking $w \to 0$.

\begin{eqnarray}
\vert m_{ee}\vert^2 & \approx & m_4^2 w^4 + m_2^2 x^4+m_1^2
\left(1-x^2\right)^2+m_3^2 y^4 \nonumber \\ 
&+& 2 \, m_1 m_4\left(1-x^2\right) w^2 \cos \left[2
  \left(\lambda_1-\lambda_4\right)\right] \nonumber \\ 
&+& 2 \, m_1 m_2 \, x^2 \left(1-x^2\right) \cos \left(2 \lambda_1\right) 
\nonumber \\ 
&+& 2 \, m_1 m_3 \, y^2 \left(1-x^2\right) \cos \left[2
  \left(\delta -\lambda_1+\lambda_3\right)\right] \nonumber \\ 
&+& 2\, m_2 m_4 \, x^2 w^2 \cos \left(2 \lambda _4\right) \nonumber \\ 
&+& 2 \, m_2 m_3 \, x^2 y^2 \cos \left[2 \left(\delta
  +\lambda_3\right)\right] \nonumber \\ 
&+& 2 \, m_3 m_4 \, y^2 w^2\cos \left[2 \left(\delta +\lambda_3-\lambda_4 \right) \right]
\label{eq:mee}
\end{eqnarray}

\begin{eqnarray}
\vert m_{e\mu}\vert^2 &\approx & \left(1-x^2\right) x^2 (1-z^2)(m_1^2 + m_2^2) 
+ \left [ m_1^2 \, \left(1-x^2\right)^2 y^2 + m_2^2 \, x^4 y^2 + m_3^2 \, y^2 \right ] z^2 \nonumber \\
&+& \left[ 2 \, m_1 m_2 \, x^2 y^2 z^2 \left(1-x^2\right) - 2 \, m_1 m_2 \, x^2   \left(1-z^2\right) \left(1-x^2\right)  
\right] \cos \left(2 \lambda_1\right) \nonumber \\
&-& 2 \, m_1 m_3 \, y^2 z^2  (1-x^2) \cos \left[2 \left(\delta -\lambda_1+\lambda_3\right)\right] \nonumber \\
&+ & \left [ 2 \, m_1^2 \, x y z \left(1-x^2\right)^{3/2} \sqrt{1-z^2} - 2 \, m_2^2  \, x^{3} y z \sqrt{1-x^2} \sqrt{1-z^2}  \right] \cos \delta  \nonumber \\
&-& 2 \, m_1 m_2 \, x y z \left(1-x^2\right)^{3/2} \sqrt{1-z^2} \cos \left(\delta -2 \lambda_1\right) \nonumber \\
&-& 2 \, m_2 m_3 \, x^2 y^2 z^2 \cos \left[ 2\left(\delta + \lambda_3\right)\right] \nonumber \\
&+& 2 \, m_2 m_3 \, x y z \sqrt{1-x^2} \sqrt{1-z^2} \cos \left(\delta +2 \lambda_3\right) \nonumber \\
&- & 2 \, m_1 m_3 \, x y z \sqrt{1-x^2} \sqrt{1-z^2} \cos \left(\delta -2 \lambda_1 + 2 \lambda_3\right) \nonumber \\
&+ & 2 \, m_1 m_2 \, x^3 y z \sqrt{1-x^2} \sqrt{1-z^2} \cos \left(\delta +2 \lambda_1 \right) 
\end{eqnarray}

\begin{eqnarray}
\vert m_{e\tau}\vert^2 &\approx & \left(1-x^2\right) x^2 z^2 (m_1^2 + m_2^2) 
+ \left [ m_1^2 \, \left(1-x^2\right)^2 y^2  + m_2^2 \, x^4 y^2  
+ m_3^2 \, y^2 \right ] (1-z^2) \nonumber \\
&+& \left[ 2 \, m_1 m_2 \, x^2 y^2 (1-z^2) \left(1-x^2\right) - 2 \, m_1 m_2 \, x^2 z^2  
\left(1-x^2\right)\right] \cos \left(2 \lambda_1\right) \nonumber \\
&-& 2 \, m_1 m_3 \, y^2 (1-z^2)  (1-x^2) \cos \left[2 \left(\delta -\lambda_1+\lambda_3\right)\right] \nonumber \\
&- & \left [ 2 \, m_1^2 \, x y z \left(1-x^2\right)^{3/2} \sqrt{1-z^2} - 2 \, m_2^2  \, x^{3} y z \sqrt{1-x^2} \sqrt{1-z^2}  \right] \cos \delta  \nonumber \\
&+& 2 \, m_1 m_2 \, x y z \left(1-x^2\right)^{3/2} \sqrt{1-z^2} \cos \left(\delta -2 \lambda_1\right) \nonumber \\
&-& 2 \, m_2 m_3 \, x^2 y^2 (1-z^2) \cos \left[ 2\left(\delta + \lambda_3\right)\right] \nonumber \\
&-& 2 \, m_2 m_3 \, x y z \sqrt{1-x^2} \sqrt{1-z^2} \cos \left(\delta +2 \lambda_3\right) \nonumber \\
&+ & 2 \, m_1 m_3 \, x y z \sqrt{1-x^2} \sqrt{1-z^2} \cos \left(\delta -2 \lambda_1 + 2 \lambda_3\right) \nonumber \\
&- & 2 \, m_1 m_2 \, x^3 y z \sqrt{1-x^2} \sqrt{1-z^2} \cos \left(\delta +2 \lambda_1 \right) 
\end{eqnarray}
\begin{eqnarray}
\vert m_{\mu\mu} \vert^2 &\approx& 4 \left(1-z^2\right) x^2 y^2 z^2 (1-x^2)(m_1^2 + m_2^2) 
+ (1-z^2)^2 \left[ m_1^2 x^4 + m_2^2 (1-x^2)^2 \right] \nonumber \\
&+& \left[ m_1^2 \, \left(1-x^2\right)^2 y^4 + m_2^2 \, x^4 y^4 + m_3^2 \right] z^4\nonumber \\
&+& 2 \, m_1 m_2 \, x^2 \left(1-x^2\right) \left [ (1-z^2)^2 - 4 \, y^2 z^2   \left(1-z^2\right) + y^4 z^4 \right] 
\cos \left(2 \lambda_1\right) \nonumber \\
&+& 2 \, m_1 m_2 \, (1-z^2) x^4 y^2 z^2 \cos \left[2\left( \delta+\lambda_1\right)\right] \nonumber \\
&+& 2 \, m_1 m_2 \, (1-z^2) (1-x)^2 (1+x)^2 y^2 z^2 \cos \left[2\left( \delta-\lambda_1\right)\right] \nonumber \\
&+& 2 \, m_2 m_3 \, (1-z^2) (1-x^2)  z^2 \cos \left(2\lambda_3\right) \nonumber \\
&+& 2 \, (m_1^2 +m_2^2) \, (1-z^2) (1-x^2) x^2 y^2 z^2 \cos \left(2\delta\right) \nonumber \\
&+& 4 \, m_1 m_3 \, \sqrt{1-z^2} \sqrt{1-x^2} x y z^3 \cos \left(\delta -2\lambda_1 + 2 \lambda_3\right) \nonumber \\
&-& 4 \, m_2 m_3 \, \sqrt{1-z^2} \sqrt{1-x^2} x y z^3 \cos \left(\delta + 2 \lambda_3\right) \nonumber \\
&+& 4 \, m_1 m_2 \,x^3 y z \sqrt{1-z^2} \sqrt{1-x^2} \left[  y^2 z^2 - (1-z^2)  \right]
\cos \left(\delta + 2 \lambda_1\right) \nonumber \\
&-& 4 \, x y z \sqrt{1-x^2}  \sqrt{1-z^2} \left[m_1^2 \left(x^2 \left(\left(y^2+1\right)
   z^2-1\right)-y^2 z^2\right) \right . \nonumber \\
&+& \left . m_2^2 \left(x^2 \left(\left(y^2+1\right)
   z^2-1\right)-z^2+1\right)\right]\cos \left( \delta \right) \nonumber \\
&+& 4 \, m_1 m_2 \, x y z \sqrt{1-z^2} \sqrt{1-x^2} \, (1-x^2)
\left[1-(1+y^2) z^2\right] \cos \left( \delta -2 \lambda_1\right) \nonumber \\
&+& 2 \, m_2 m_3 \, x^2 y^2 z^4 \cos \left[ 2\left( \delta + \lambda_3\right)\right] \nonumber \\
&+& 2 \, m_1 m_3 \, (1-x^2) y^2 z^4 \cos \left[ 2\left( \delta -\lambda_1 + \lambda_3\right)\right] \nonumber \\
&+& 2 \, m_1 m_3 \, (1-z^2) x^2 z^2 \cos \left[2\left( \lambda_1-\lambda_3\right)\right] 
\end{eqnarray}

\begin{eqnarray}
\vert m_{\mu\tau} \vert^2 &\approx& (1-z^2) z^2 \left[ m_3^2 +(1-x^2)^2 (m_1^2 \, y^4 + m_2^2) + x^4 (m_1^2+ m_2^2 \, y^4)\right] \nonumber \\
&+& (1-x^2) \, x^2 y^2 \, (m_1^2+m_2^2) \left[ (1-z^2)^2 + z^4 \right] \nonumber \\
&+& 2 \, m_1 m_2 \, x^2 \left [ z^2(1-z^2) ((1+y^2)^2 -x^2)
- \, y^2 (1-x^2)(z^4+(1-z^2)^2) \right] 
\cos \left(2 \lambda_1\right) \nonumber \\
&+& 2 \, x y z (1-2 z^2)\sqrt{1-z^2}\sqrt{1-x^2}\left[ (1-x^2)(m_2^2+m_1^2 y^2) 
- x^2 (m_1^2+m_2^2 y^2) \right] \cos \delta  \nonumber \\ 
&-& 2 \, m_1 m_2 \, x y z (1-2 z^2)\sqrt{1-z^2}(1-x^2)^{3/2} (1+y^2) \cos \left( \delta -2 \lambda_1\right) \nonumber \\ 
&+& 2 \, m_1 m_3 \, x y z (1-2 z^2)\sqrt{1-z^2}\sqrt{1-x^2} \cos \left( \delta -2 \lambda_1 + 2 \lambda_3\right) \nonumber \\ 
&-& 2 \, m_2 m_3 \, x y z (1-2 z^2)\sqrt{1-z^2}\sqrt{1-x^2} \cos \left( \delta + 2 \lambda_3\right) \nonumber \\
&-& 2 \, (m_1^2 +m_2^2) \, (1-z^2) (1-x^2) x^2 y^2 z^2 \cos \left(2\delta\right) \nonumber \\ 
&-& 2 \, m_2 m_3 \, (1-z^2) (1-x^2)  z^2 \cos \left(2\lambda_3\right) \nonumber \\
&+& 2 \, m_1 m_2 \, x^3 y z (1-2 z^2)\sqrt{1-z^2}\sqrt{1-x^2}(1+y^2) \cos \left( \delta +2 \lambda_1 \right) \nonumber \\ 
&-& 2 \, m_1 m_2 \, y^2 z^2 (1-z^2)(1-x^2)^2 \cos \left[2\left( \delta - \lambda_1\right) \right] \nonumber \\ 
&-& 2 \, m_1 m_2 \, x^4 y^2 z^2 (1-z^2)  \cos \left[2\left( \delta + \lambda_1\right) \right] \nonumber \\ 
&+& 2 \, m_2 m_3 \, x^2 y^2 z^2 (1-z^2)  \cos \left[2\left( \delta + \lambda_3\right) \right] \nonumber \\ 
&+&  m_1 m_3 \, y^2 (1-x^2)  \cos \left[2\left( \delta -\lambda_1 + \lambda_3\right) \right] \nonumber \\ 
&-& 2 \, m_1 m_3 \, (1-z^2) x^2 z^2 \cos \left[2\left( \lambda_1-\lambda_3\right)\right] 
\end{eqnarray}

\begin{eqnarray}
\vert m_{\tau\tau} \vert^2 &\approx&
4 \left(1-z^2\right) x^2 y^2 z^2 (1-x^2)(m_1^2 + m_2^2) 
+ z^4 \left[ m_1^2 x^4 + m_2^2 (1-x^2)^2 \right] \nonumber \\
&+& \left[ m_1^2 \, \left(1-x^2\right)^2 y^4 + m_2^2 \, x^4 y^4 + m_3^2 \right] (1-z^2)^2\nonumber \\
&+& 2 \, m_1 m_2 \, x^2 \left(1-x^2\right) \left [ y^4 (1-z^2)^2 - 4 \, y^2 z^2   \left(1-z^2\right) + z^4 \right] 
\cos \left(2 \lambda_1\right) \nonumber \\
&+& 2 \, m_1 m_2 \, (1-z^2) x^4 y^2 z^2 \cos \left[2\left( \delta+\lambda_1\right)\right] \nonumber \\
&+& 2 \, m_1 m_2 \, (1-z^2) (1-x)^2 (1+x)^2 y^2 z^2 \cos \left[2\left( \delta-\lambda_1\right)\right] \nonumber \\
&+& 2 \, m_2 m_3 \, (1-z^2) (1-x^2)  z^2 \cos \left(2\lambda_3\right) \nonumber \\
&+& 2 \, (m_1^2 +m_2^2) \, (1-z^2) (1-x^2) x^2 y^2 z^2 \cos \left(2\delta\right) \nonumber \\
&-& 4 \, m_1 m_3 \, \sqrt{1-z^2} \sqrt{1-x^2} x y z (1-z^2) \cos \left(\delta -2\lambda_1 + 2 \lambda_3\right) \nonumber \\
&+& 4 \, m_2 m_3 \, \sqrt{1-z^2} \sqrt{1-x^2} x y z (1-z^2) \cos \left(\delta + 2 \lambda_3\right) \nonumber \\
&+& 4 \, m_1 m_2 \,x^3 y z \sqrt{1-z^2} \sqrt{1-x^2} \left[  y^2 z^2 - y^2 + z^2 \right]
\cos \left(\delta + 2 \lambda_1\right) \nonumber \\
&-& 4 \, x y z \sqrt{1-x^2}  \sqrt{1-z^2} \left[ m_1^2 \left(\left(1-x^2\right) y^2
   \left(1- z^2\right)+x^2 z^2\right) \right . \nonumber \\
&-& \left . m_2^2 \left(x^2 y^2
   \left(1-z^2\right) +z^2(1-x^2 ) \right)\right] \cos \left( \delta \right) \nonumber \\
&+& 4 \, m_1 m_2 \, x y z \sqrt{1-z^2} \sqrt{1-x^2} \, (1-x^2)
\left[(1-z^2) y^2 -z^2\right] \cos \left( \delta -2 \lambda_1\right) \nonumber \\
&+& 2 \, m_2 m_3 \, x^2 y^2 (1-z)^2 (1+z)^2 \cos \left[ 2\left( \delta + \lambda_3\right)\right] \nonumber \\
&+& 2 \, m_1 m_3 \, (1-x^2) y^2 (1-z)^2 (1+z)^2 \cos \left[ 2\left( \delta -\lambda_1 + \lambda_3\right)\right] \nonumber \\
&+& 2 \, m_1 m_3 \, (1-z^2) x^2 z^2 \cos \left[2\left( \lambda_1-\lambda_3\right)\right] 
\end{eqnarray}

\begin{eqnarray}
\vert m_{se} \vert^2 &\approx& w^2 \left[m_2^2 x^4 + m_1^2 \left(1-x^2\right)^2 +m_3^2 y^4
+ m_4^2\right] \nonumber \\
&-& 2 \, m_3 m_4 \, w^2 y^2    \cos \left[ 2\left(\delta +\lambda_3-\lambda_4\right) \right]
\nonumber \\
&+& 2 \, m_1 m_2 \, w^2 x^2 \left(1-x^2\right) \cos \left(2\lambda_1\right) \nonumber \\
&+& 2 \, m_1 m_3 \, w^2 y^2 \left(1-x^2\right) \cos \left[2 \left(\delta -\lambda_1+\lambda_3\right) \right] \nonumber \\
&-& 2 \, m_1 m_4 \, w^2 \left(1-x^2\right) \cos \left[2 \left(\lambda_1-\lambda_4\right) \right] \nonumber \\
&+& 2 \, m_2 m_3 \, w^2 x^2 y^2 \cos \left[2 \left(\delta +\lambda_3\right)\right] \nonumber \\
&-& 2 \, m_2 m_4 \, w^2 x^2 \cos \left(2 \lambda_4\right)
\end{eqnarray}

\begin{eqnarray}
\vert m_{s\mu} \vert^2 &\approx&
w^2 \left[m_1^2 \left(x^2-1\right) \left(x^2 \left(\left(y^2+1\right) z^2-1\right)-y^2
   z^2\right)+m_2^2 x^2 \left(x^2 \left(\left(y^2+1\right) z^2-1\right)-z^2+1\right)+m_3^2
   y^2 z^2\right]\nonumber \\
&+& 2 \, w^2 x y z \sqrt{1-x^2} \sqrt{1-z^2} \left[ 
(1-x^2) m_1^2 - x^2 m_2^2 \right] \cos \delta  \nonumber \\
&-& 2 \, m_1 m_2 \, w^2 x y z (1-x^2)^{3/2} \sqrt{1-z^2} \cos \left(\delta -2 \lambda_1\right)\nonumber \\
&+& 2 \, m_1 m_2 \, w^2 x^3 y z \sqrt{1-x^2} \sqrt{1-z^2} \cos \left(\delta +2 \lambda_1\right)\nonumber \\
&-& 2 \, m_1 m_3 \, w^2 x y z \sqrt{1-x^2} \sqrt{1-z^2} \cos \left(\delta -2 \lambda_1 + 2 \lambda_3\right) \nonumber \\
&+& 2 \, m_2 m_3 \, w^2 x y z \sqrt{1-x^2} \sqrt{1-z^2} \cos \left(\delta +2 \lambda_3\right)\nonumber \\
&-& 2 \, m_2 m_3 \, w^2 x^2 y^2 z^2 \cos \left[2 \left(\delta +\lambda_3\right)\right] \nonumber \\
&+& 2 \, m_1 m_2 \, w^2 x^2 (1-x^2) \left[(1+y^2) z^2 -1 \right]\ cos \left(2 \lambda_1\right) \nonumber \\
&-& 2\, m_1 m_3 \, (1-x^2) \, w^2 y^2 z^2 \cos \left[2 \left(\delta -\lambda_1+\lambda_3\right)\right] 
\end{eqnarray}

\begin{eqnarray}
\vert m_{s\tau} \vert^2 &\approx&
w^2 \left[m_1^2 \left(1-x^2\right) \left(\left(1-x^2\right) y^2
   \left(1-z^2\right)+x^2 z^2\right) \right . \nonumber \\
&+& \left . m_2^2 \left(x^2 z^2+x^4 \left(y^2 \left(1-z^2\right)-z^2\right)\right)+m_3^2 y^2 \left(1-z^2\right)\right] \nonumber \\
&-& 2 \, w^2 x y z \sqrt{1-x^2} \sqrt{1-z^2} \left[ 
(1-x^2) m_1^2 - x^2 m_2^2 \right] 
\cos \delta  \nonumber \\
&+& 2 \, m_1 m_2 \, w^2 x y z (1-x^2)^{3/2} \sqrt{1-z^2} 
\cos \left(\delta -2 \lambda_1\right)\nonumber \\
&-& 2 \, m_1 m_2 \, w^2 x^3 y z \sqrt{1-x^2} \sqrt{1-z^2} 
\cos \left(\delta +2 \lambda_1\right)\nonumber \\
&+& 2 \, m_1 m_3 \, w^2 x y z \sqrt{1-x^2} \sqrt{1-z^2} 
\cos \left(\delta -2 \lambda_1 + 2 \lambda_3\right) \nonumber \\
&-& 2 \, m_2 m_3 \, w^2 x y z \sqrt{1-x^2} \sqrt{1-z^2} 
\cos \left(\delta +2 \lambda_3\right)\nonumber \\
&-& 2 \, m_2 m_3 \, w^2 x^2 y^2 (1-z^2) 
\cos \left[2 \left(\delta +\lambda_3\right)\right] \nonumber \\
&-& 2 \, m_1 m_2 \, w^2 x^2 (1-x^2) \left[z^2 -(1-z^2) y^2  \right]
\ cos \left(2 \lambda_1\right) \nonumber \\
&-& 2\, m_1 m_3 \, (1-x^2) \, w^2 y^2 (1-z^2) 
\cos \left[2 \left(\delta -\lambda_1+\lambda_3\right)\right] 
\end{eqnarray}

\begin{eqnarray}
\vert m_{ss} \vert^2 &\approx& w^4 \left[ (1-x^2)^2 m_1^2 + x^4 m_2^2 + y^4 m_3^2\right] + m_4^2  \nonumber \\
& +& 2 \, m_1 m_2 \, w^4 x^2 (1-x^2) \cos (2 \lambda_1 ) \nonumber \\
& +& 2 \, m_2 m_4 \, w^4 x^2 \cos (2 \lambda_4 ) \nonumber \\
& +& 2 \, m_1 m_3 \, w^4 y^2 (1-x^2) \cos \left[2 \left( \delta -\lambda_1 + \lambda_3 \right) \right] \nonumber \\
& +& 2 \, m_1 m_4 \, w^2 (1-x^2) \cos \left[ 2\left( \lambda_1-\lambda_4 \right) \right] \nonumber \\
& +& 2 \, m_2 m_3 \, w^4 x^2 y^2 \cos \left[ 2\left( \delta +\lambda_3 \right) \right] \nonumber \\
& +& 2 \, m_3 m_4 \, w^2 y^2 \cos \left[ 2\left( \delta +\lambda_3 - \lambda_4 \right) \right] 
\end{eqnarray}

\section{Complete Set of Correlation Plots for the Matrix Elements}
\label{appendixB}

\begin{figure}[p]
\begin{center}
 \includegraphics[width=0.3\textwidth]{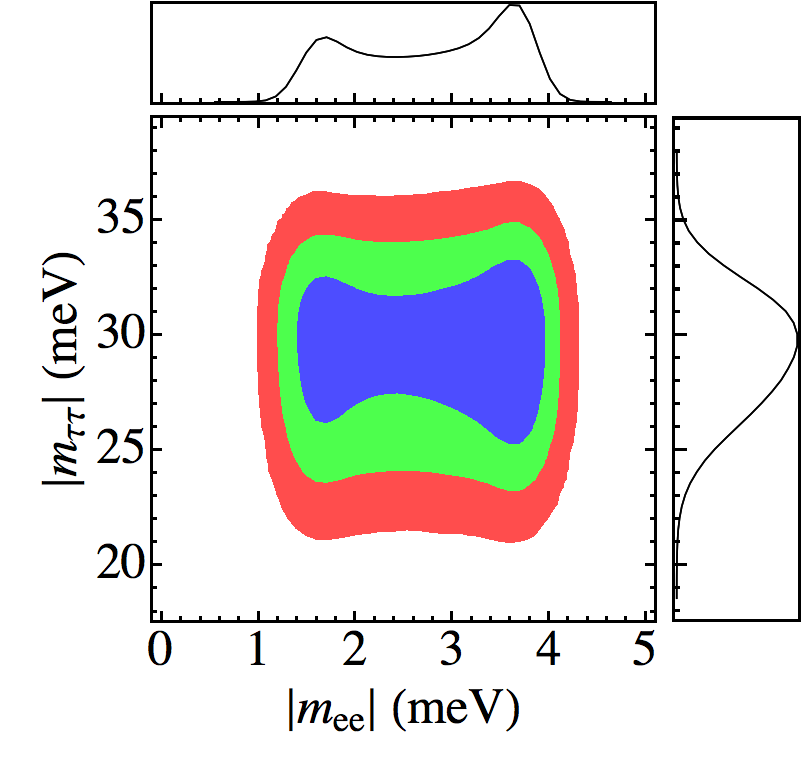}
 \includegraphics[width=0.3\textwidth]{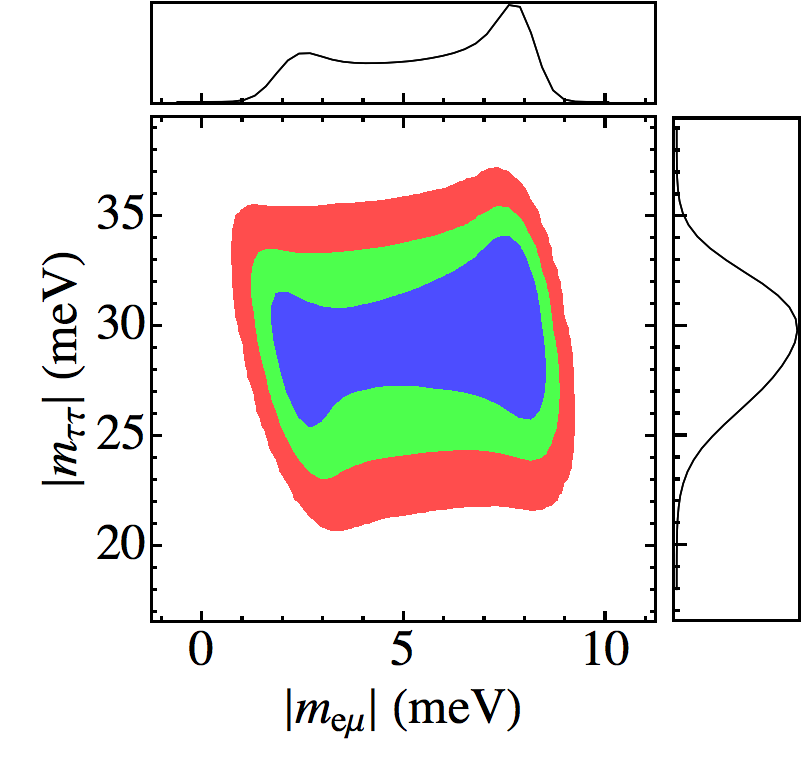}
 \includegraphics[width=0.3\textwidth]{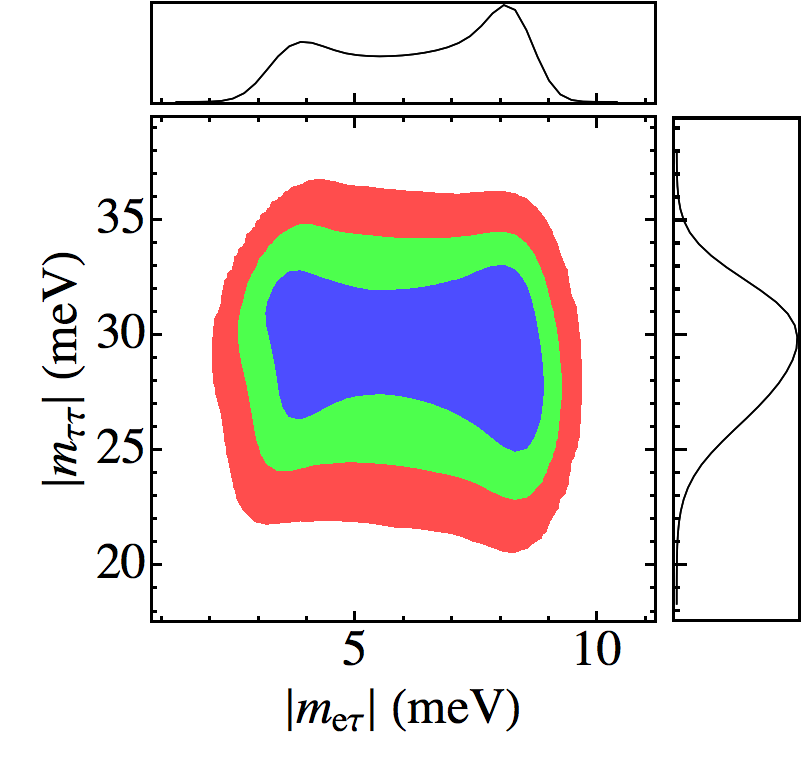}
 \includegraphics[width=0.3\textwidth]{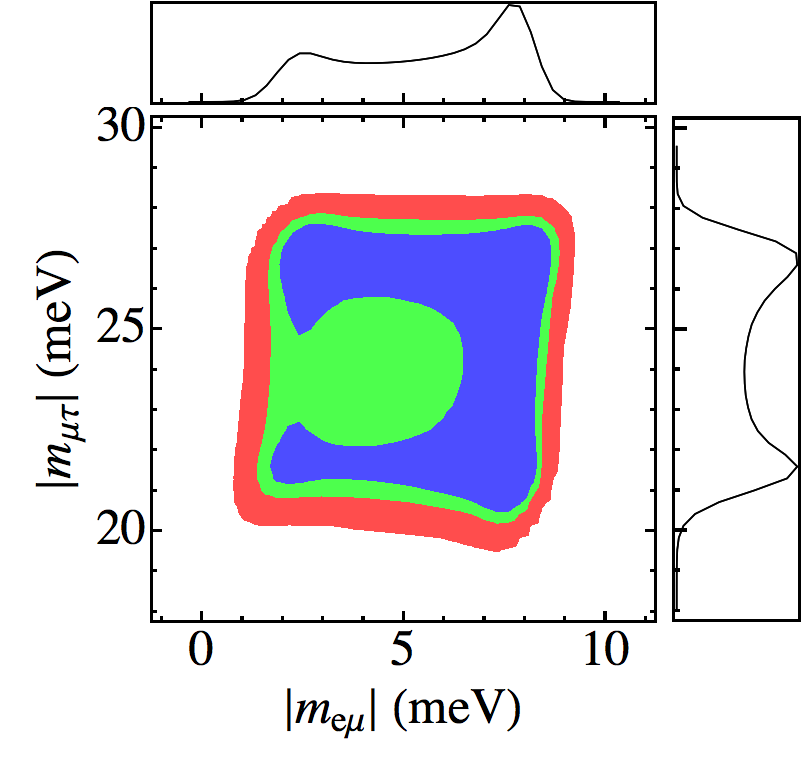}
 \includegraphics[width=0.3\textwidth]{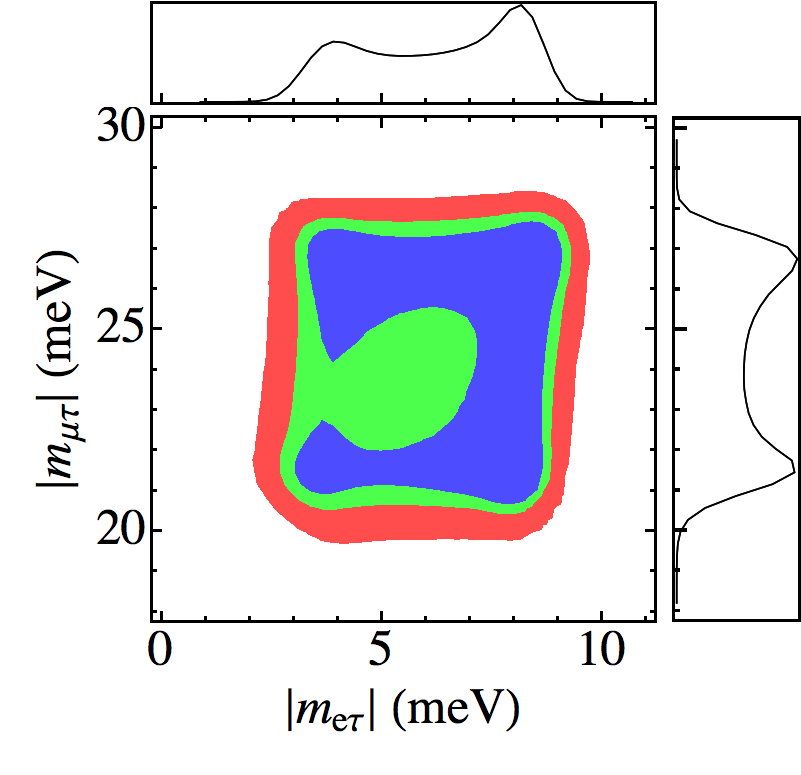}
 \includegraphics[width=0.3\textwidth]{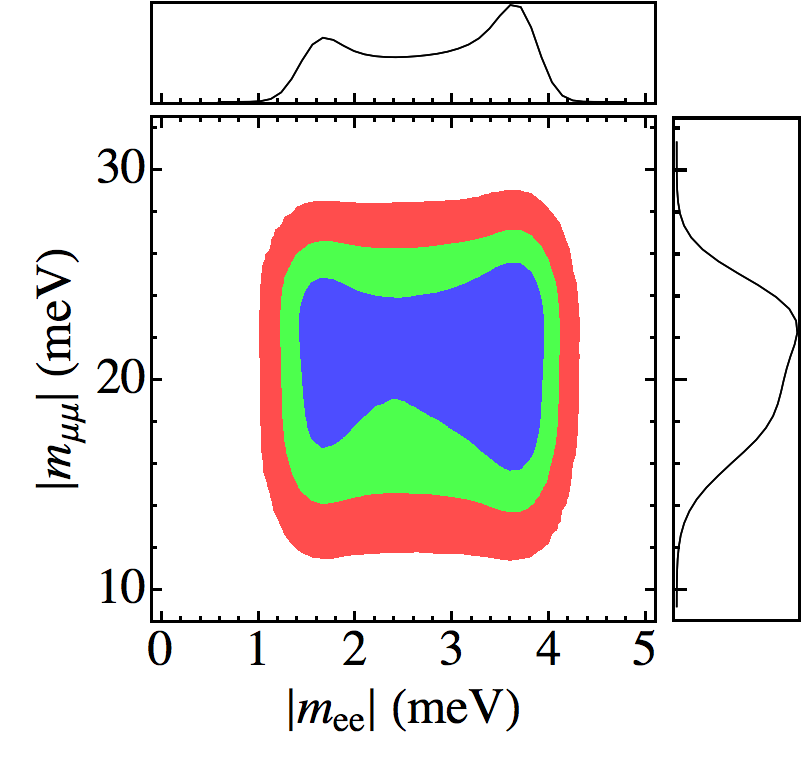}
 \includegraphics[width=0.3\textwidth]{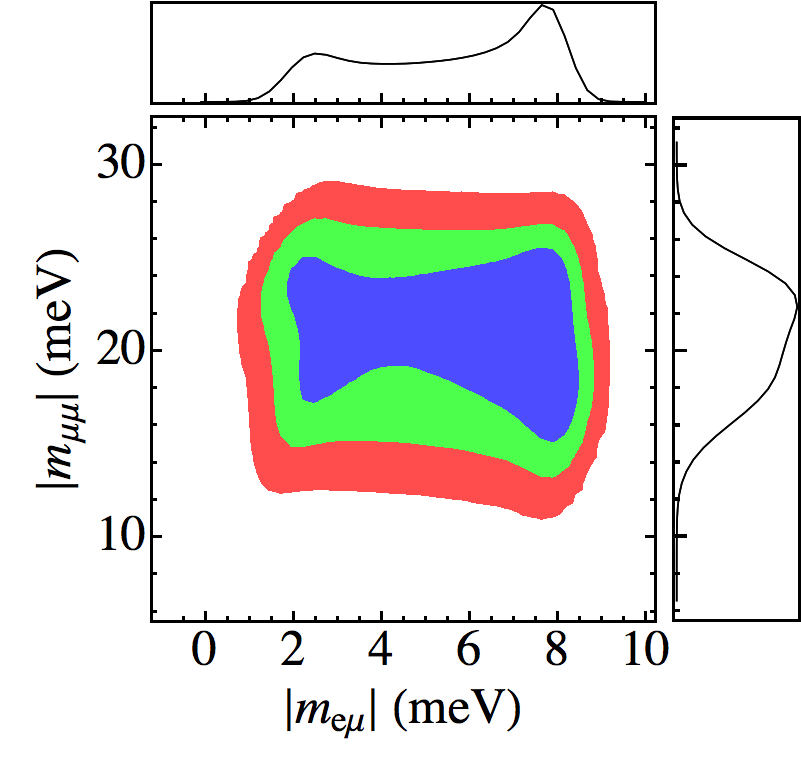}
 \includegraphics[width=0.3\textwidth]{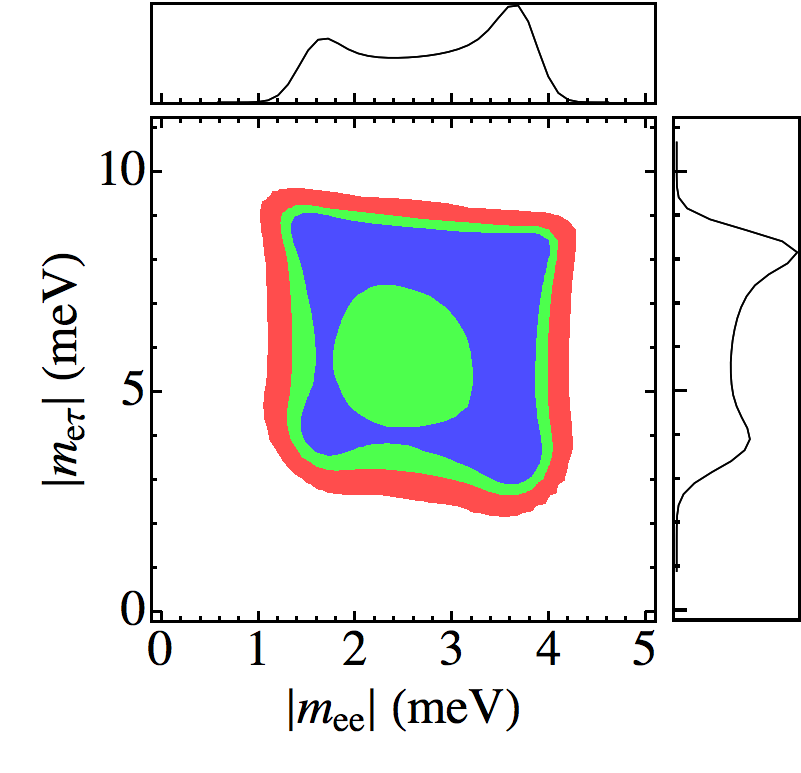}
 \includegraphics[width=0.3\textwidth]{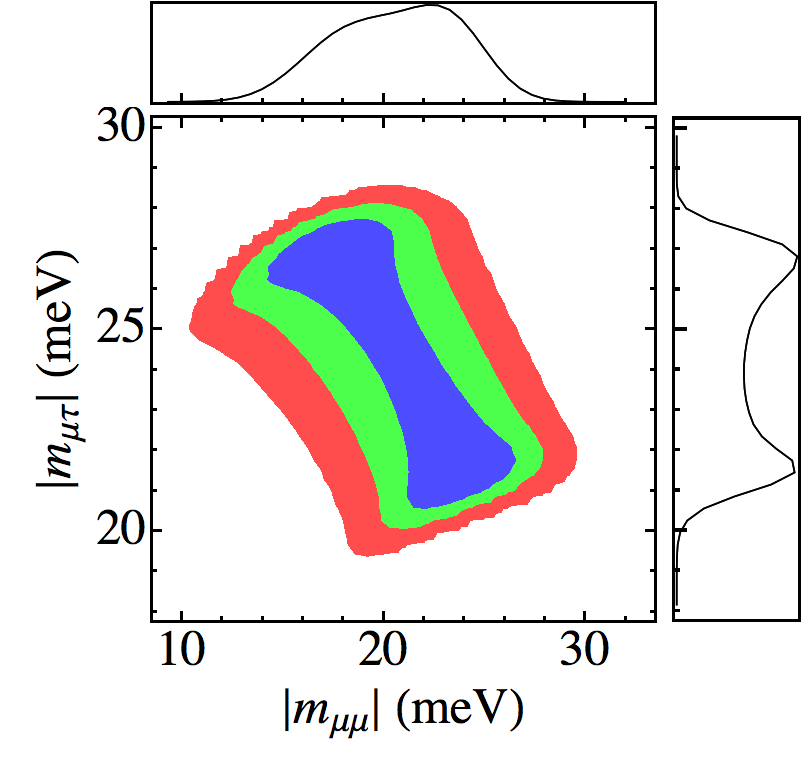}
\vspace{-2mm}
\end{center}
\vspace{-0.1cm}
\caption{PDFs for the distribution of the absolute value of several 
pairs of matrix elements. We use blue, green and
  red for the allowed region at 68.27\%, 95.45\% and 99.73\% CL,
  respectively. Here $m_1 \to 0$ and $\theta_{23}$ is 
assumed to be in the first octant.}
\label{fig:nh-1st}
\end{figure}

\begin{figure}[htb]
\begin{center}
 \includegraphics[width=0.28\textwidth]{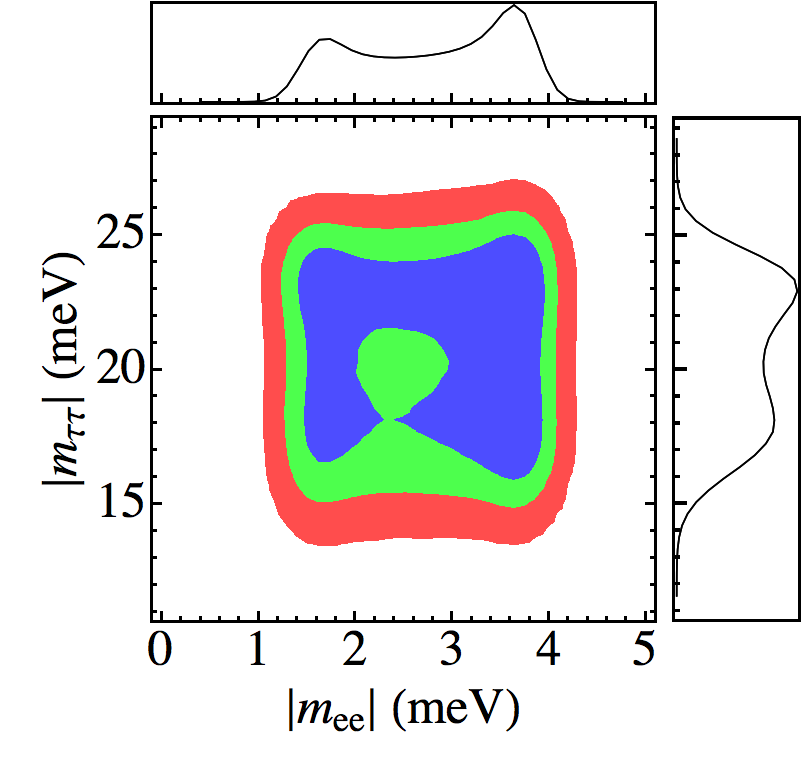}
 \includegraphics[width=0.28\textwidth]{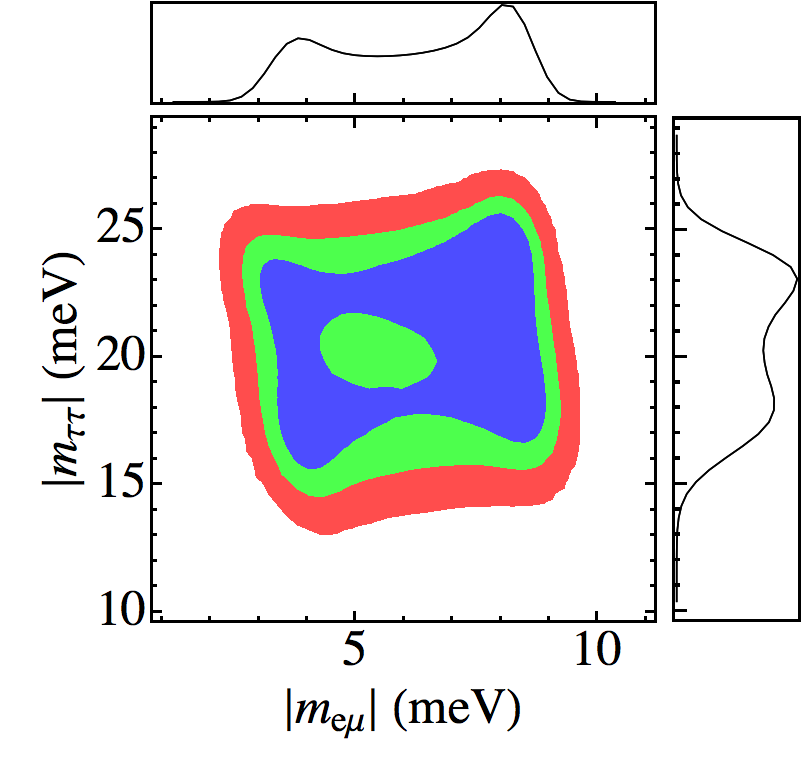}
 \includegraphics[width=0.28\textwidth]{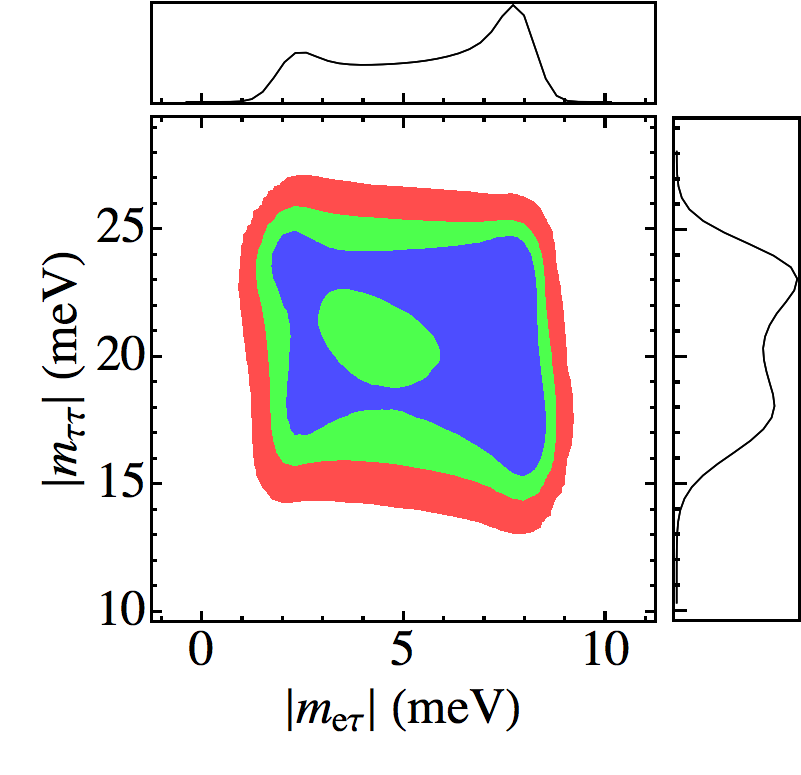}
 \includegraphics[width=0.28\textwidth]{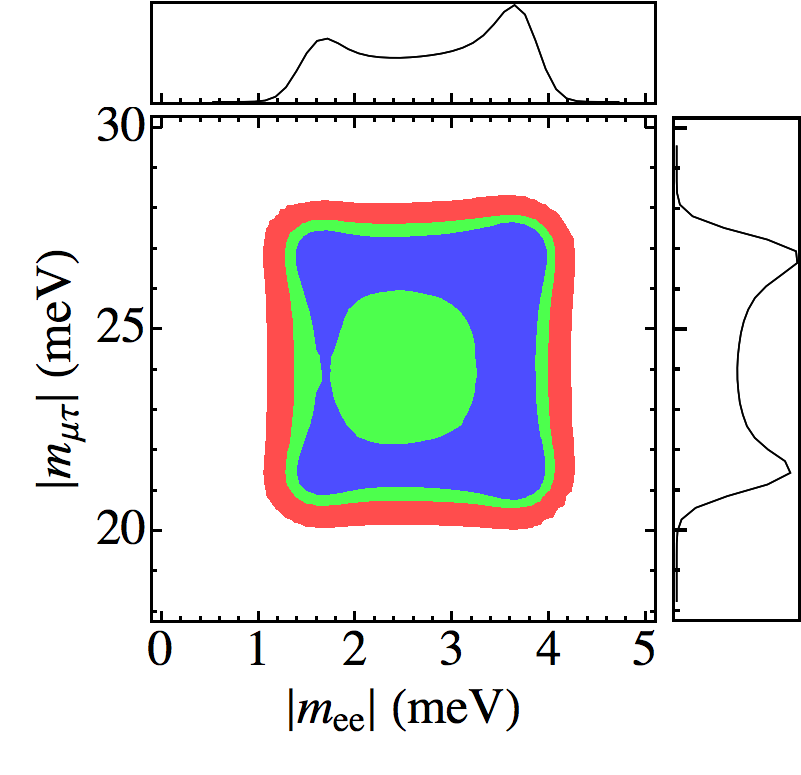}
 \includegraphics[width=0.28\textwidth]{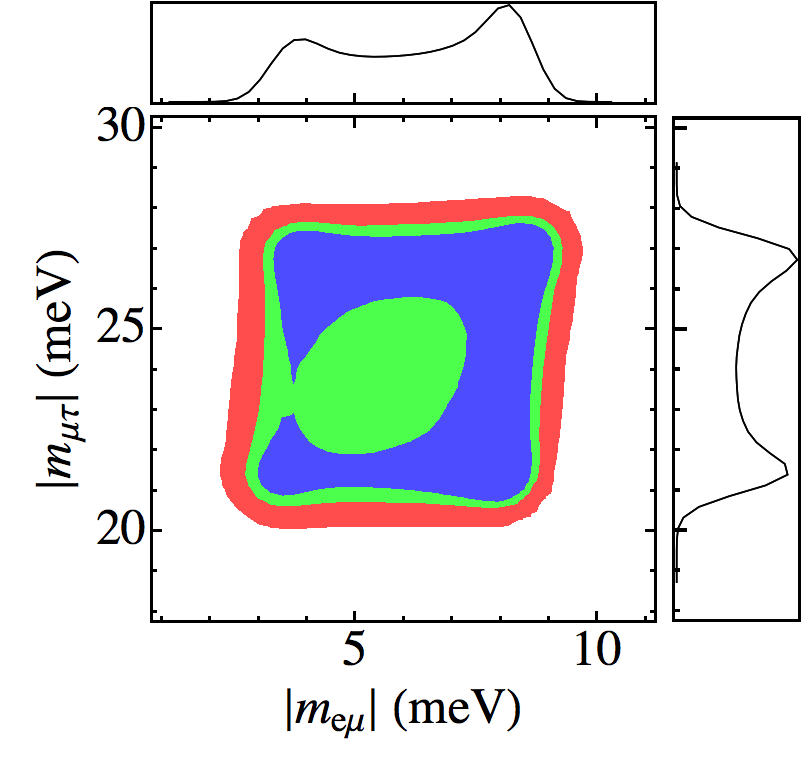}
 \includegraphics[width=0.28\textwidth]{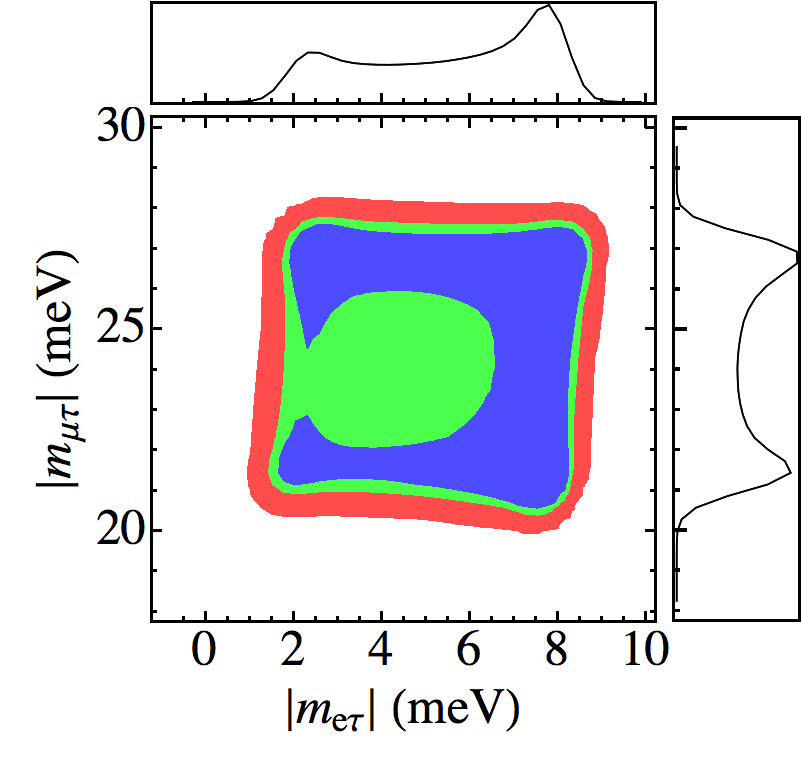}
 \includegraphics[width=0.28\textwidth]{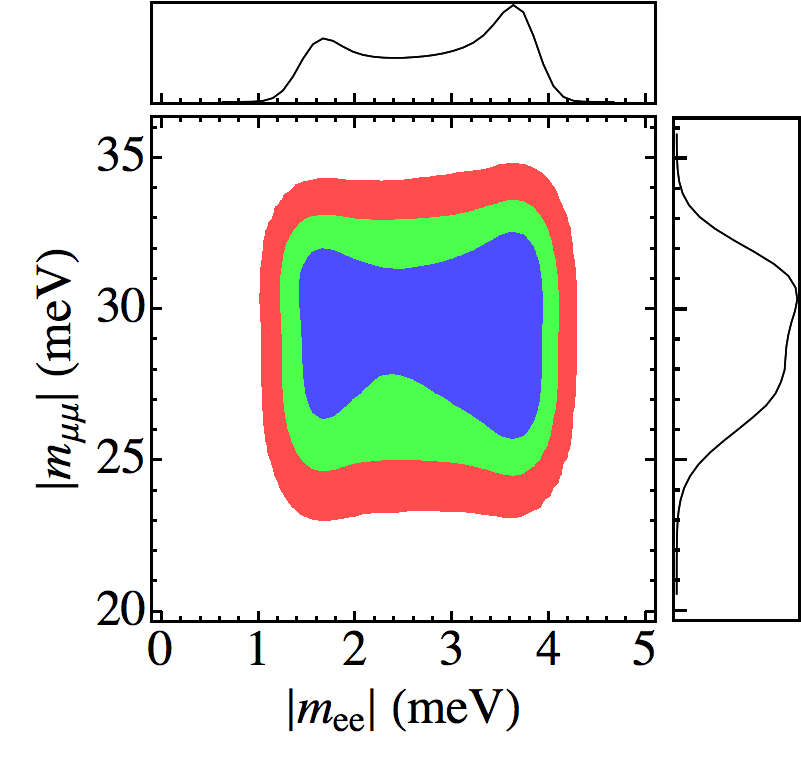}
 \includegraphics[width=0.28\textwidth]{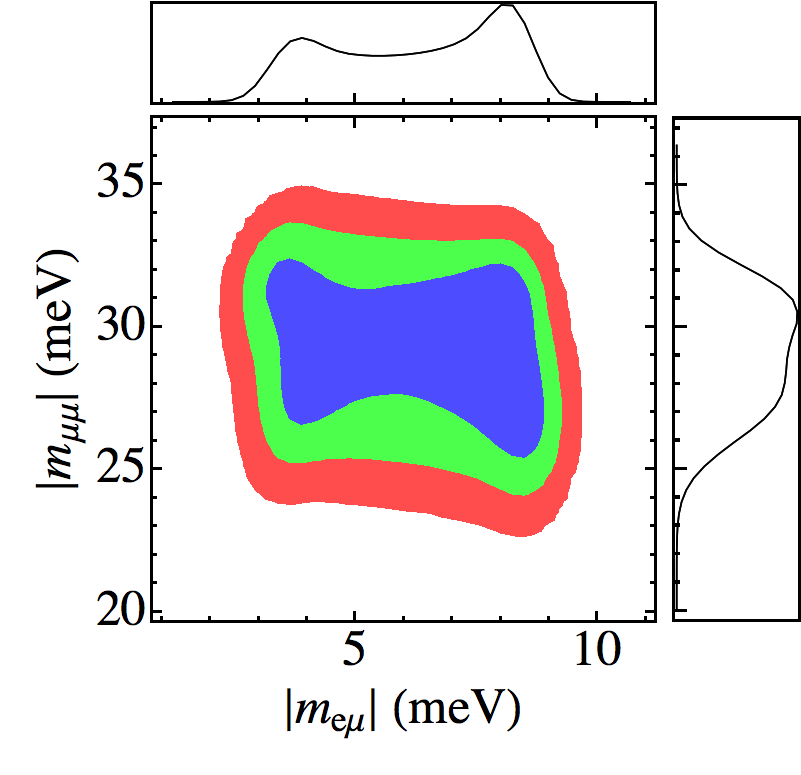}
 \includegraphics[width=0.28\textwidth]{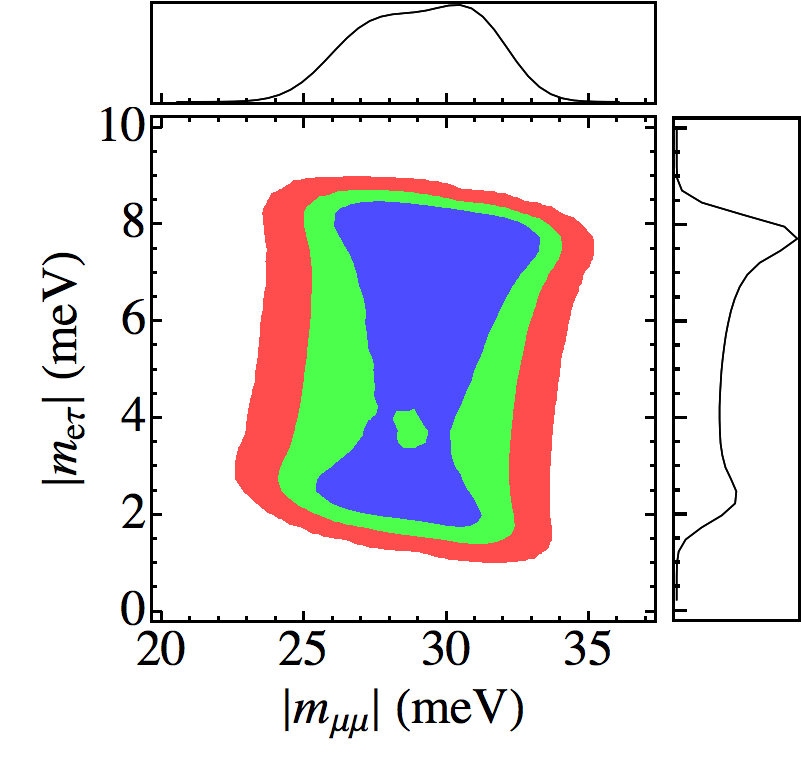}
 \includegraphics[width=0.28\textwidth]{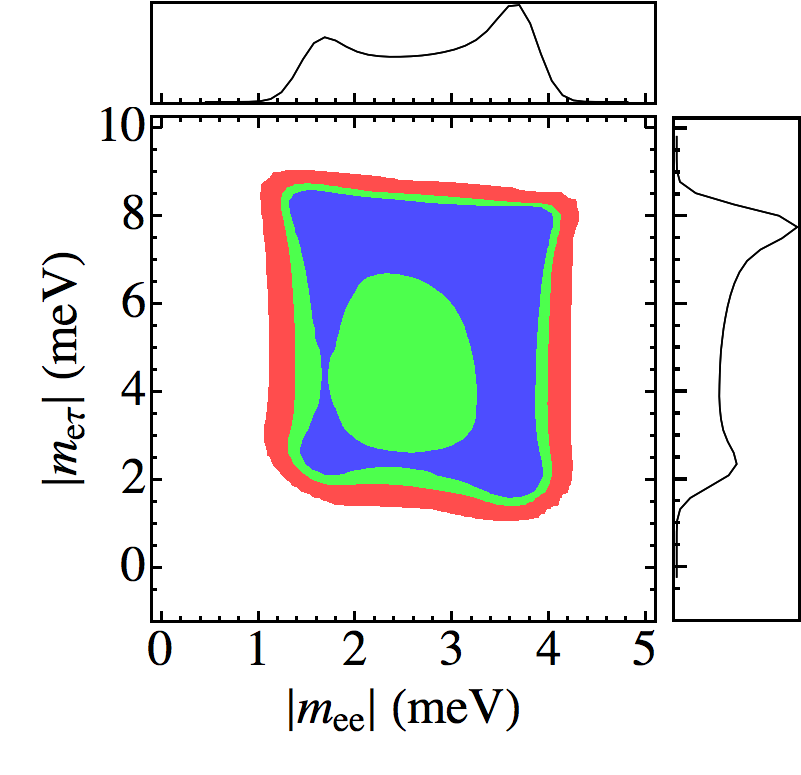}
 \includegraphics[width=0.28\textwidth]{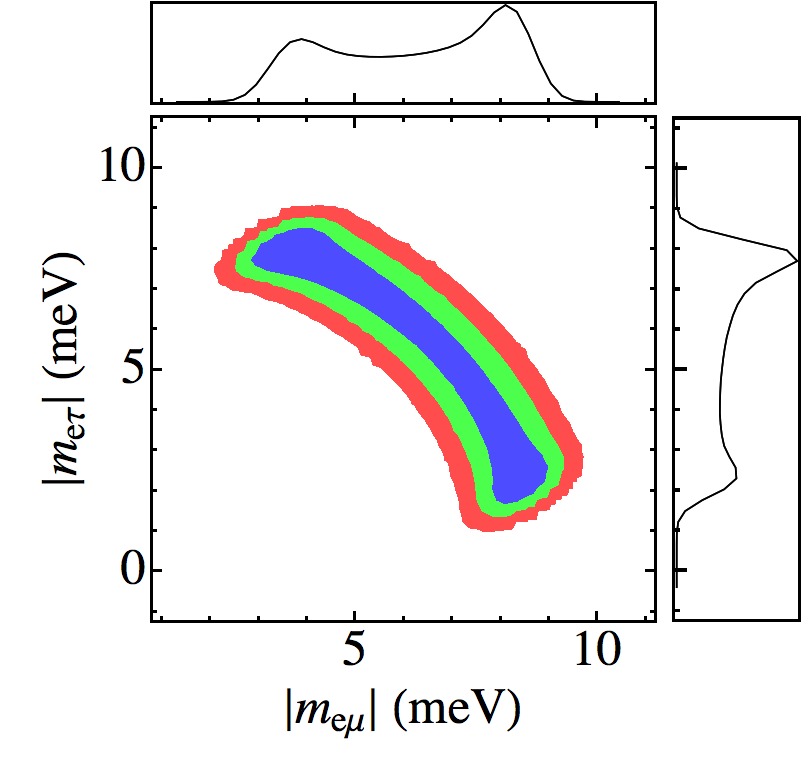}
 \includegraphics[width=0.28\textwidth]{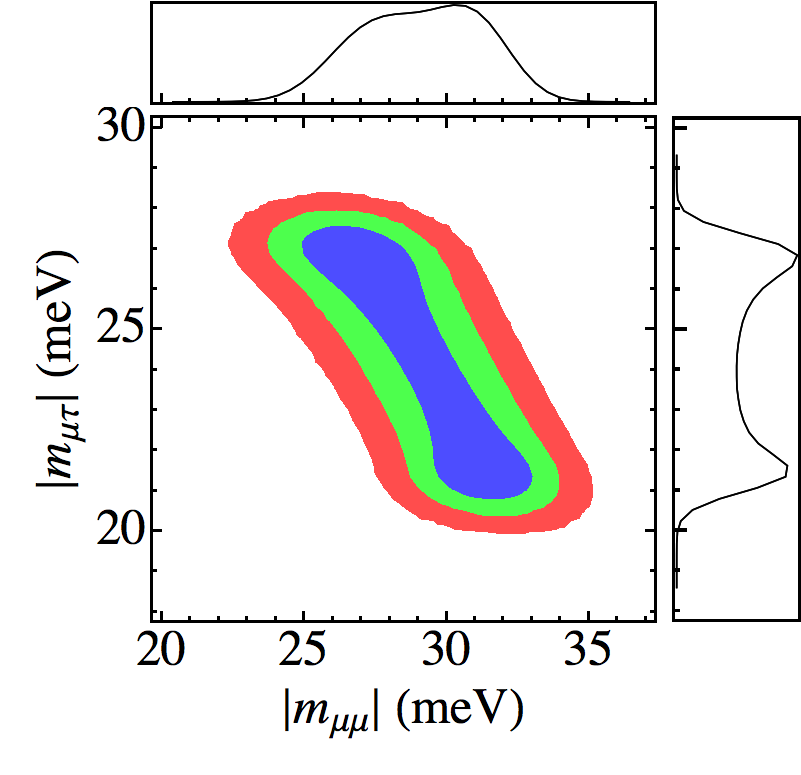}
 \includegraphics[width=0.28\textwidth]{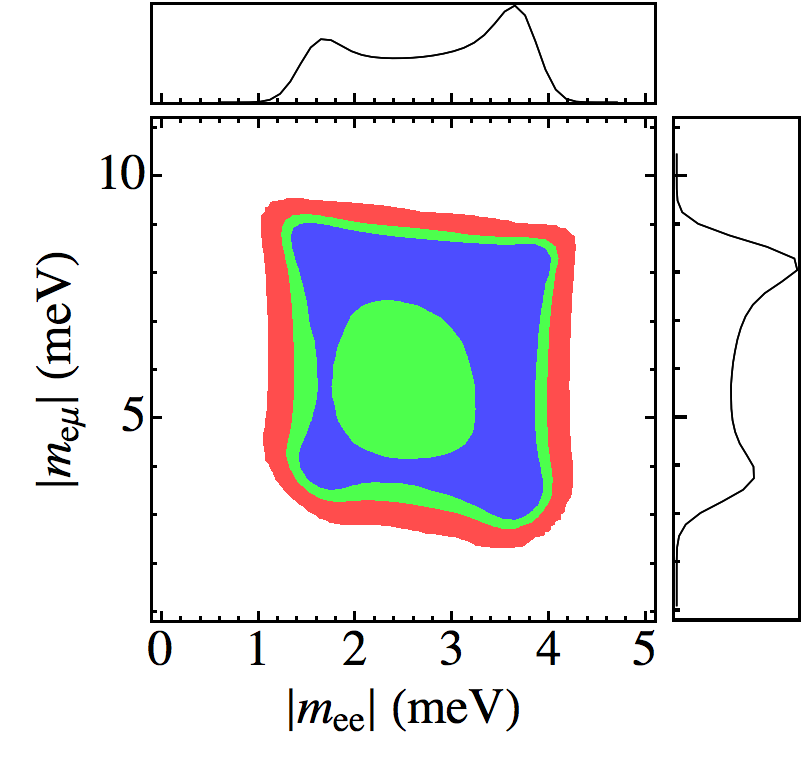}
 \includegraphics[width=0.28\textwidth]{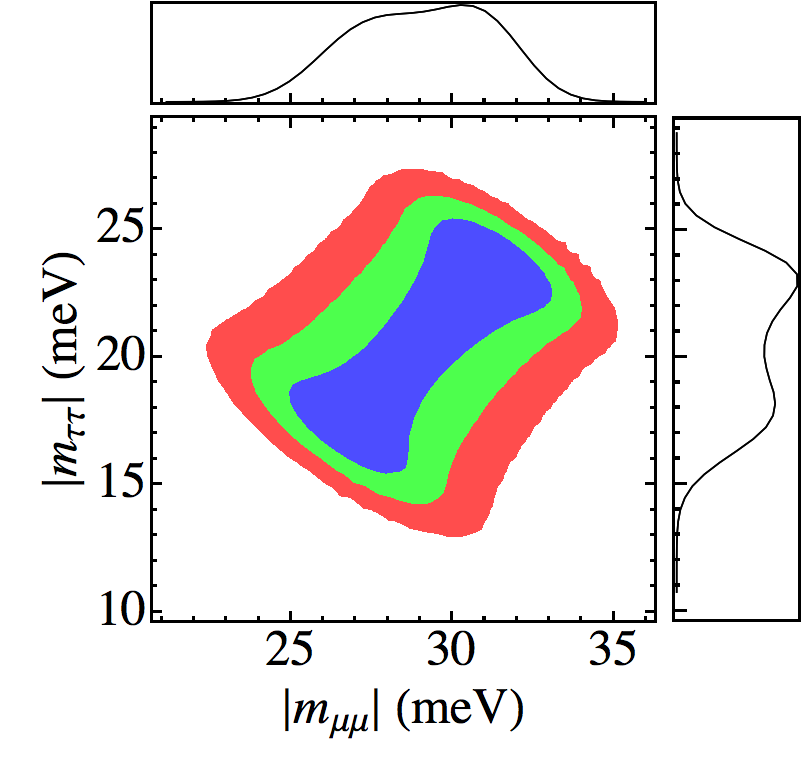}
 \includegraphics[width=0.28\textwidth]{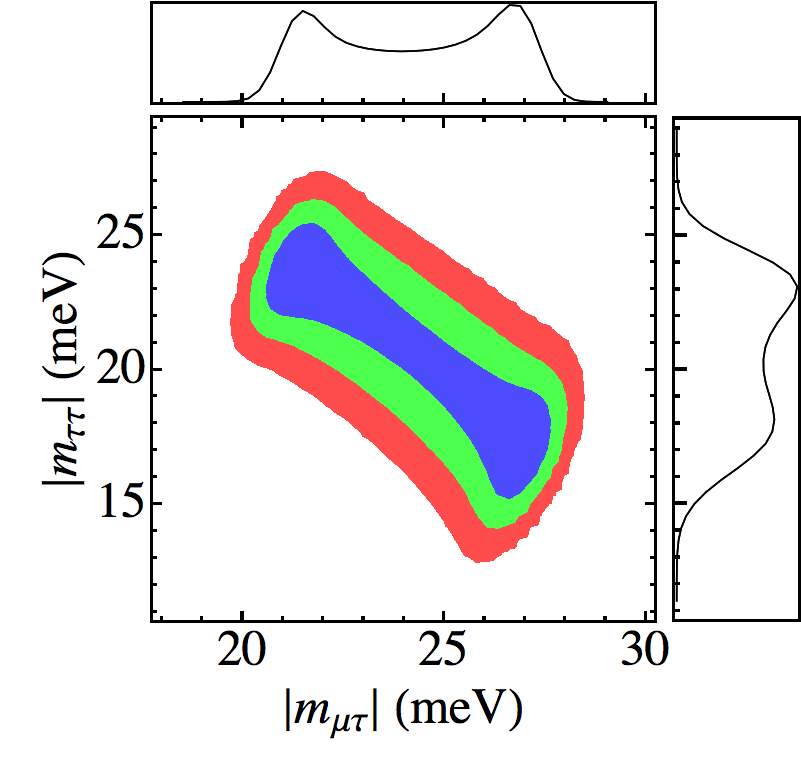}
\vspace{-2mm}
\end{center}
\vspace{-0.1cm}
\caption{Same as Fig.\ref{fig:nh-1st} but for $\theta_{23}$ in the second 
octant.}
\label{fig:nh-2nd}
\end{figure}

\begin{figure}[htb]
\begin{center}
 \includegraphics[width=0.3\textwidth]{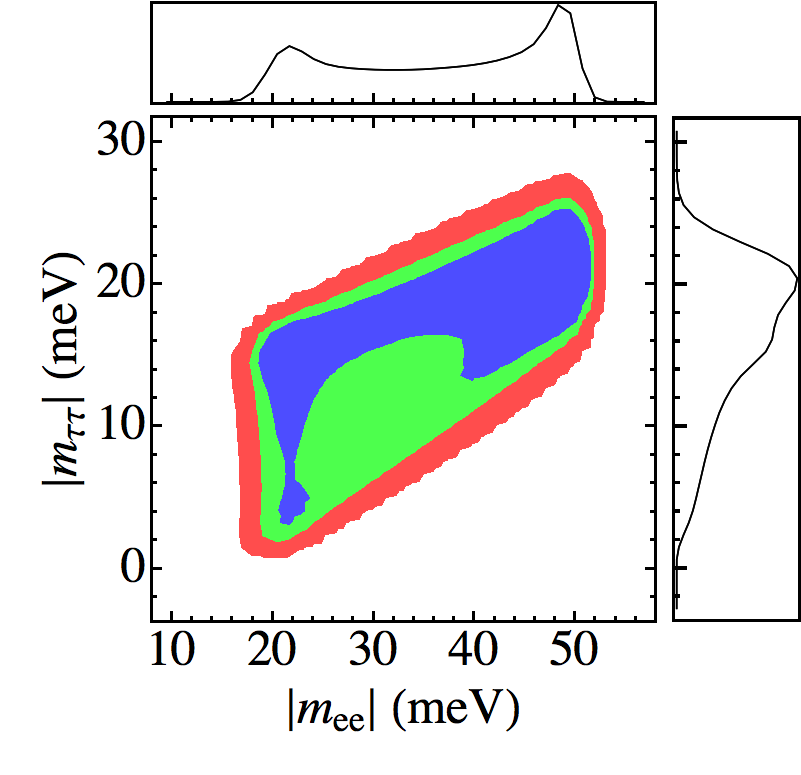}
 \includegraphics[width=0.3\textwidth]{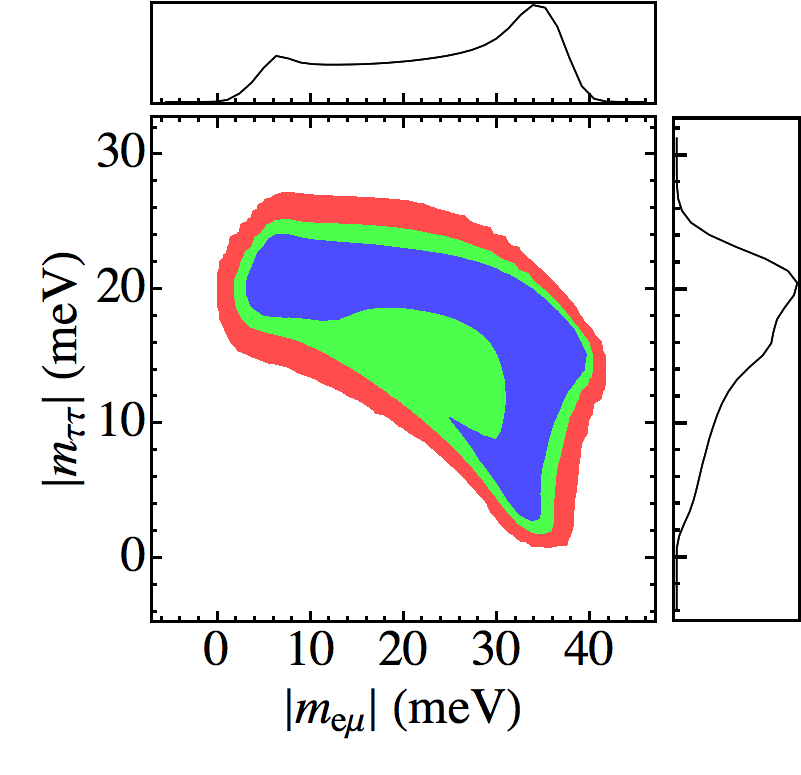}
 \includegraphics[width=0.3\textwidth]{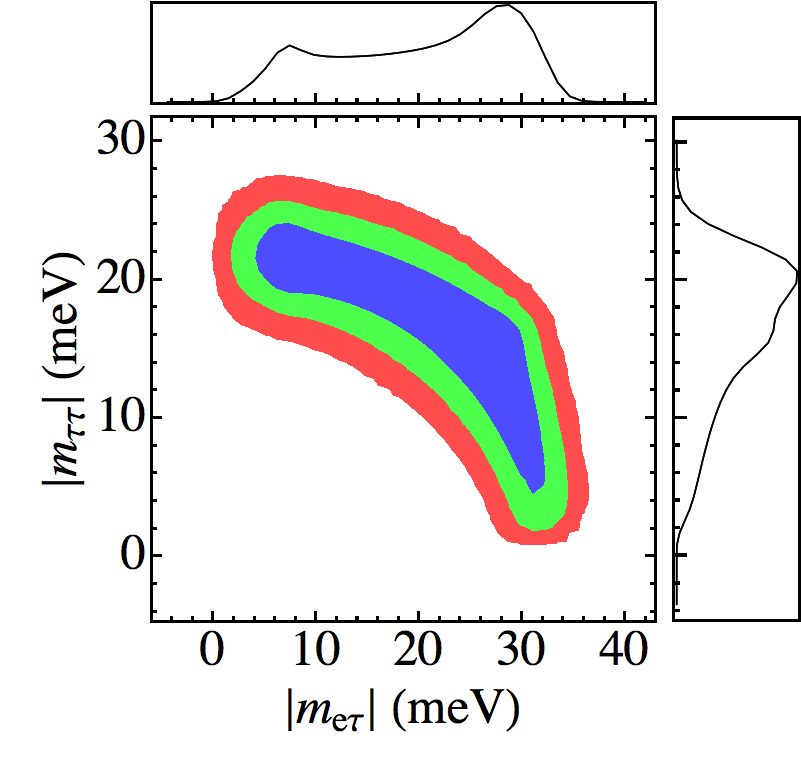}
 \includegraphics[width=0.3\textwidth]{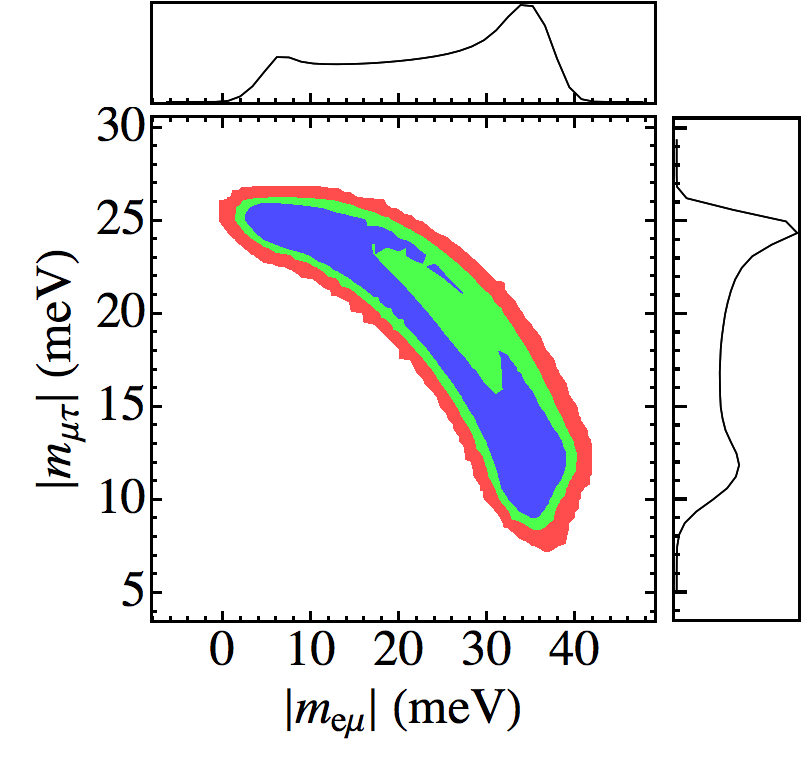}
 \includegraphics[width=0.3\textwidth]{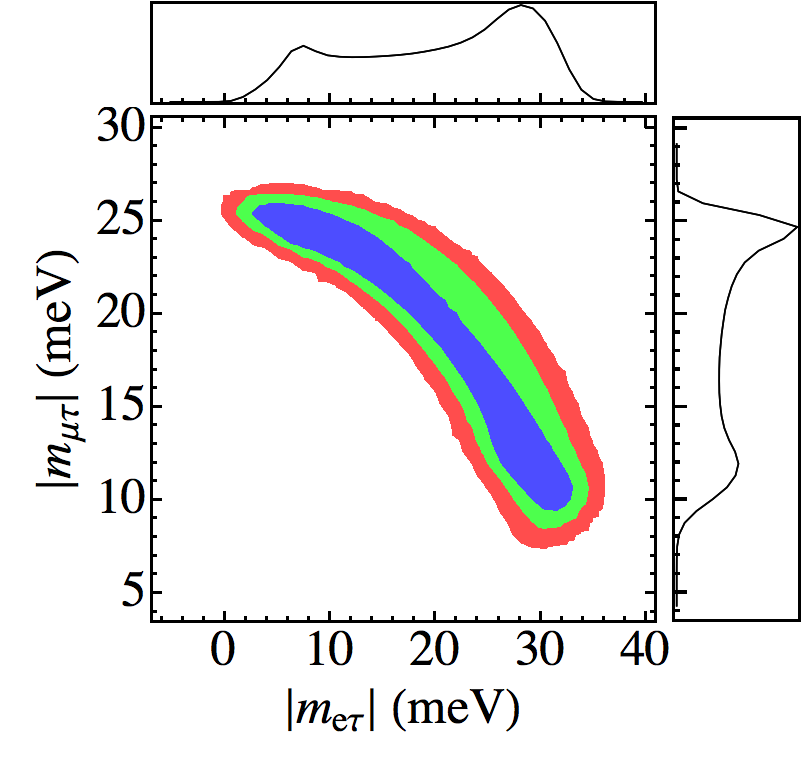}
 \includegraphics[width=0.3\textwidth]{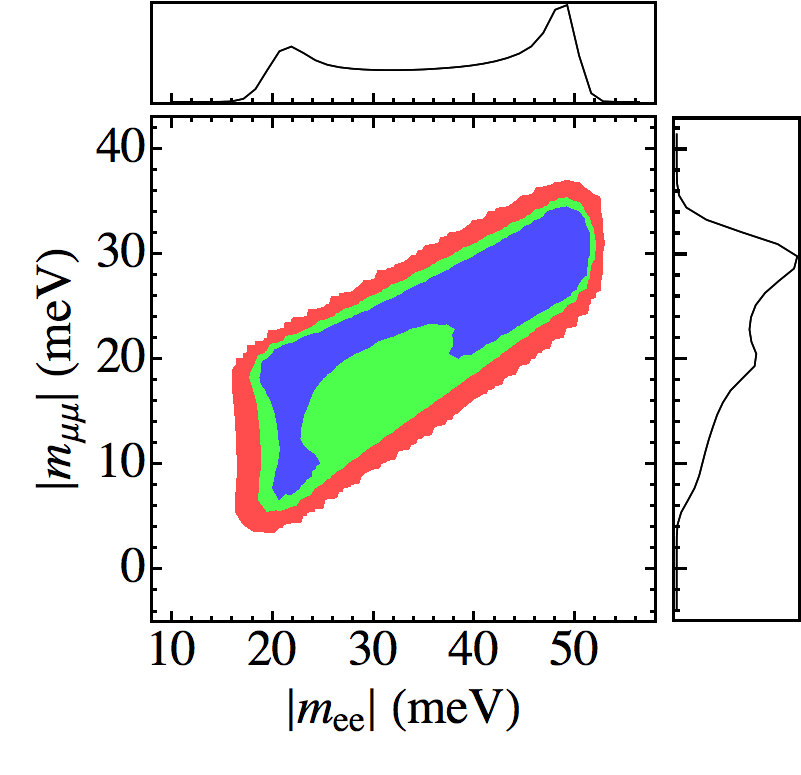}
 \includegraphics[width=0.3\textwidth]{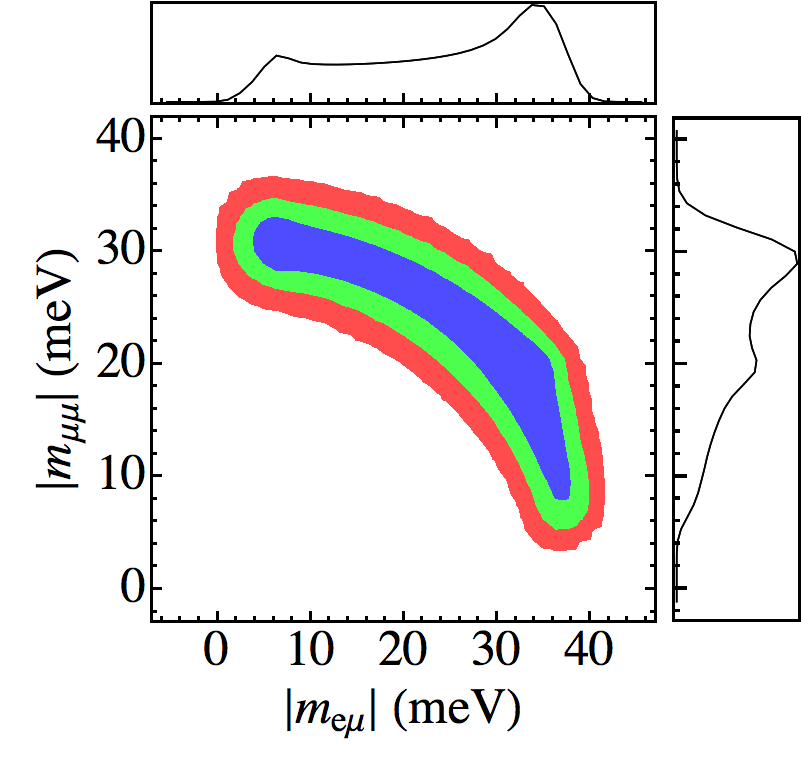}
 \includegraphics[width=0.3\textwidth]{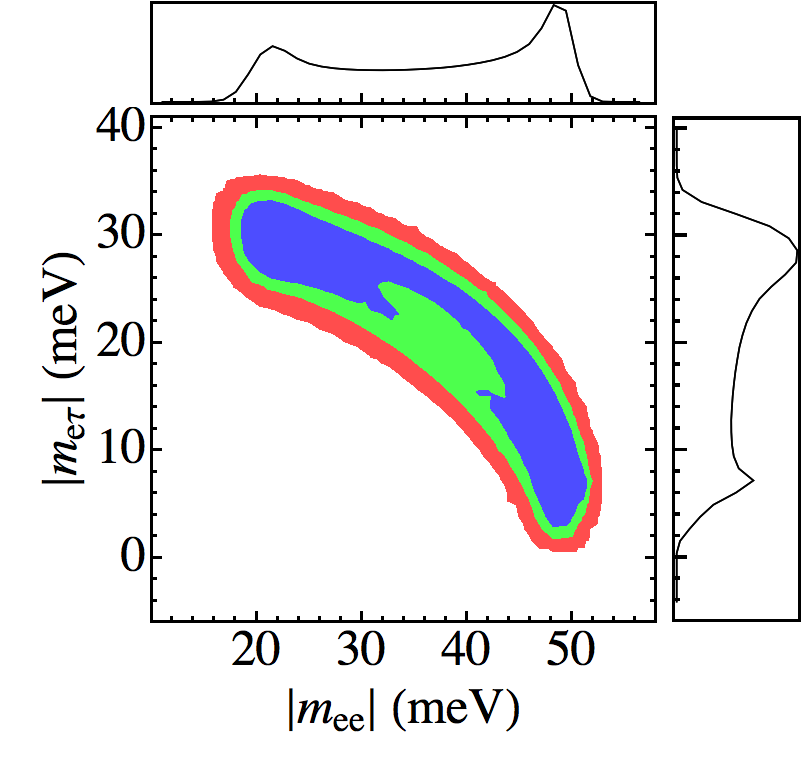}
 \includegraphics[width=0.3\textwidth]{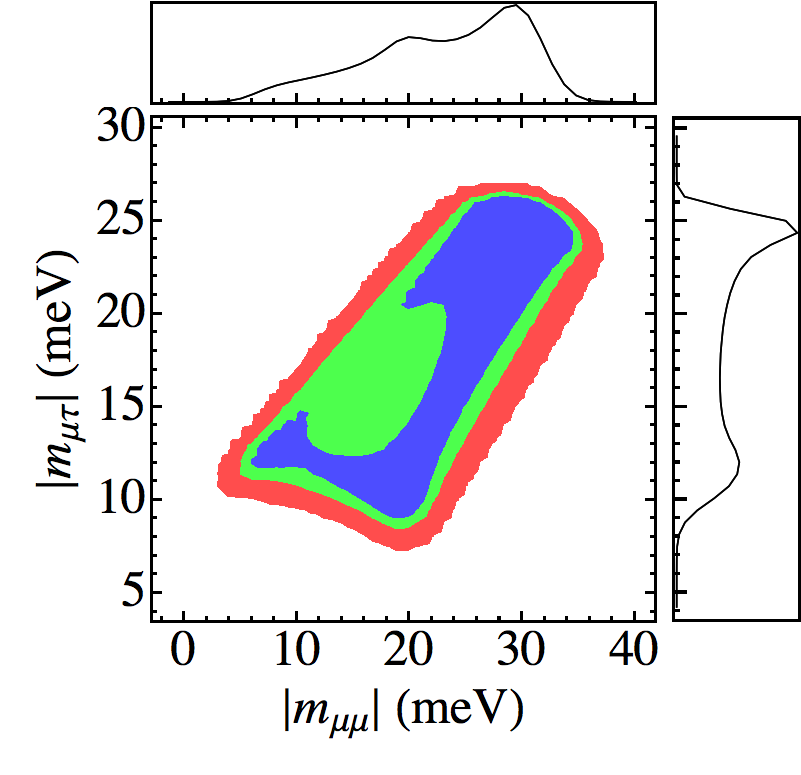}
\vspace{-2mm}
\end{center}
\vspace{-0.1cm}
\caption{PDFs for the distribution of the absolute value of several 
pairs of matrix elements. We use blue, green and
  red for the allowed region at 68.27\%, 95.45\% and 99.73\% CL,
  respectively. Here $m_3 \to 0$ and $\theta_{23}$ is 
assumed to be in the first octant.}
\label{fig:ih-1st}
\end{figure}

\begin{figure}[htb]
\begin{center}
 \includegraphics[width=0.28\textwidth]{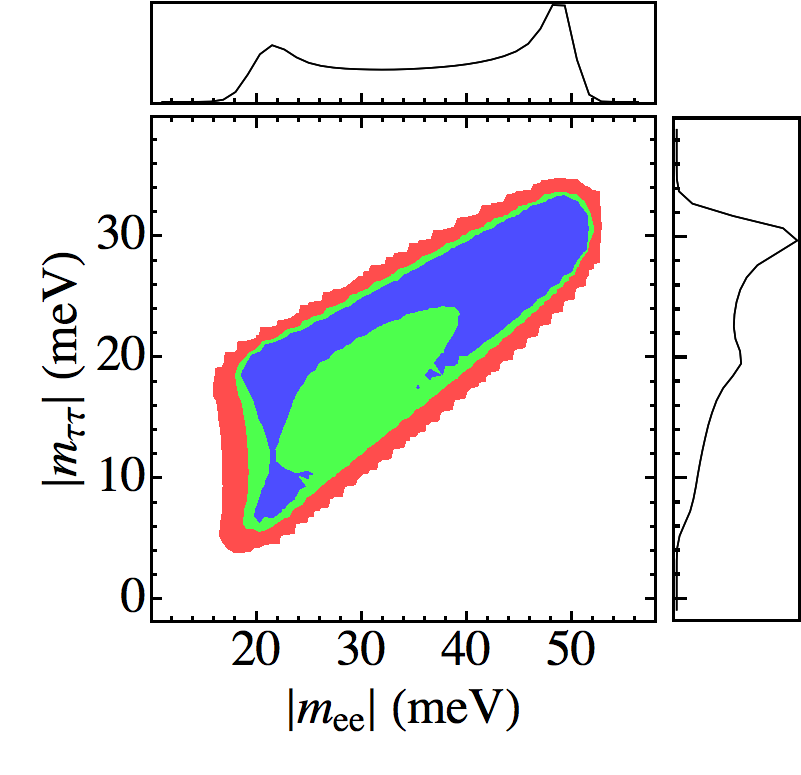}
 \includegraphics[width=0.28\textwidth]{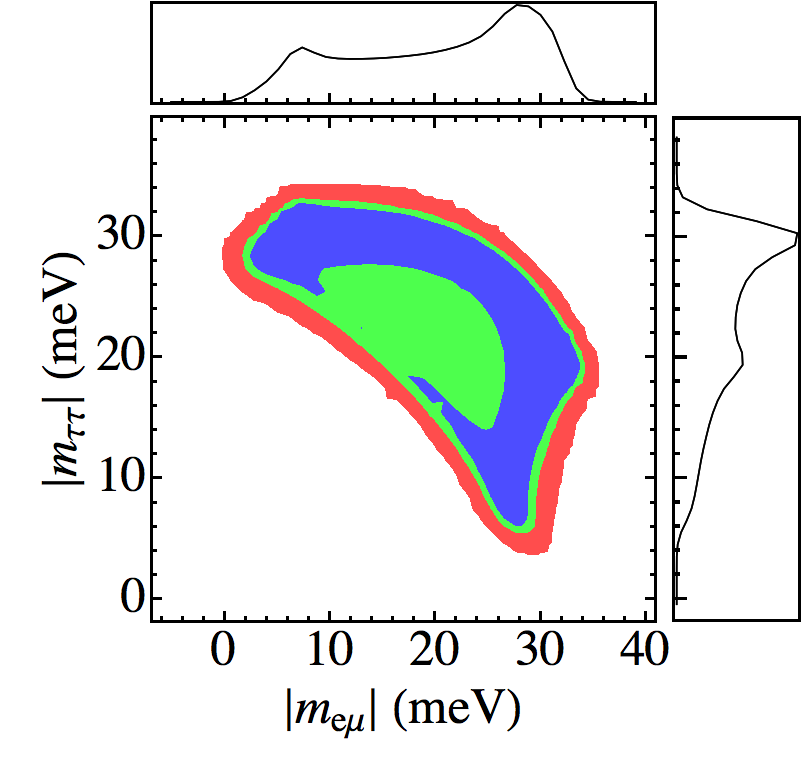}
 \includegraphics[width=0.28\textwidth]{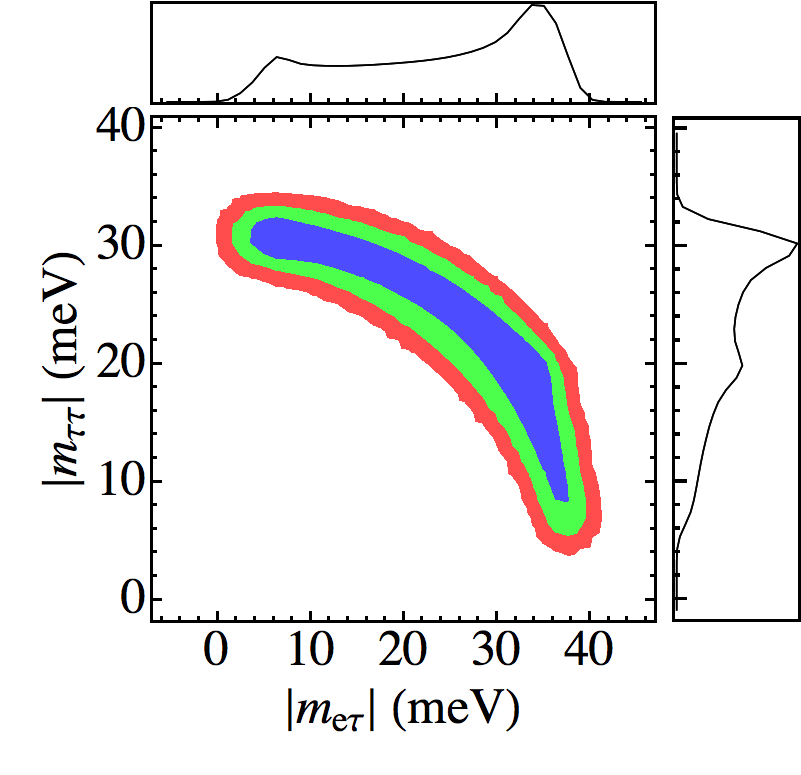}
 \includegraphics[width=0.28\textwidth]{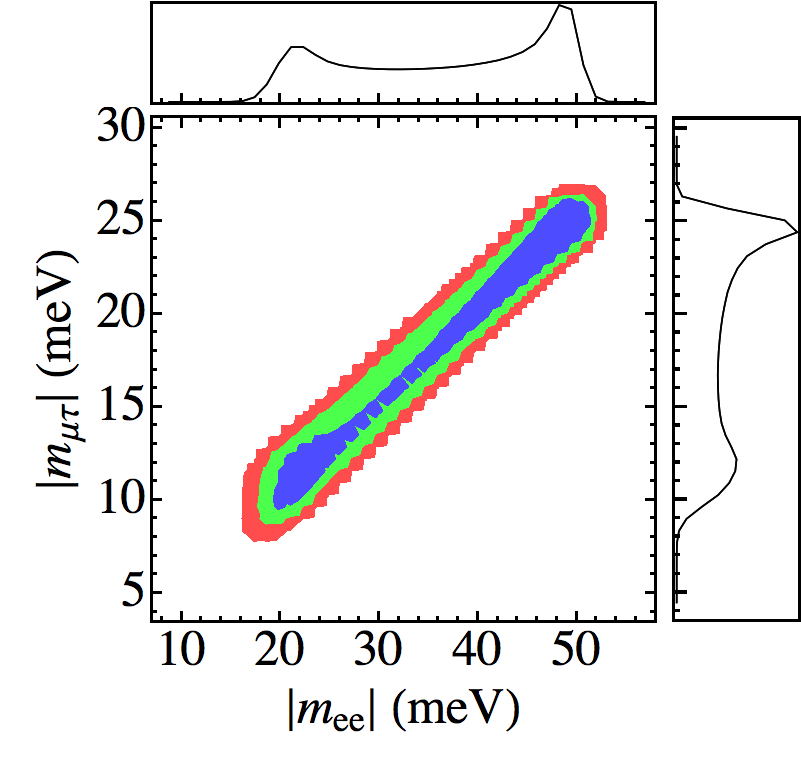}
 \includegraphics[width=0.28\textwidth]{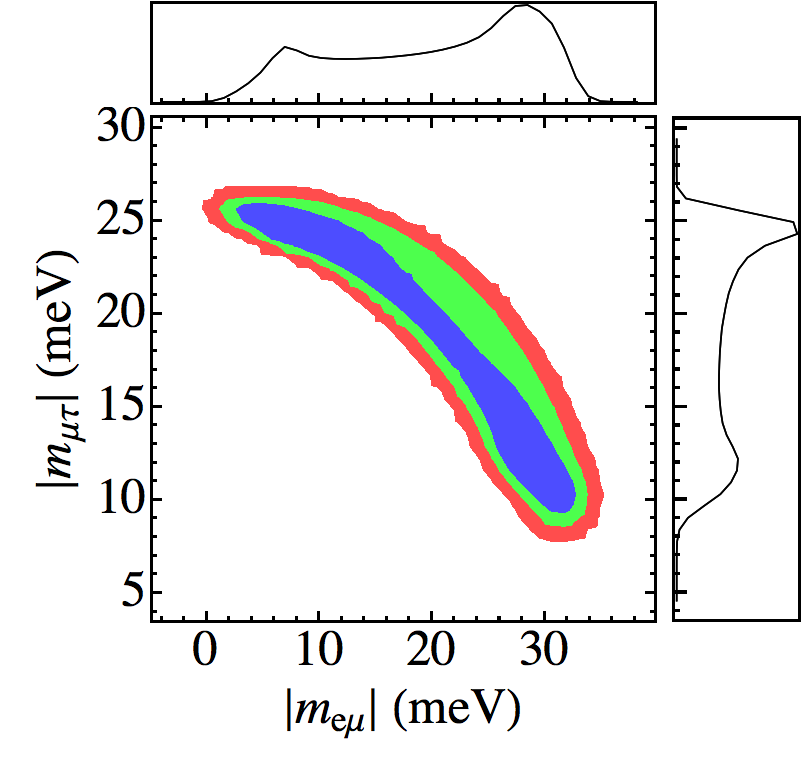}
 \includegraphics[width=0.28\textwidth]{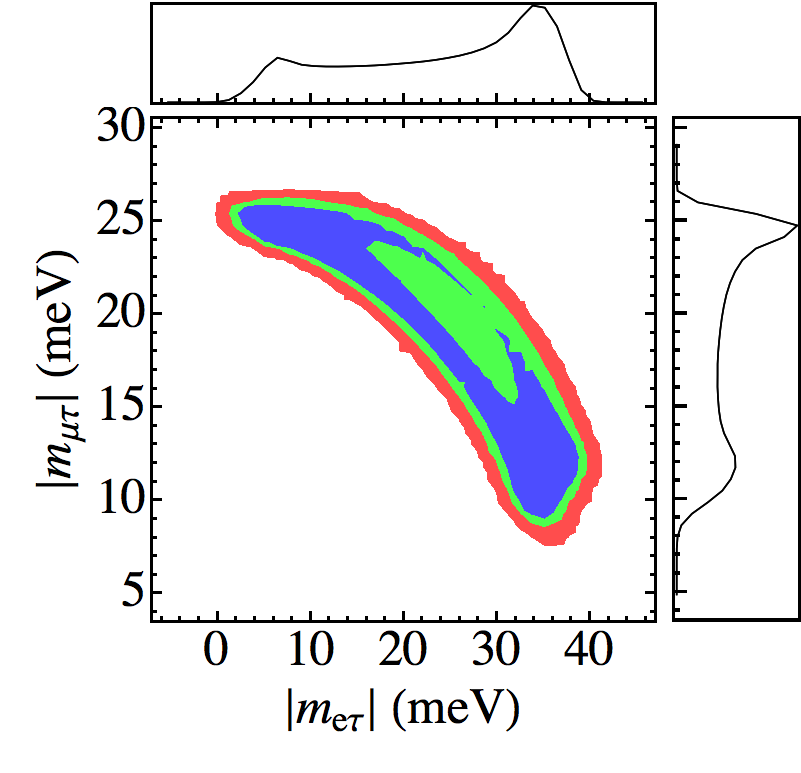}
 \includegraphics[width=0.28\textwidth]{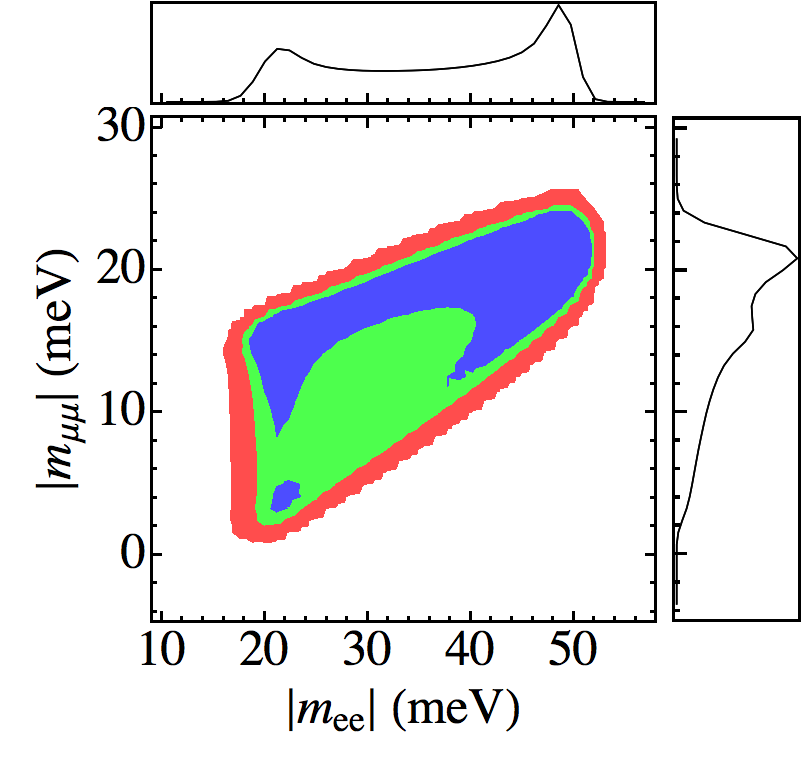}
 \includegraphics[width=0.28\textwidth]{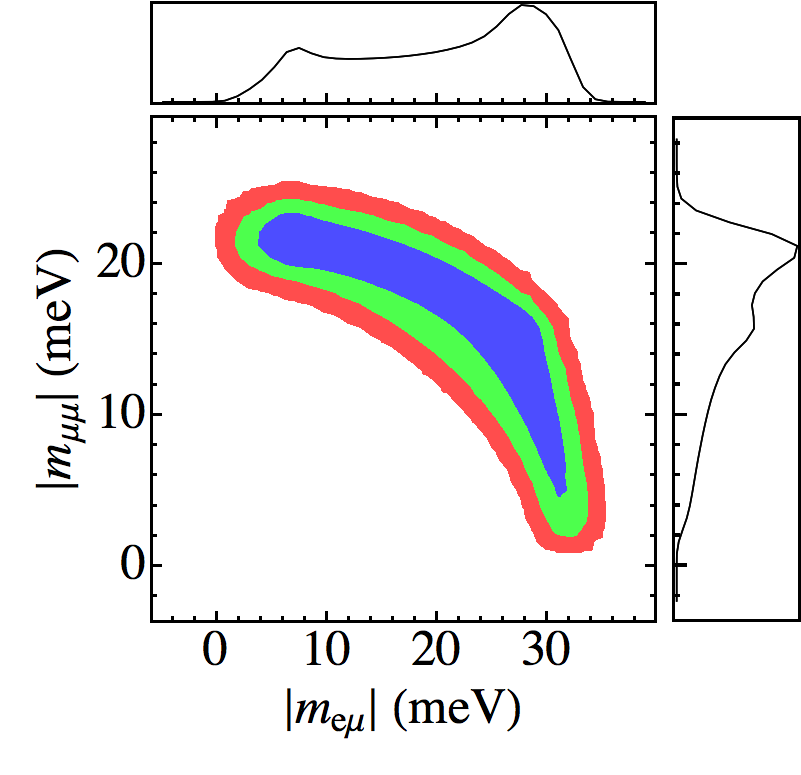}
 \includegraphics[width=0.28\textwidth]{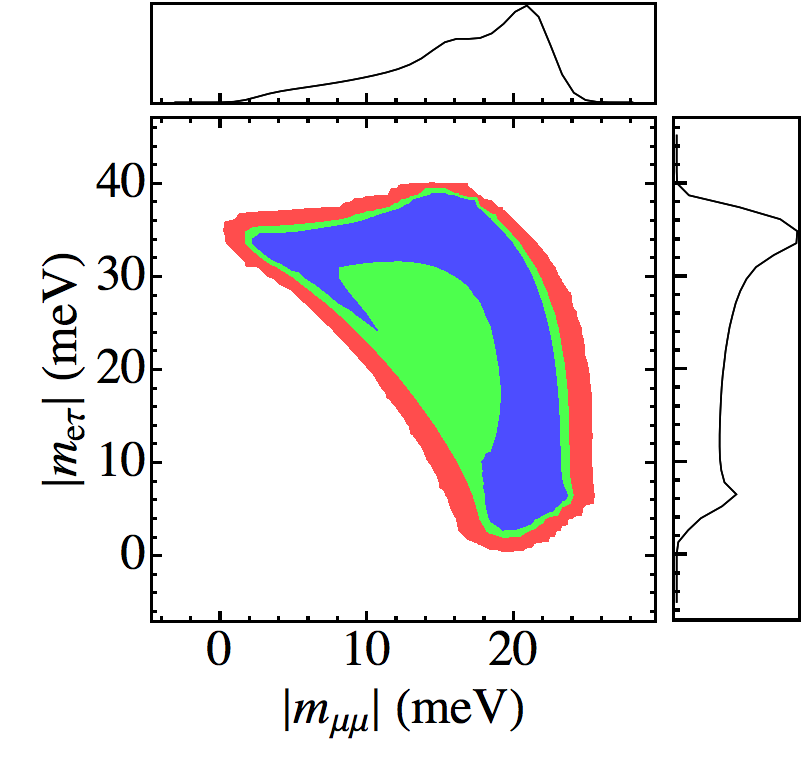}
 \includegraphics[width=0.28\textwidth]{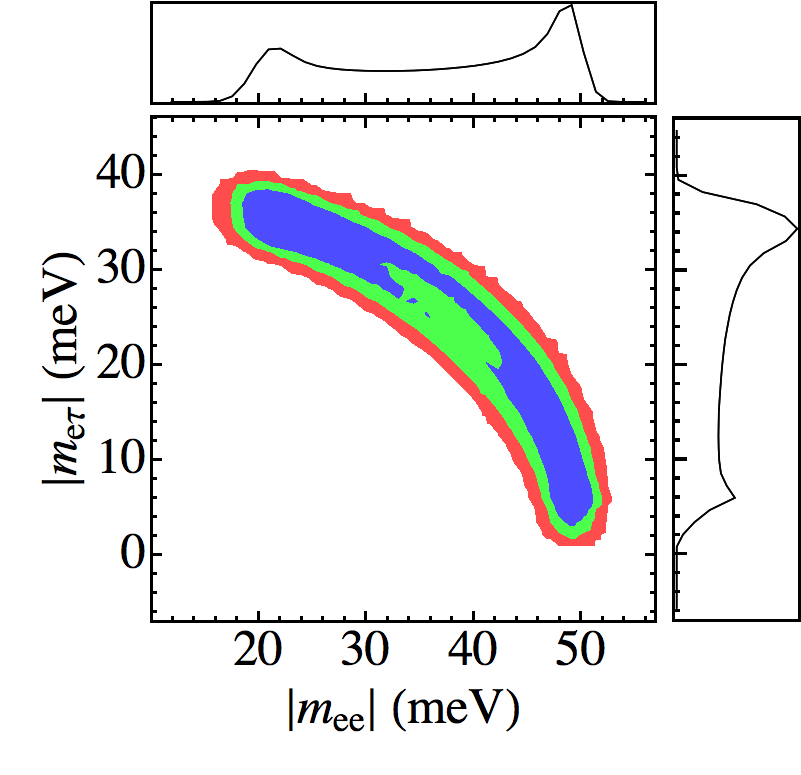}
 \includegraphics[width=0.28\textwidth]{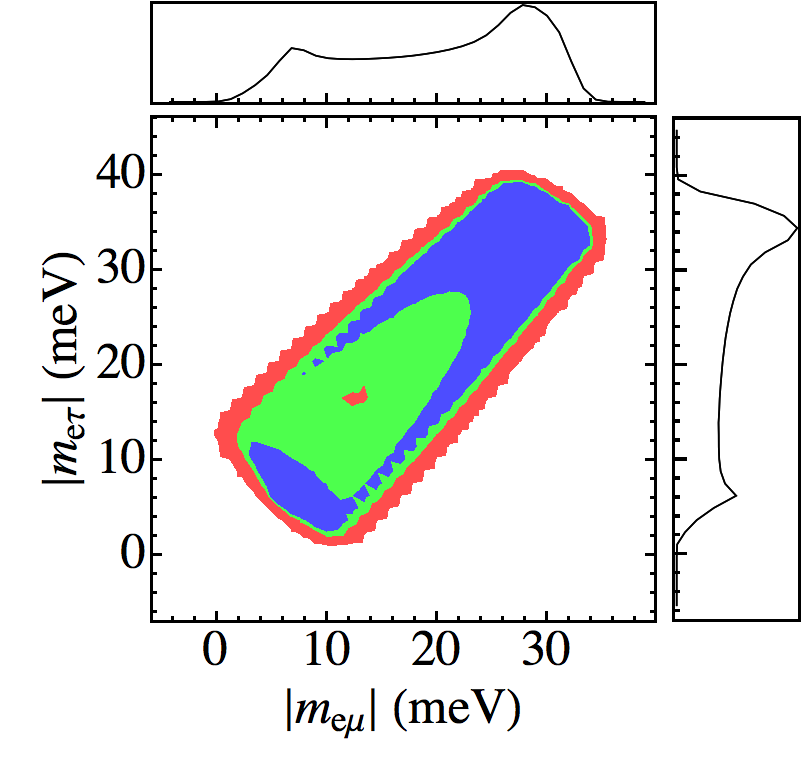}
 \includegraphics[width=0.28\textwidth]{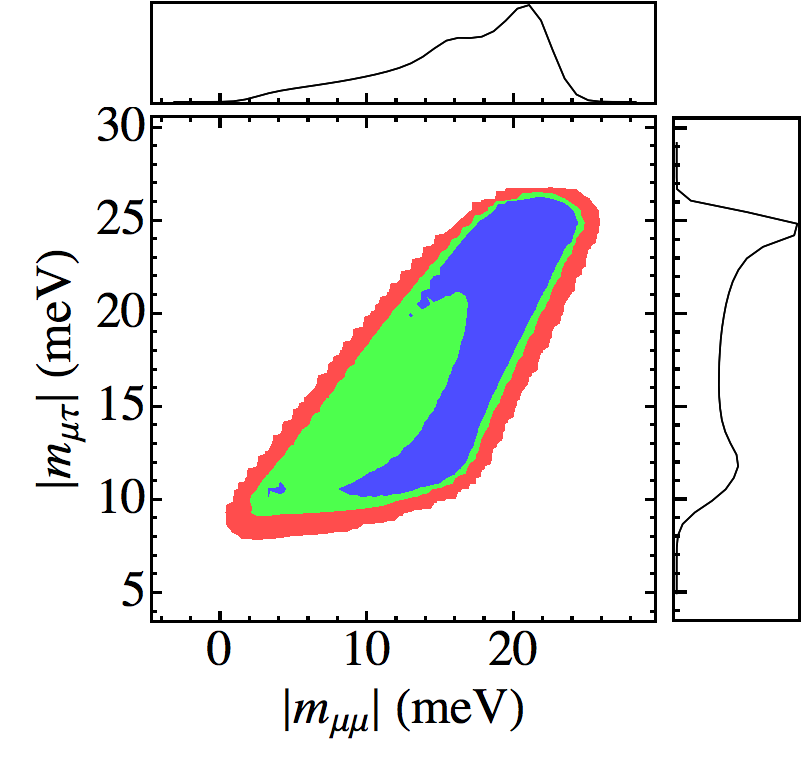}
 \includegraphics[width=0.28\textwidth]{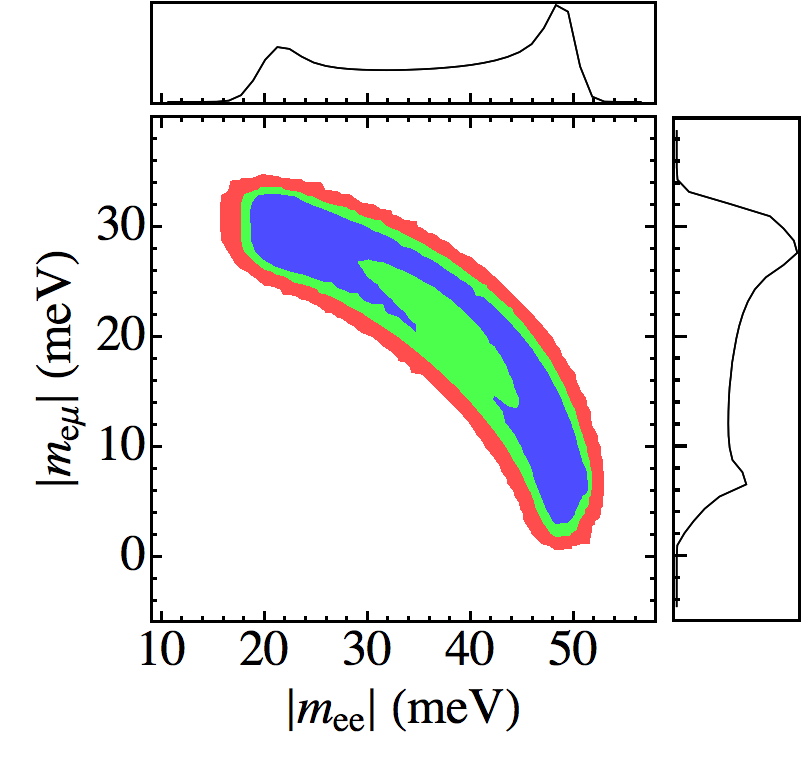}
 \includegraphics[width=0.28\textwidth]{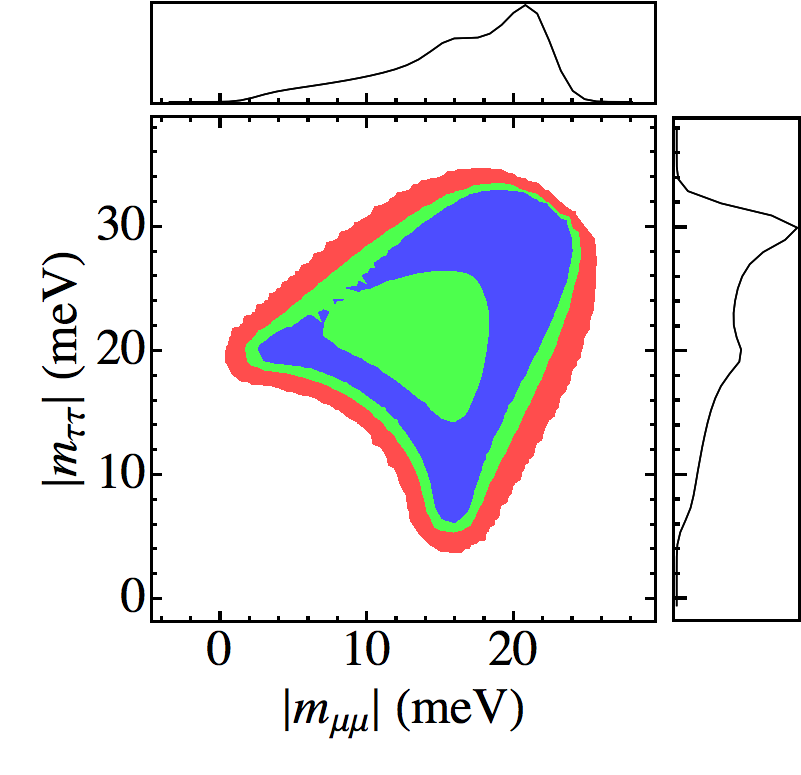}
 \includegraphics[width=0.28\textwidth]{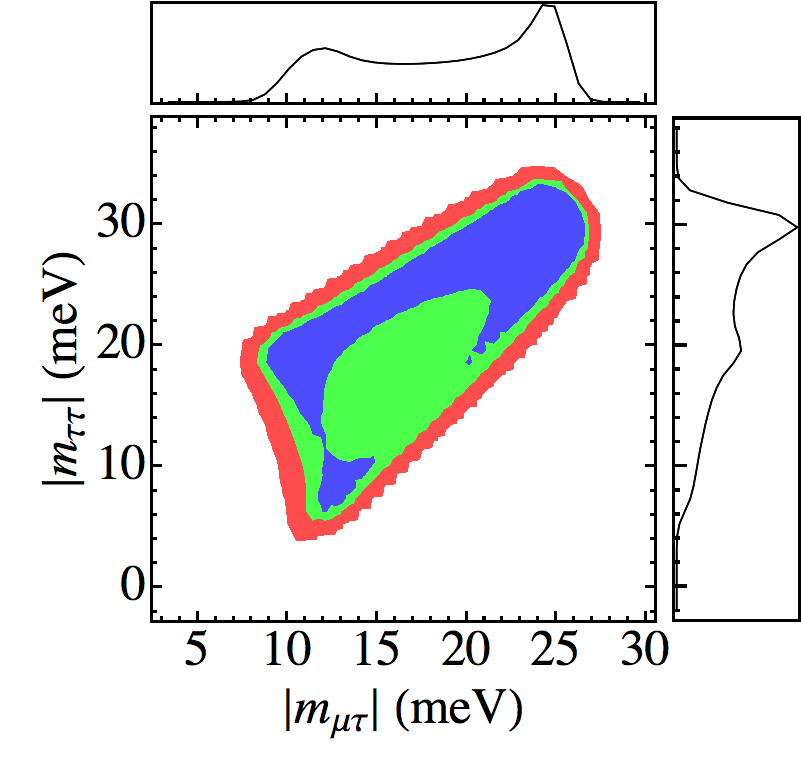}
\vspace{-2mm}
\end{center}
\vspace{-0.1cm}
\caption{Same as Fig.\ref{fig:ih-1st} but for $\theta_{23}$ in the second 
octant.}
\label{fig:ih-2nd}
\end{figure}

\begin{figure}[htb]
\begin{center}
 \includegraphics[width=0.3\textwidth]{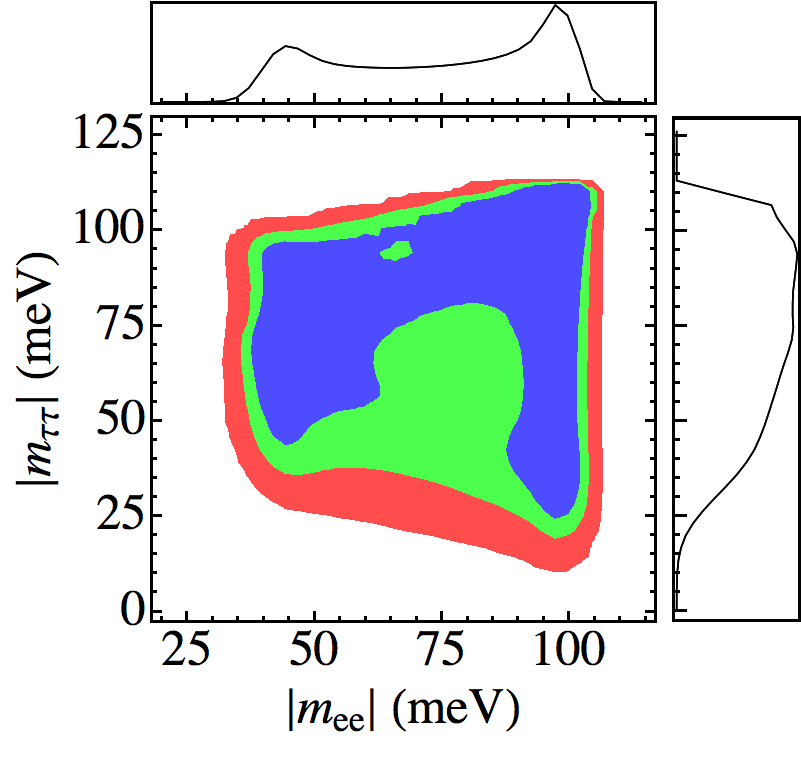}
 \includegraphics[width=0.3\textwidth]{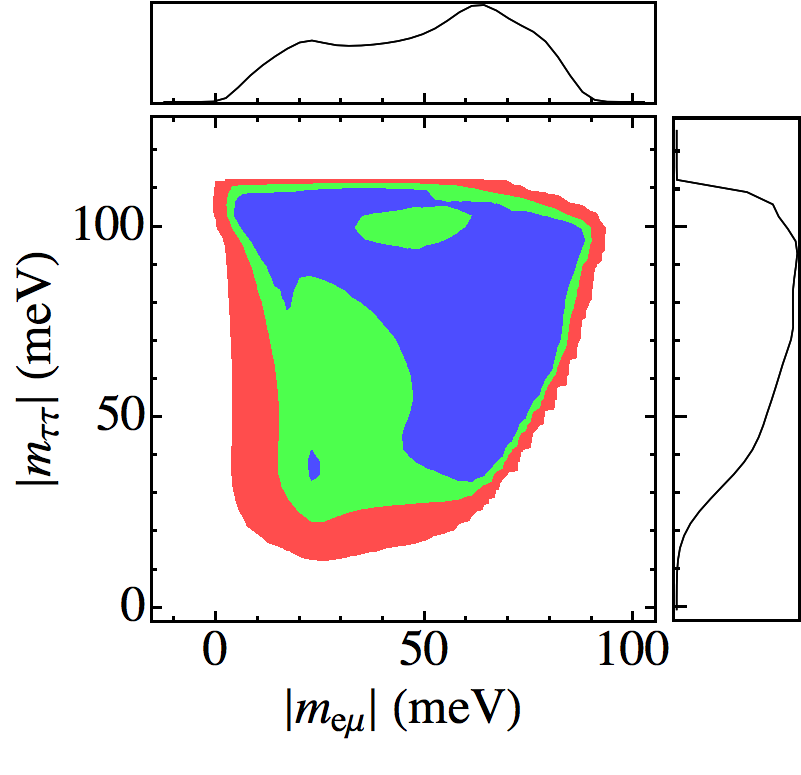}
 \includegraphics[width=0.3\textwidth]{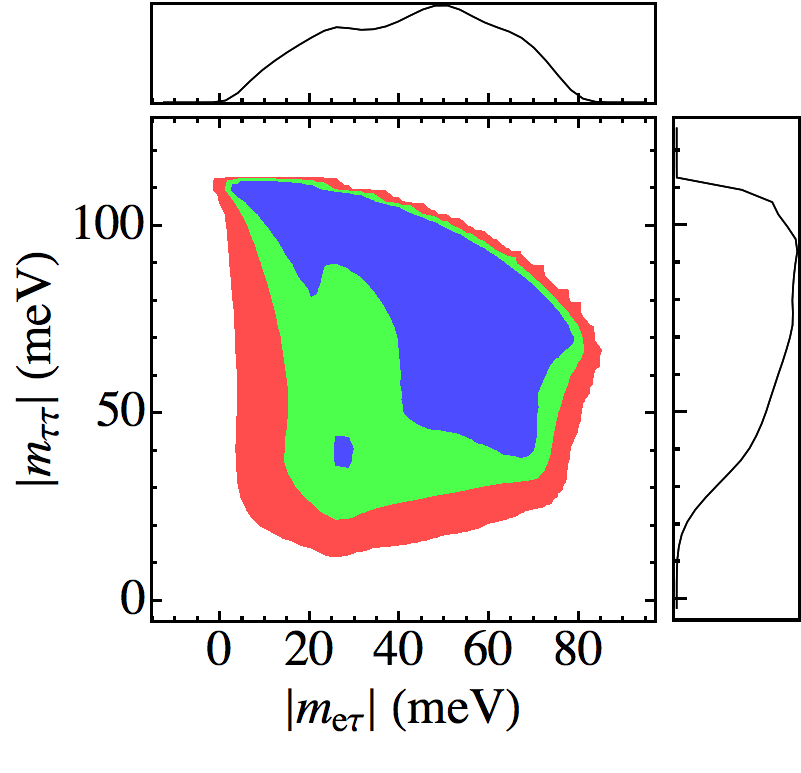}
 \includegraphics[width=0.3\textwidth]{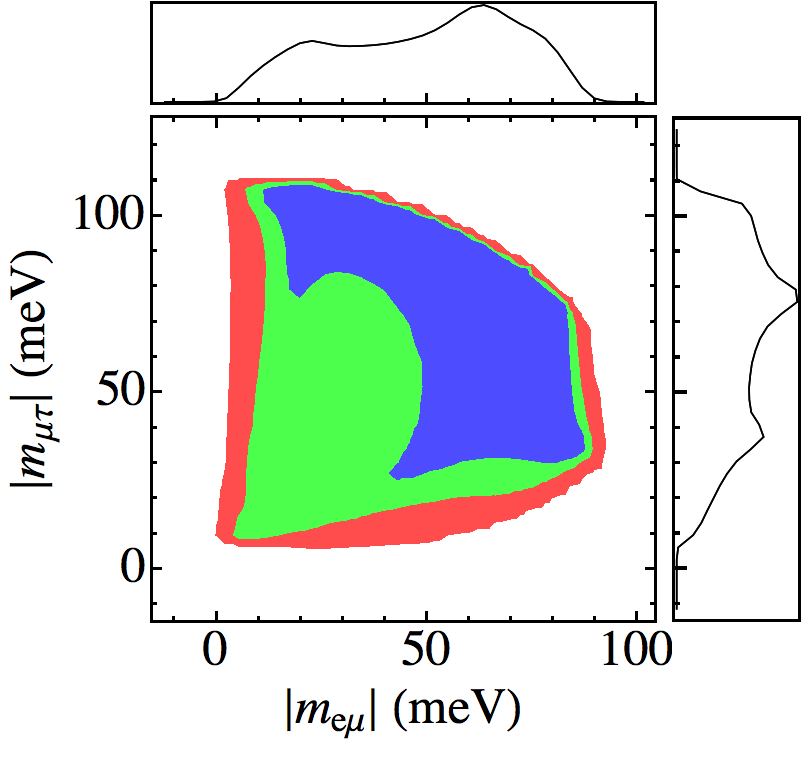}
 \includegraphics[width=0.3\textwidth]{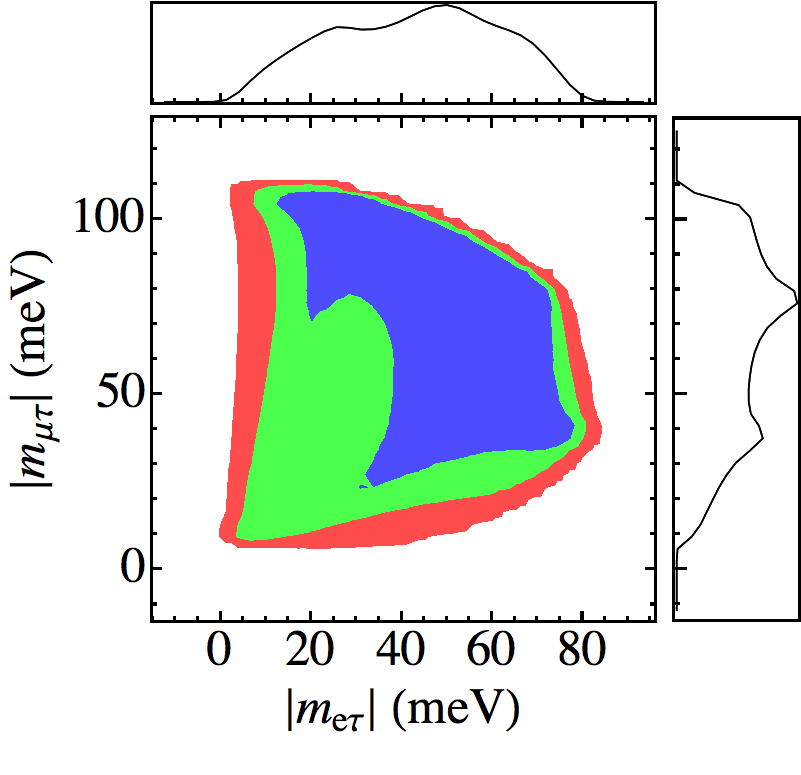}
 \includegraphics[width=0.3\textwidth]{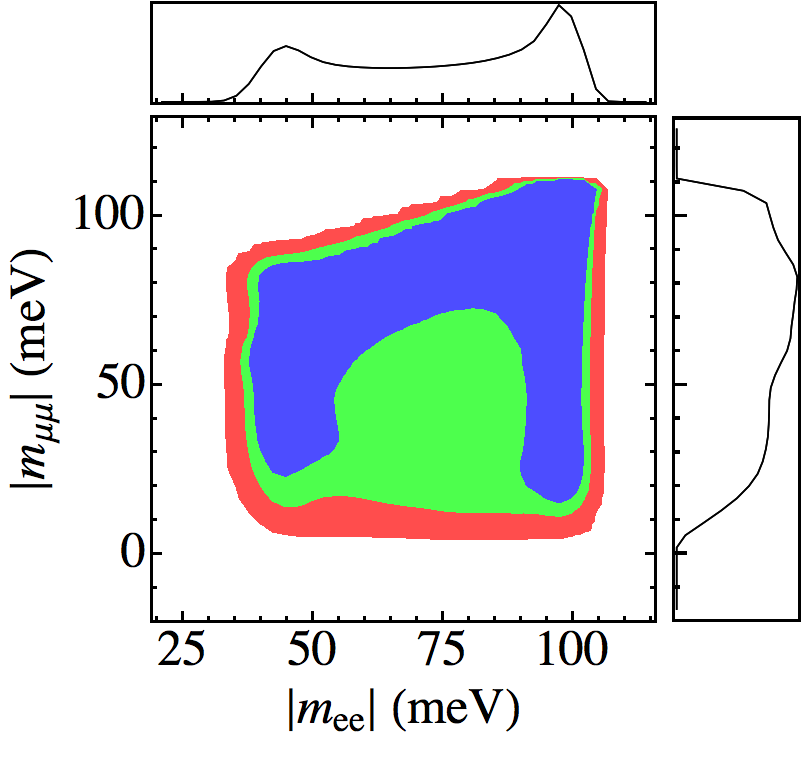}
 \includegraphics[width=0.3\textwidth]{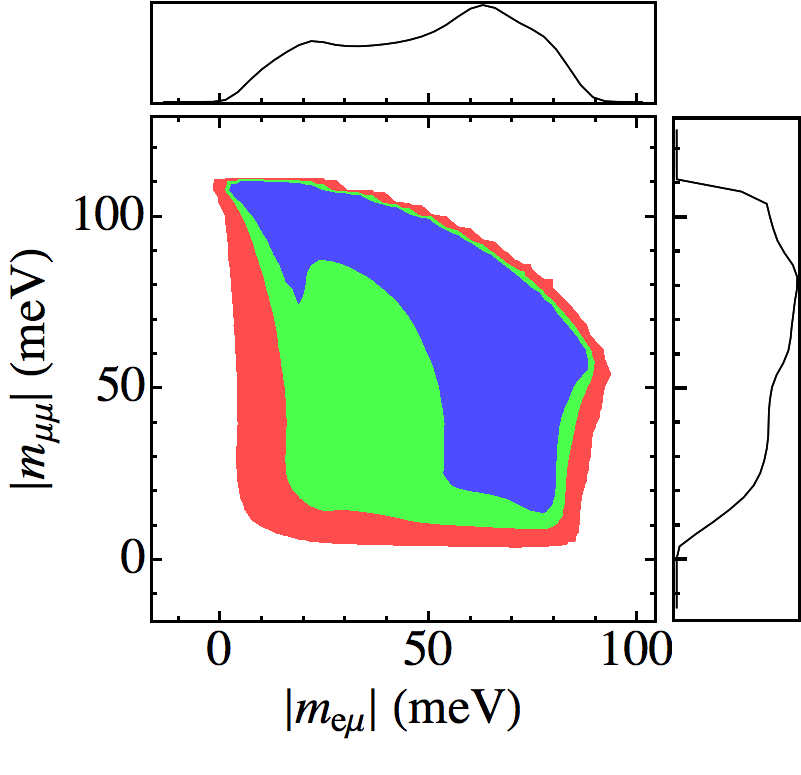}
 \includegraphics[width=0.3\textwidth]{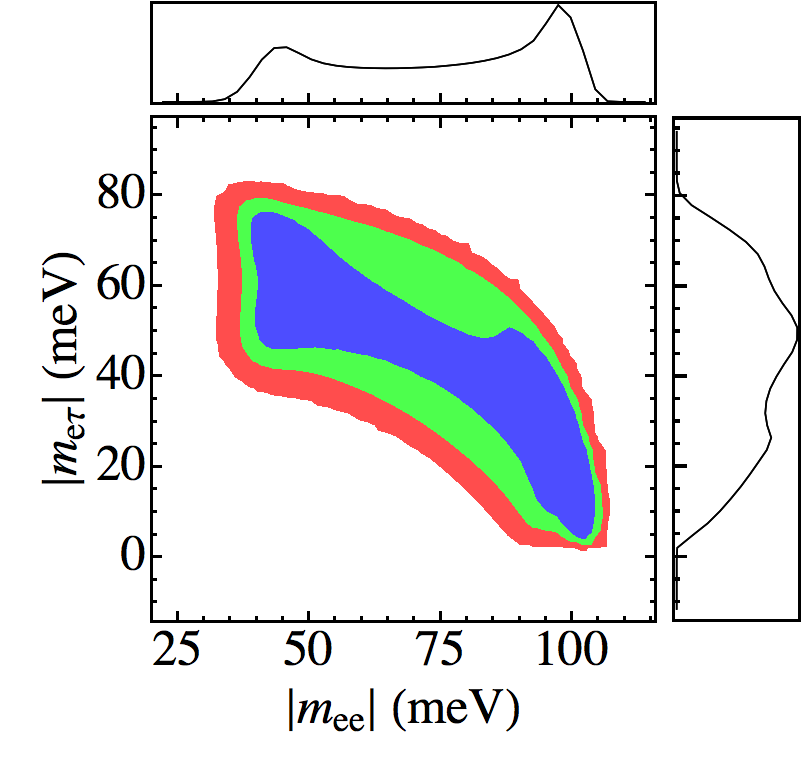}
 \includegraphics[width=0.3\textwidth]{./figs/m0-01/corr-1st-NH-2233-deg.png}
\vspace{-2mm}
\end{center}
\vspace{-0.1cm}
\caption{PDFs for the distribution of the absolute value of several 
pairs of matrix elements. We use blue, green and
  red for the allowed region at 68.27\%, 95.45\% and 99.73\% CL,
  respectively. Here $m_1 = 0.1$ eV, we impose the normal mass ordering 
and $\theta_{23}$ is assumed to be in the first octant.}
\label{fig:nh-1st-m01}
\end{figure}

\begin{figure}[htb]
\begin{center}
 \includegraphics[width=0.28\textwidth]{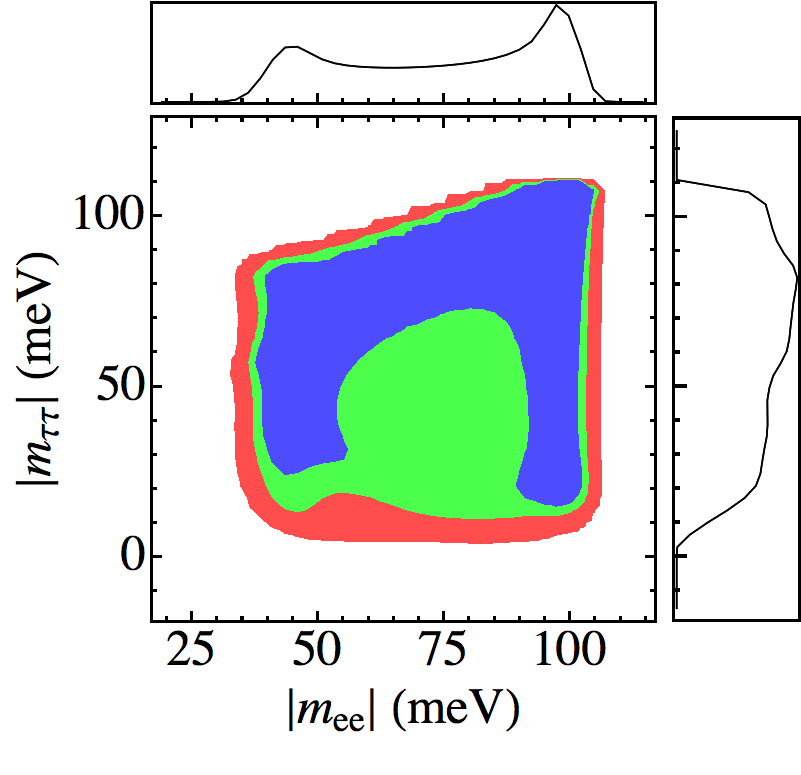}
 \includegraphics[width=0.28\textwidth]{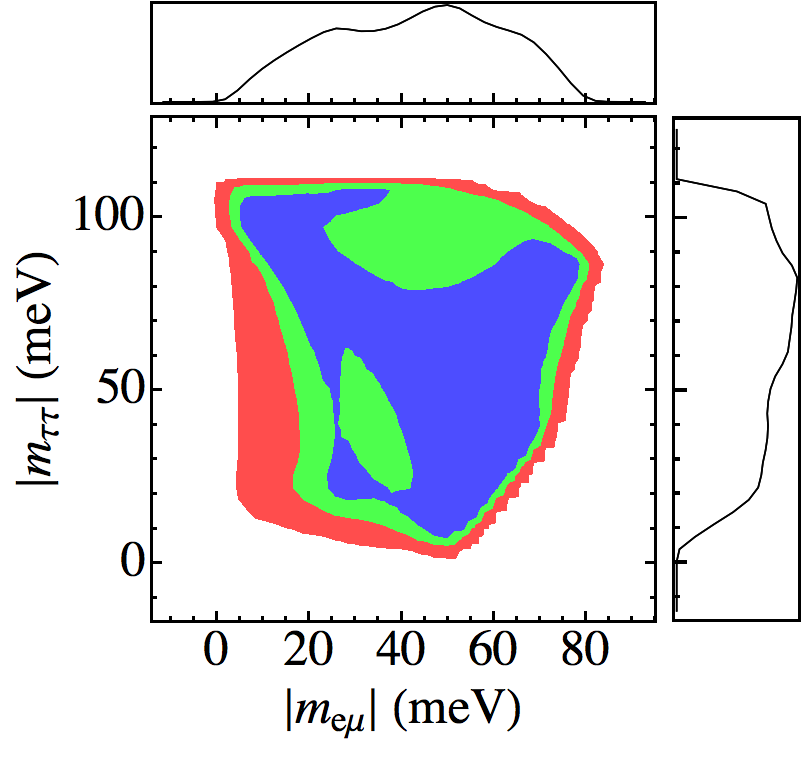}
 \includegraphics[width=0.28\textwidth]{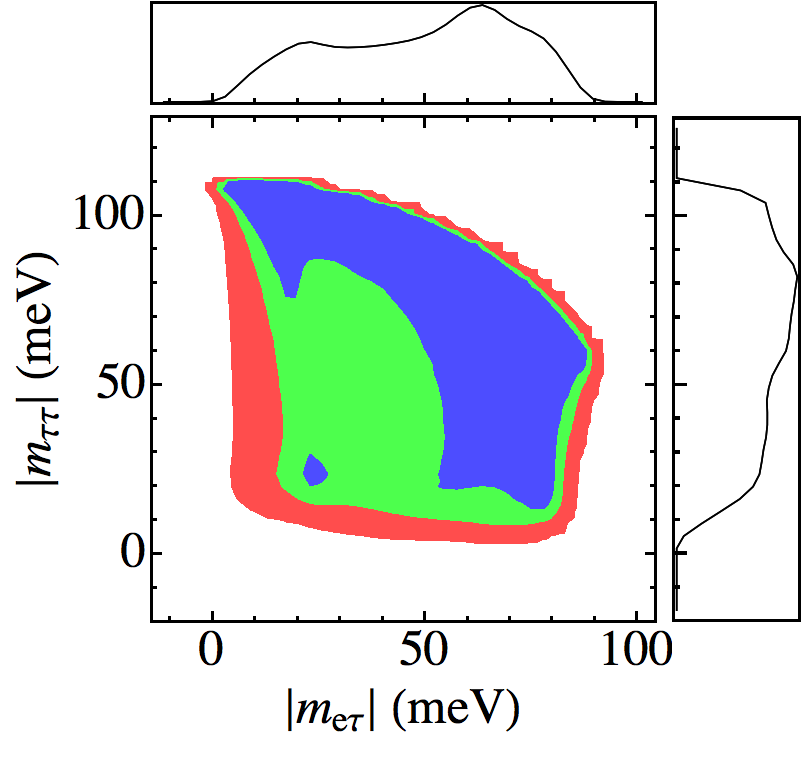}
 \includegraphics[width=0.28\textwidth]{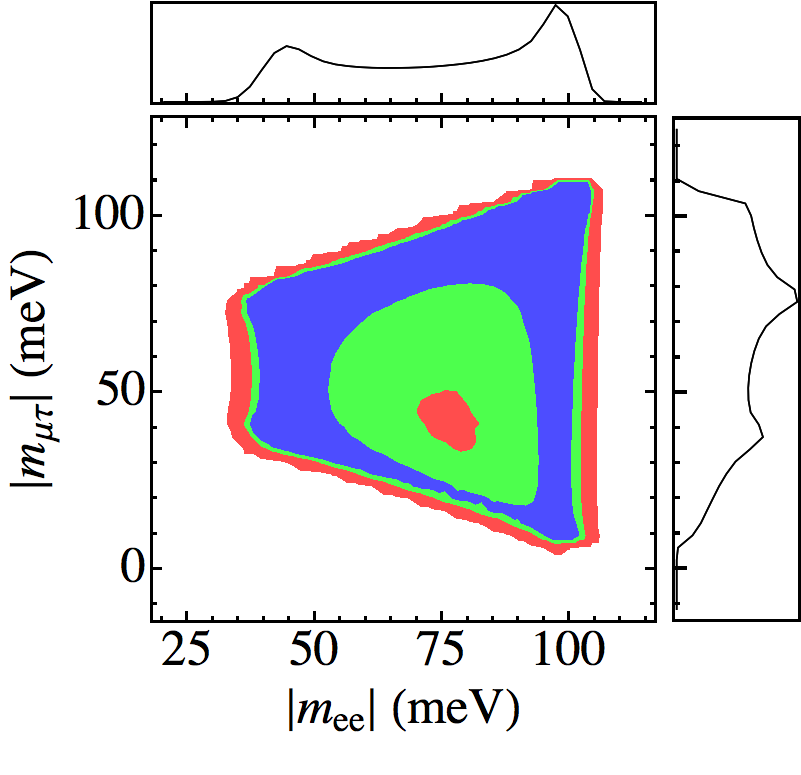}
 \includegraphics[width=0.28\textwidth]{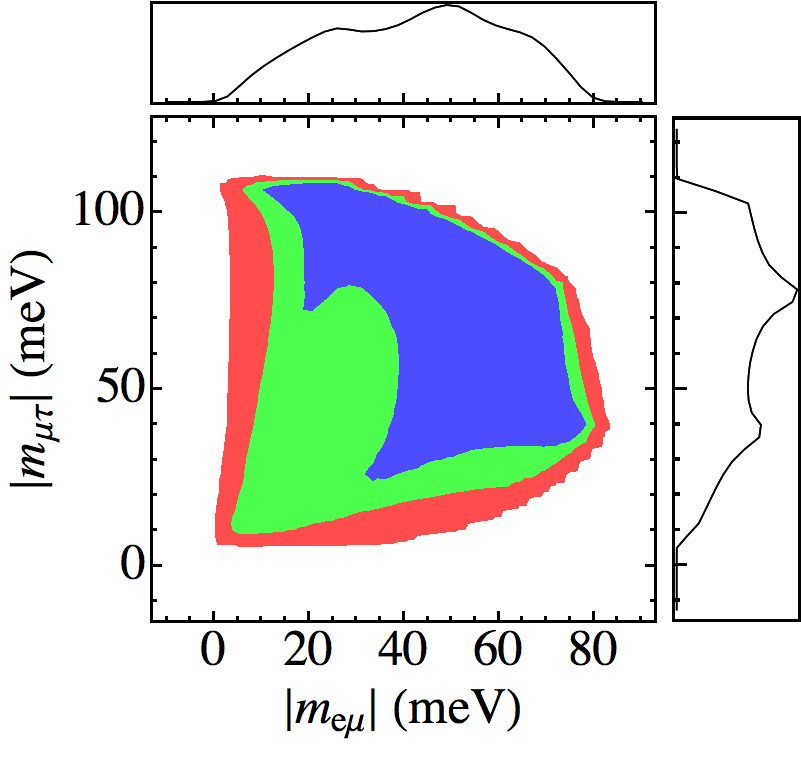}
 \includegraphics[width=0.28\textwidth]{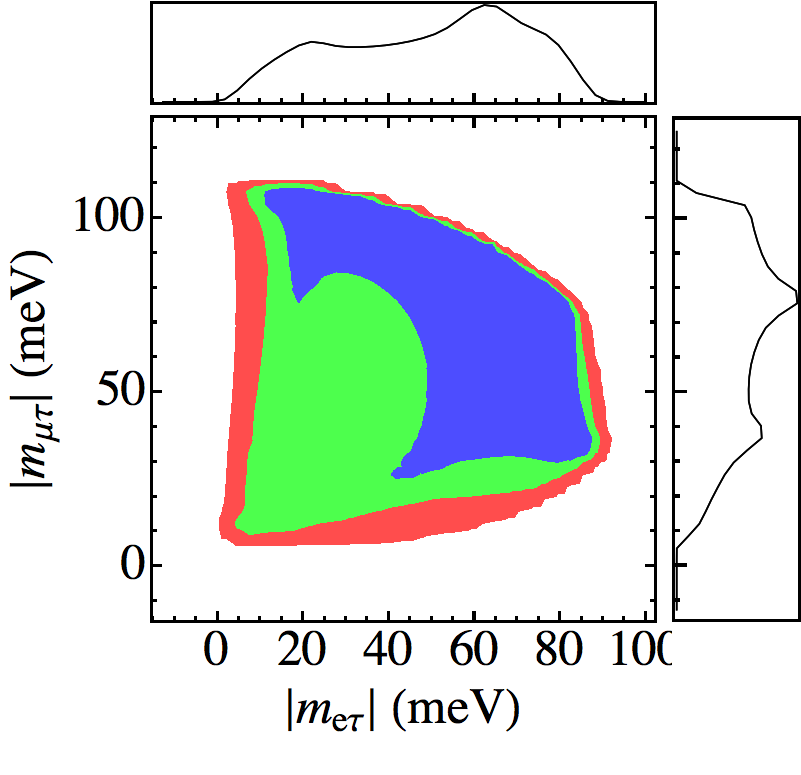}
 \includegraphics[width=0.28\textwidth]{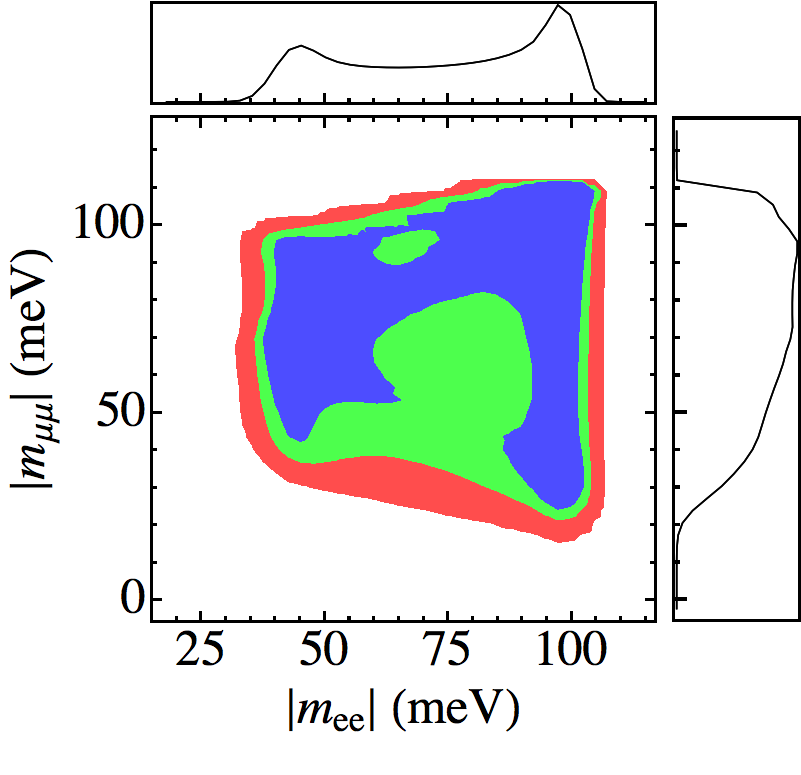}
 \includegraphics[width=0.28\textwidth]{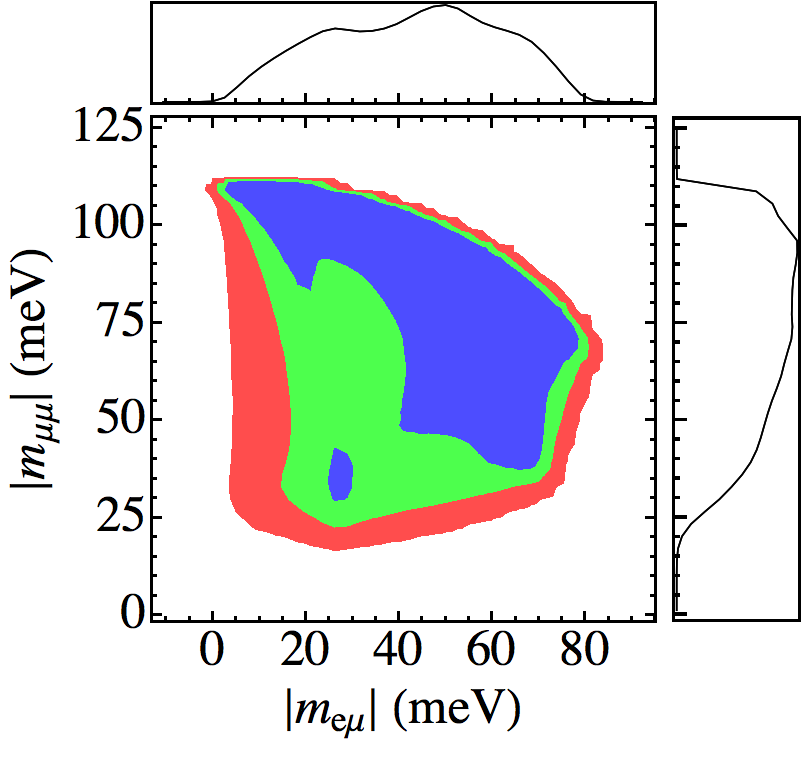}
 \includegraphics[width=0.28\textwidth]{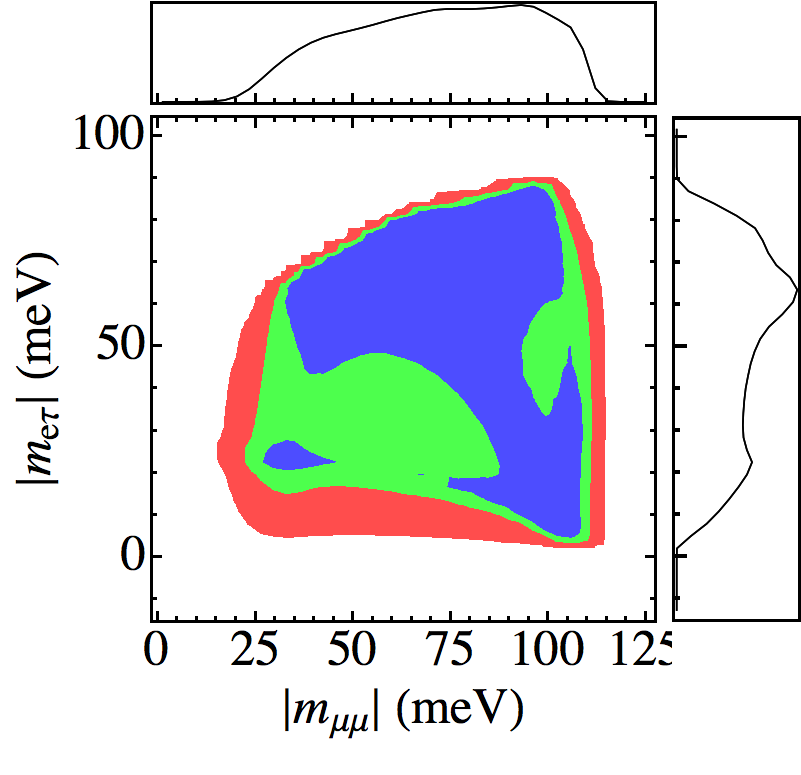}
 \includegraphics[width=0.28\textwidth]{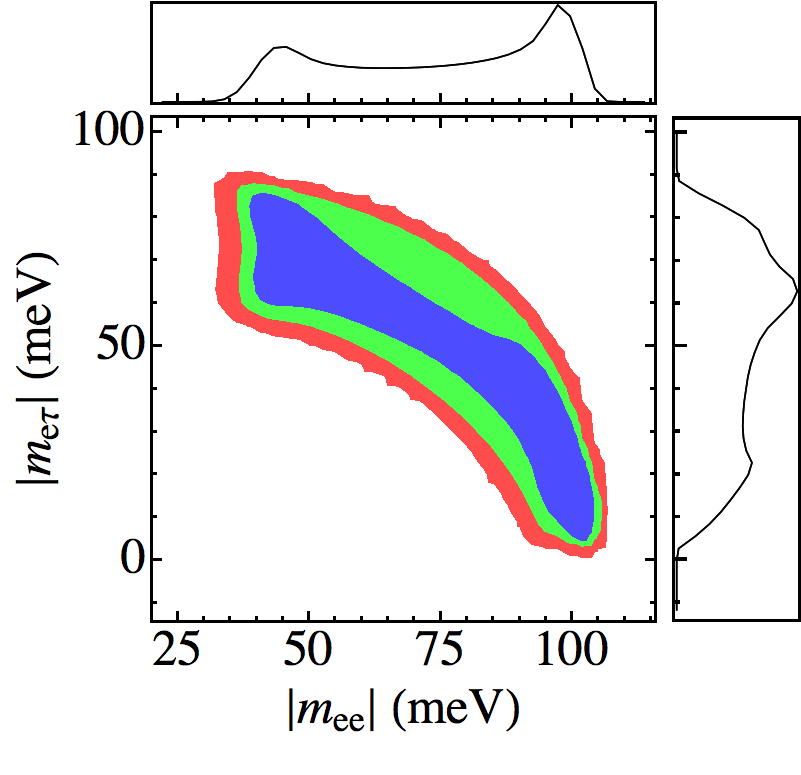}
 \includegraphics[width=0.28\textwidth]{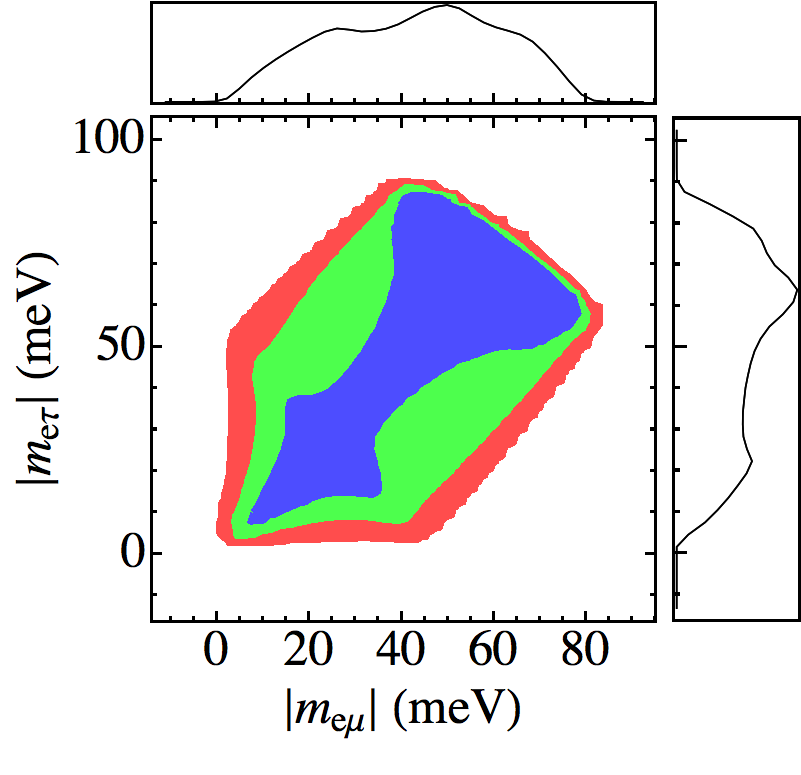}
 \includegraphics[width=0.28\textwidth]{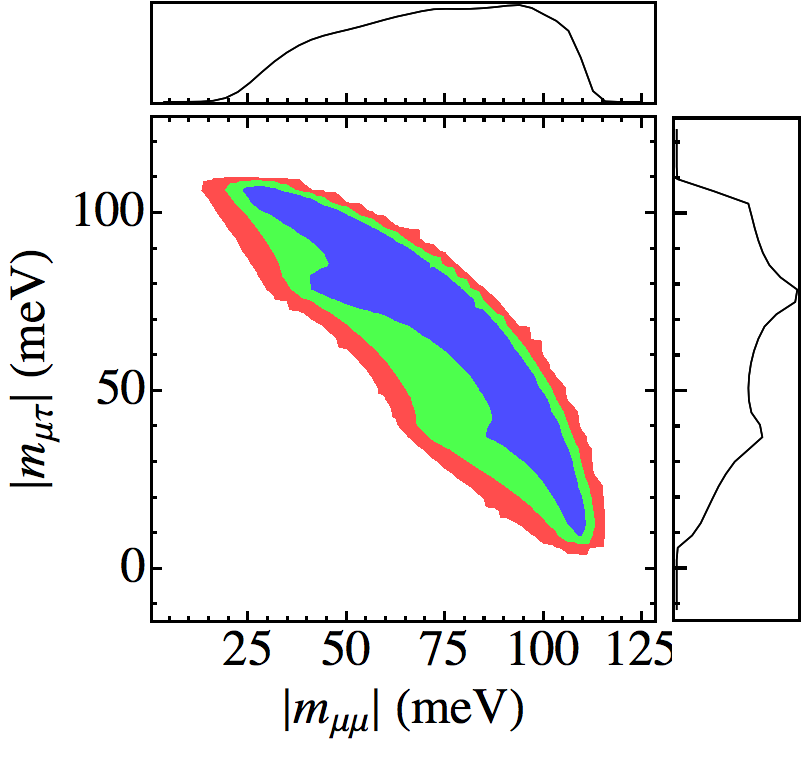}
 \includegraphics[width=0.28\textwidth]{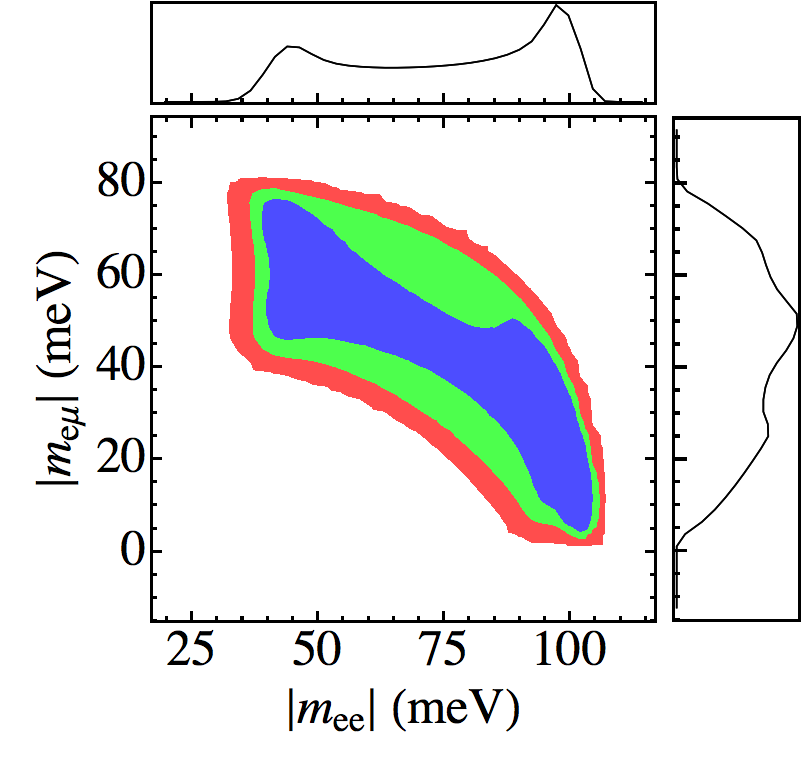}
 \includegraphics[width=0.28\textwidth]{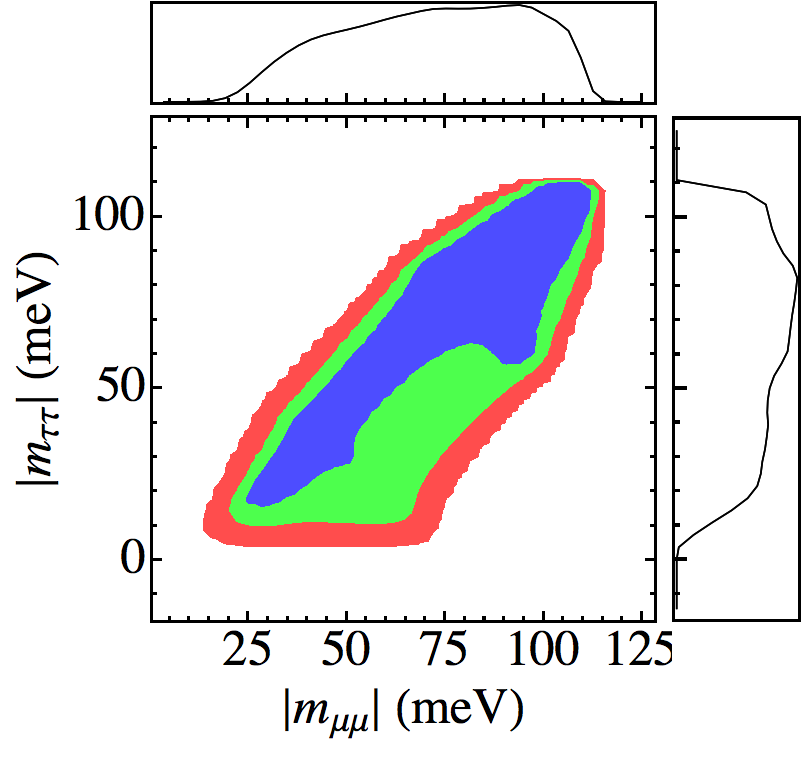}
 \includegraphics[width=0.28\textwidth]{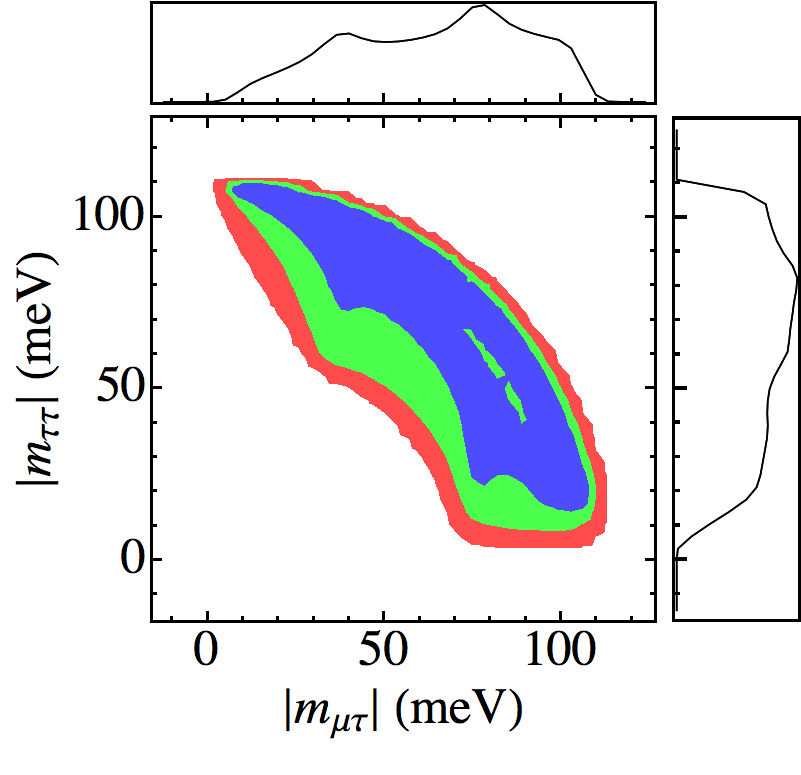}
\vspace{-2mm}
\end{center}
\vspace{-0.1cm}
\caption{Same as Fig.\ref{fig:nh-1st-m01} but for $\theta_{23}$ in the second 
octant.}
\label{fig:nh-2nd-m01}
\end{figure}

\end{document}